\documentclass[10pt,twocolumn,twoside]{IEEEtran} 
\usepackage{multirow}
\usepackage{graphicx}
\usepackage{amsmath}
\usepackage[psamsfonts]{amssymb}
\usepackage{amsxtra}
\usepackage{threeparttable}
\usepackage{subfigure}
\usepackage{caption}
\usepackage{epstopdf}
\usepackage{stfloats}
\usepackage{algorithm}
\usepackage{algorithmicx}
\usepackage{algpseudocode}

\ifCLASSINFOpdf
\else
\fi
\begin{document}
%
\title{Halftone Image Watermarking by Content Aware Double-sided Embedding Error Diffusion}
%
%
%

\author{Yuanfang Guo,~\IEEEmembership{Member,~IEEE,}
        Oscar C. Au,~\IEEEmembership{Fellow,~IEEE,}
        Rui Wang,~\IEEEmembership{Member,~IEEE,}
        Lu Fang,~\IEEEmembership{Member,~IEEE,}
        Xiaochun Cao,~\IEEEmembership{Senior Member,~IEEE.}
\thanks{Yuanfang Guo, Rui Wang and Xiaochun Cao are with the State Key Laboratory of Information Security, Institute of Information Engineering, Chinese Academy of Sciences, Beijing 100093, China (email: eeandyguo@connect.ust.hk, wangrui@iie.ac.cn, caoxiaochun@iie.ac.cn).

Oscar C. Au and Lu Fang were/are with the Department of Electronic and Computer Engineering, The Hong Kong University of Science and Technology, Hong Kong (email: eeau90@gmail.com, fanglu922@gmail.com).

Xiaochun Cao is also with the University of Chinese Academy of Sciences.

}
}
\maketitle

\begin{abstract}


In this paper, we carry out a performance analysis from a probabilistic perspective to introduce the EDHVW methods' expected performances and limitations. Then, we propose a new general error diffusion based halftone visual watermarking (EDHVW) method, Content aware Double-sided Embedding Error Diffusion (CaDEED), via considering the expected watermark decoding performance with specific content of the cover images and watermark, different noise tolerance abilities of various cover image content and the different importance levels of every pixel (when being perceived) in the secret pattern (watermark). To demonstrate the effectiveness of CaDEED, we propose CaDEED with expectation constraint (CaDEED-EC) and CaDEED-NVF\&IF (CaDEED-N\&I). Specifically, we build CaDEED-EC by only considering the expected performances of specific cover images and watermark. By adopting the noise visibility function (NVF) and proposing the importance factor (IF) to assign weights to every embedding location and watermark pixel, respectively, we build the specific method CaDEED-N\&I. In the experiments, we select the optimal parameters for NVF and IF via extensive experiments. In both the numerical and visual comparisons, the experimental results demonstrate the superiority of our proposed work.
\end{abstract}


%
\IEEEpeerreviewmaketitle

%
%


\section{Introduction}\label{intro}

Printed material, such as newspapers, books, magazines etc., has been widely employed in the distribution of multimedia content for multiple decades. Although distributing multimedia content in digital versions has been very popular in recent years, printed material is still playing an important role. Due to the explosive usage of digital and printed multimedia content, multimedia security issues have arisen quickly in recent years. In general, to protect a multimedia content, not only should the digital versions of the multimedia content be protected, but also the printed versions.
Since most of the techniques, which protects the digital versions, cannot be directly applied to the printed versions, halftone image watermarking has been specially developed for protecting the printed versions of multimedia content over the past two decades. Numerous issues still exist and the proposed work is developed for resolving some of the problems.

The paper is organized as follows. Section \ref{introwh} introduces watermarking and halftoning. Section \ref{review} presents a literature review. Section \ref{contributions} briefs the reader about the motivations and contributions of the paper. Section \ref{background} recalls three highly related methods. Section \ref{analysis} presents the performance analysis to the error diffusion based halftone visual watermarking (EDHVW) methods from a probabilistic perspective. Section \ref{method} proposes the new method. Section \ref{results} gives the experimental results. Finally, the conclusion is drawn in Section \ref{conclusion}.

\subsection{Introduction of Watermarking and Halftoning}\label{introwh}

First of all, watermarking and halftoning (and the halftone image) will be introduced respectively.

In watermarking, a message called a watermark is embedded into another message called the cover message to generate another message called a stego message, which is similar to the cover message. The watermark can typically be decoded by processing the stego message. The message may be an image, video, audio, speech or other media content. This paper is focused on image watermarking. In most applications, image watermarks are invisible in the stego images, though they can be visible sometimes. Some watermarks are designed to be fragile [\ref{sharma2002fw}-\ref{cao2015rdhei}], such that they are easily broken when edited, as is typically required for authentication. Others are designed to be robust [\ref{ward2011rwmgdq}-\ref{zong2015rhsw}], such that they can not be easily removed when edited, as is typically required in copyright protection. Still more are designed for data hiding with no specific fragileness or robustness requirements, as in steganography. Further watermarks are designed to be private where the original cover message is needed during watermark decoding, while some are public as the original cover message is not needed in watermark decoding.

Halftoning is a special image processing technique to print greyscale images on printed material. Usually only two tones are available in printed material: white from the color of the paper and black from the ink. Halftoning uses just 1 bit (black and white) to approximate the original 8-bit (greyscale) image such that, when viewed from a certain distance, the halftone image will resemble the original greyscale image. Nowadays, there are several classes of halftoning methods: ordered dithering [\ref{bayers1973}], error diffusion [\ref{steinberg1976}-\ref{li2006edgeed}], dot diffusion [\ref{knuth1987}-\ref{guo2013ncdd}] and direct binary search [\ref{allebach2002}-\ref{guo2013dbslt}].

\subsection{Previous Work}\label{review}

Most regular watermarking methods for greyscale images, such as Least Significant Bit embedding [\ref{moulin2005dhc}], cannot be effectively performed on halftone images, because of the 1-bit nature of halftone images mentioned above. Therefore, researchers pay various attentions to this special area, halftone image watermarking.


In general, halftone image watermarking methods can be classified into two classes. The Class 1 methods [\ref{au2000dhspt}-\ref{guo2011omes}] embed a secret binary bitstream into a single halftone image (with both the cover and stego images being halftone images) and the embedded bitstream can be extracted later by applying a certain algorithm (with respective to the embedding algorithm) to the stego image. The Class 2 methods [\ref{guo2011dhcdd}-\ref{guo2015wemed}] embed a secret pattern (watermark) into multiple halftone images (with $N$ stego halftone images obtained from $N$ cover halftone images, $N>1$, $N$ usually equals $2$), such that when the stego halftone images are overlaid or the not-exclusive-or operation is performed, the secret pattern will be revealed. As mentioned in [\ref{guo2016hvw}], halftone visual watermarking (HVW) is employed to represent the Class 2 methods. Since HVW methods usually perform embedding during the halftoning process, most of the HVW methods are designed based on different halftoning methods and thus they can be further classified into different categories. Among the HVW methods, error diffusion based HVW (EDHVW) methods drew most of the researchers' attentions due to the fact that error diffusion, which generates halftone images with good visual quality while remains simple implementation procedures, are widely adopted during the past forty years.

Among the EDHVW methods, in 2001, Fu and Au in [\ref{au2001dhsed}] proposed the first method called DHSED (Data Hiding by Stochastic Error Diffusion) to embed a secret pattern in two halftone images generated using stochastic error diffusion. In 2003, Fu and Au proposed another method called Data Hiding by Conjugate Error Diffusion (DHCED) [\ref{au2003dhced}], which imposes conjugate properties in the two stego halftone images according to the watermark pattern and improves the previous poor performance [\ref{au2001dhsed}] significantly. Pei and Guo in [\ref{pei2003nbed}] proposed to shift the quantization threshold, which provided a similar performance compared to [\ref{au2003dhced}]. In 2006, Chang et al. in [\ref{chang2006}] proposed to firstly adjust the dynamic range of the original greyscale image and then employ pattern lookup and pixel swapping to perform embedding. The contrast of the revealed secret pattern was low and the visual quality of the stego images was quite poor compared to the original images, though the details of the embedded halftone image was preserved with this method. In 2008, Yang et al. found that when applying the gradient attacks to the stego images generated by DHCED [\ref{au2003dhced}], some visual boundaries of the secret pattern appeared in the edge map. Thus, they proposed in [\ref{yang2008}] to extend traditional visual cryptography. However, the visual quality of the second generated stego image dropped obviously compared to their first generated image. Also, the contrast of the revealed secret pattern in [\ref{yang2008}] was much lower than that in [\ref{au2003dhced}], which makes it inconvenient for the users to distinguish the secret pattern from the background. In 2011, Guo and Tsai [\ref{guo2011anbed}] improves the method in [\ref{pei2003nbed}] by adaptively shifting the quantization threshold. Unfortunately, their performance is highly depending on the lookup table training results. Meanwhile, we analyzed the advantages and weaknesses of the approach in [\ref{au2003dhced}] and then proposed Data Hiding by Dual Conjugate Error Diffusion (DHDCED) in [\ref{guo2011dhdced}], which makes amendments to both stego halftone images rather than only the second halftone image. The experimental results in [\ref{guo2011dhdced}] show that DHDCED improved the performance significantly. Recently, we proposed to tackle the HVW problems from a theoretical approach in [\ref{guo2016hvw}] by proposing a general formulation for HVW problems. Then the general formulation is applied to two specific EDHVW problems by proposing two general methods, named Single-sided Embedding Error Diffusion (SEED) and Double-sided Embedding Error Diffusion (DEED). If the L-2 Norm is selected in DEED, DEED(L-2) can achieve better performances compared to previous methods. Still, DEED possesses weaknesses. It actually assumes that the identical amount of embedding distortions for different pixel locations are equally perceived and every pixel in the secret pattern is equally important.

\subsection{Motivations and Contributions}\label{contributions}

Although many EDHVW methods are developed in the past years as introduced in Section \ref{review}, no general performance analysis has been performed except the specific analysis to DHSED and DHCED in [\ref{au2003dhced}]. Inspired by [\ref{au2003dhced}], we give a general analysis to the typical EDHVW methods in this paper. According to our analysis, the performance of EDHVW methods are limited by the cover image content. More embedding distortions, which are less demanded intuitively, are likely required when the EDHVW methods embed the secret patterns into the bright/dark image regions to achieve a high correct decoding rate.

Besides of the problem above, the assumption of DEED also exists in most of the previous methods, while in reality, identical distortions may be perceived quite differently due to the different cover image content. Also, the watermark pixels, which contains more structural information, tends to be more important than others when considering the effect of human perception in the HVW decoding procedure, i.e. the structural information tends to be more sensitive to the human eyes intuitively. If identical certain amount of watermark decoding distortions is considered, distortions to the structural pixels (especially for the pixels in the fine texture regions) may significantly degrade the human perceived watermark information, while the distortions at the flat regions usually affect less when human perceives the decoded image. Under such circumstances, we propose a new general EDHVW method called Content aware Double-sided Embedding Error Diffusion (CaDEED). The detailed contributions are as follows:

\begin{description}
  \item[1:] ~ We theoretically analyze the expected performances and limitations of the EDHVW methods from a probabilistic perspective. The analysis indicates that the performances of EDHVW methods are highly related to the content of the cover images.
  \item[2:] ~ According to the analysis, we propose to formulate the watermark decoding distortions, i.e. the differences between the extracted and reference watermarks, with a linear combination of two models. One is the traditional difference between the decoded and original watermark. The other is our proposed model, the difference between the decoded and expected watermark calculated from the cover images' content and the original watermark.
  \item[3:] ~ With the new model of the watermark decoding distortions, we propose a new general EDHVW method called CaDEED by further assigning different weights to the embedding distortions according to the different cover image content and the differences between the reference and revealed secret pattern. With the problem formulation of CaDEED, we solve the optimization problem to obtain a relaxed optimal solution.
  \item[4:] ~ To better demonstrate CaDEED, we propose two specific schemes, CaDEED with expectation constraint (CaDEED-EC) and CaDEED-NVF\&IF (CaDEED-N\&I). CaDEED-EC only considers the expected performances with specific content of the cover images and watermark. CaDEED-N\&I adopts the classical noise visibility function (NVF) [\ref{volo2000nvf}] to calculate the weights for the embedding distortions. We also propose a simple yet effective method to calculate the weights for the watermark decoding distortions, i.e. the importance factors (IF) for each pixel in the secret pattern.
  \item[5:] ~ In the experiments, the parameters in CaDEED-N\&I are firstly selected via extensive experiments. Then we propose a new measurement which considers the watermark content and better quantifies the differences between the proposed method and previous methods. For better measuring the performances of the EDHVW methods, we also modify the traditional image distortion measure Peak Signal-to-Noise Ratio (PSNR) via considering the cover image content. By employing both the traditional and new measurements, we demonstrate that our proposed work outperforms both classical and latest EDHVW methods obviously.
\end{description}

\section{Backgrounds}\label{background}

In this section, three previous important EDHVW methods, DHCED, DHDCED and DEED, will be introduced respectively.

Here some common notations, which will be employed throughout this paper, is introduced. Let $X_1$ and $X_2$ be the original grey-scale images, which can be identical or different. Let $Y_1$ and $Y_2$ be the generated stego halftone images, $W$ be the secret binary pattern to be embedded, $W_w$ be the collection of locations of the white pixels in $W$, and $W_b$ be the collection of locations of the black pixels in $W$.
Let $\otimes$ represent the binary AND operation, and let $\odot$ be the binary not-exclusive-or (XNOR) operation.

The goal of DHCED, DHDCED and DEED is to generate $Y_1$ and $Y_2$ from $X_1$ and $X_2$, such that, when $Y_1$ and $Y_2$ are overlaid (performed XNOR operation between them) to reveal $Y_1 \otimes Y_2$, it will resemble $W$.

\subsection{DHCED}

\begin{figure}
\begin{center}
\includegraphics[width=70mm]{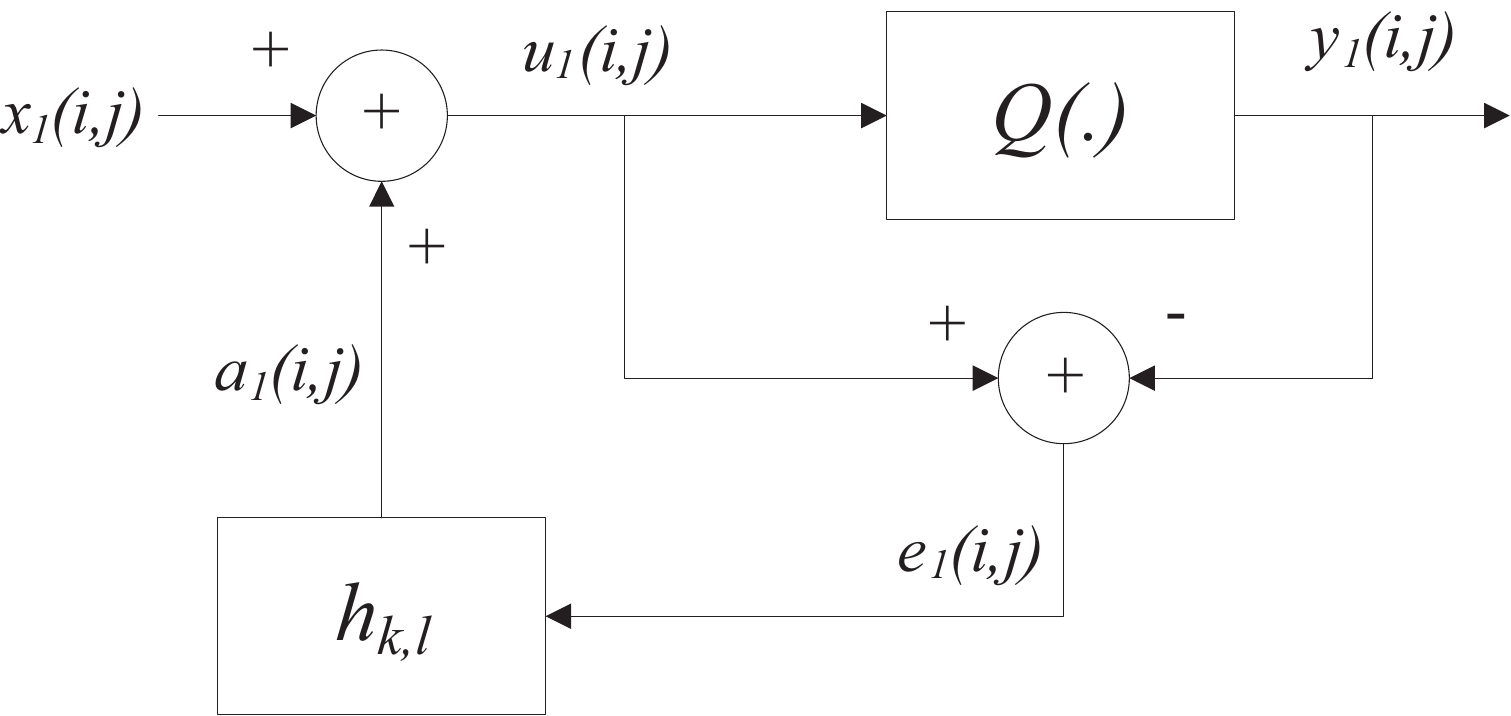}
\captionsetup{justification=centering}
\caption{Error Diffusion system. \label{fig:redprocess}}
\end{center}
\vspace{-0.5cm}
\end{figure}

In DHCED, $Y_1$ is generated by performing regular error diffusion on $X_1$, as shown in Fig. \ref{fig:redprocess} and the regular error diffusion process is described by Eqs. \ref{eq:ed1}-\ref{eq:ed3}.
\begin{equation} \label{eq:ed1}
    u_{1}(i,j)=x_{1}(i,j)+\sum h(k,l)\times e_{1}(i-k,j-l),
\end{equation}
\begin{eqnarray} \label{eq:ed2}
y_{1}(i,j) = \left\{
  \begin{array}{ll}
    0, & \hbox{ $u_{1}(i,j)$ $<$ $128$} \\
    255, & \hbox{ $u_{1}(i,j)$ $\geq$ $128$}
  \end{array}
\right.
\end{eqnarray}
\begin{equation} \label{eq:ed3}
    e_{1}(i,j)=u_{1}(i,j)-y_{1}(i,j),
\end{equation}
where $u_1(i,j)$ is the sum of the current pixel value $x_1(i,j)$ and the past error (diffused from past pixels ) to be carried by the current pixel, $h(k,l)$ is the error diffusion kernel and $e_1(i,j)$ is the error generated when processing the current pixel. Note that the to-be-diffused error $e_1(i,j)$, i.e. the quantization error, is defined as the difference between $u_1(i,j)$ and $y_1(i,j)$. Two common error diffusion kernels are the Steinberg kernel [\ref{steinberg1976}] and Jarvis kernel [\ref{jarvis1976}].

\begin{figure}
\begin{center}
\includegraphics[width=70mm]{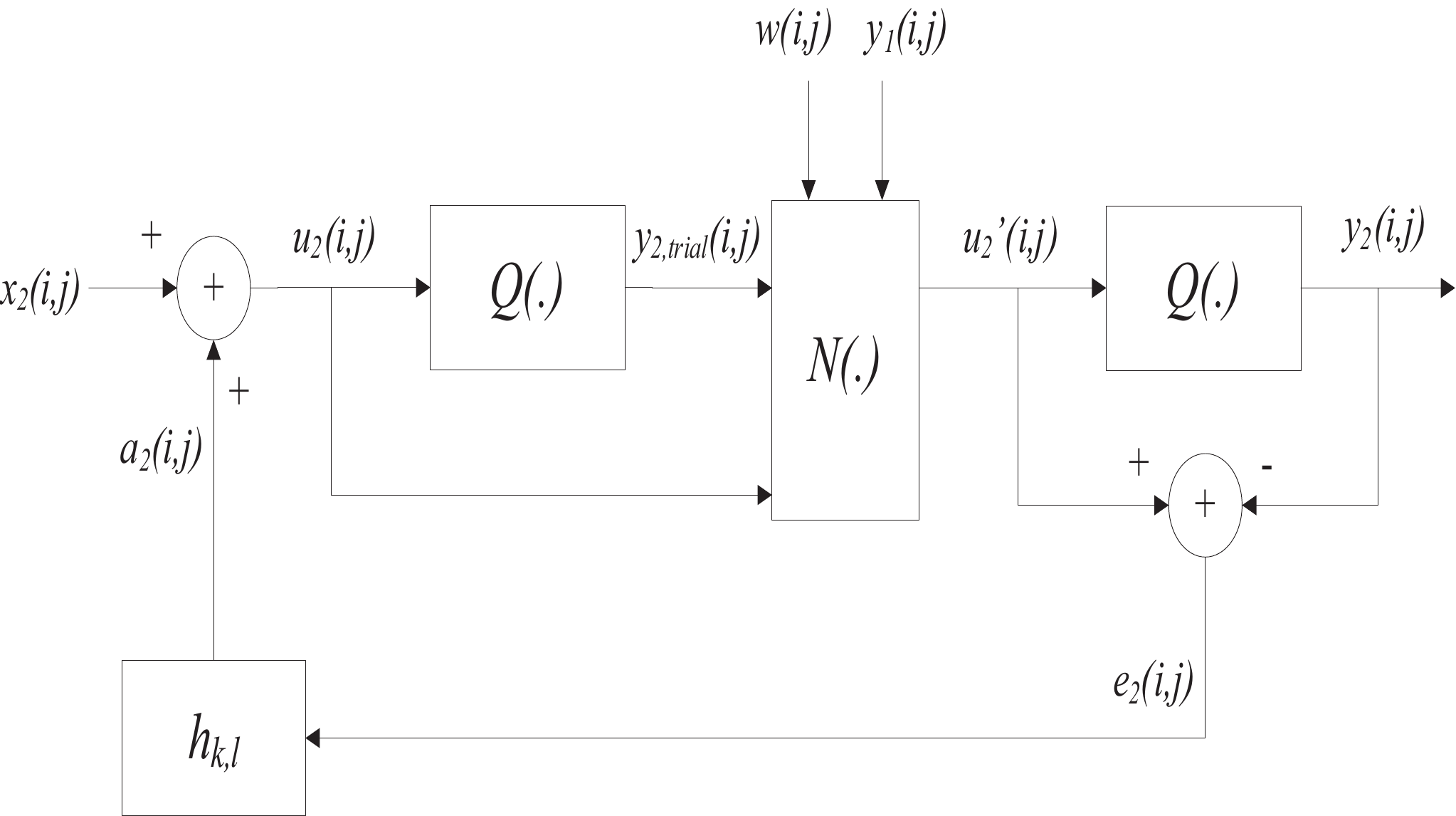}
\caption{DHCED system. \label{fig:dhcedprocess}}
\end{center}
\vspace{-0.5cm}
\end{figure}

After obtaining $Y_1$, the second halftone image $Y_2$ is generated by applying the DHCED system referencing to $X_2$, $Y_1$ and $W$, and the DHCED system is shown in Fig. \ref{fig:dhcedprocess}. For $(i,j)\in W_b$, $y_2(i,j)$ will be `favored' to be conjugate to $y_1(i,j)$, and the `favor' mechanism will be explained later. For $(i,j)\in W_w$, if $X_1=X_2$, then $y_2(i,j)=y_1(i,j)$ will be forced to be carried out. If $X_1\neq X_2$, then $y_2(i,j)$ will be favored to be identical to $y_1(i,j)$.

The `favor' mechanism works as follows. The DHCED system will perform a trial quantization on $u_2(i,j)$. Then if $y_2(i,j)\neq y_{1}(i,j)\oplus \overline{w(i,j)}$, where $y_{1}(i,j)\oplus \overline{w(i,j)}$ is the favored value of $y_2(i,j)$, the minimum distortion $\Delta u(i,j)$ to toggle the current pixel is calculated. If the minimum distortion is acceptable, i.e. $\Delta u(i,j)\leq T$, then the toggling will be performed. The threshold $T$ controls the tradeoff between the contrast of the revealed secret pattern and the visual quality of the stego halftone image $Y_2$. As $T$ increases, the contrast of the revealed secret pattern increases and the visual quality of the stego halftone image $Y_2$ decreases.

\subsection{DHDCED}

Although DHCED gives good performance, it still possesses weaknesses. A certain type of artifacts called boundary artifact will appear in $Y_2$, and are mainly located in the flat regions at the bottom and right boundaries of the locations where $(i,j)$ is co-located in $W_b$ when $X_1=X_2$. If the Sobel filter is applied to $Y_2$, the boundary artifacts will be more obvious on the edge map obtained.

\begin{figure}
\begin{center}
\includegraphics[width=70mm]{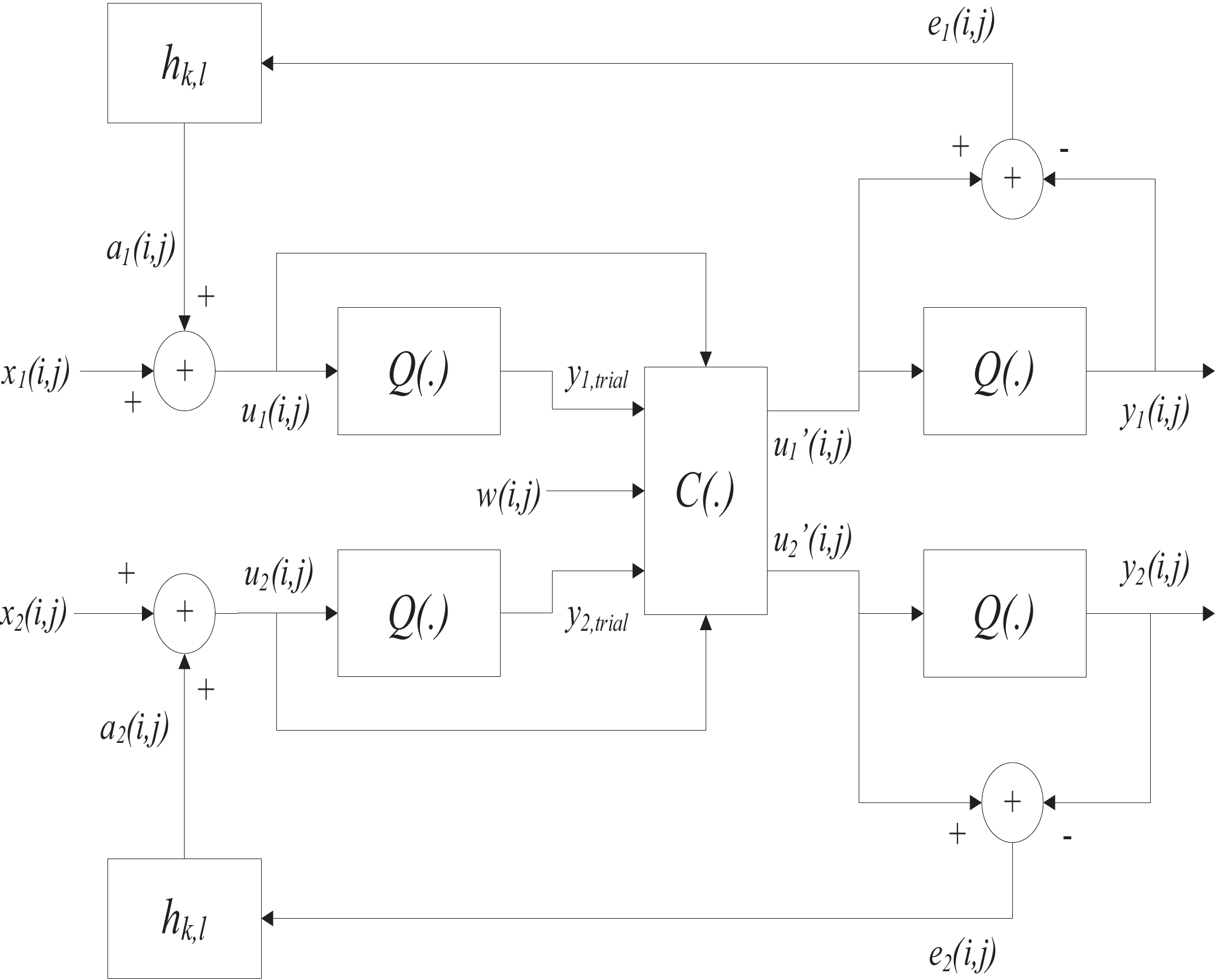}
\caption{DHDCED system. \label{fig:dhdcedprocess}}
\end{center}
\vspace{-0.5cm}
\end{figure}

To reduce the boundary artifacts and improve the performance of the watermarking technique, we proposed DHDCED which perform amendments when generating both $Y_1$ and $Y_2$.
The DHDCED system is shown in Fig. \ref{fig:dhdcedprocess}.

In DHDCED, $Y_1$ and $Y_2$ are generated simultaneously. For $(i,j)\in W_b/W_w$, DHDCED will firstly perform trial quantization on $u_{1}(i,j)$ and $u_2(i,j)$. Then two minimum distortions will be computed according to the strategies below.

\begin{description}
  \item[Strategy 1:] ~\\ Obtain $\Delta u_2(i,j)$ when $y_2(i,j)$ is favored to be conjugate/identical to $y_1(i,j)$.
  \item[Strategy 2:] ~\\ Obtain $\Delta u_1(i,j)$ when $y_1(i,j)$ is favored to be conjugate/identical to $y_2(i,j)$.
\end{description}

After $\Delta u_1(i,j)$ and $\Delta u_2(i,j)$ have been calculated, the strategy causing the smaller distortion is selected. Similar to DHCED, $T$ is also exploited during the calculations of $\Delta u_1(i,j)$ and $\Delta u_2(i,j)$ to control the distortion to be acceptable by the user. Note that when $X_1=X_2$, DHDCED will force $y_1(i,j)$ and $y_2(i,j)$ to be identical.

\subsection{DEED}

In [\ref{guo2016hvw}], we proposed a general formulation for general HVW problems as Eq. \ref{eq:genform} shows.
\begin{equation} \label{eq:genform}
    \underset{}{\operatorname{min}}\text{ } D_h+\lambda*D_w,
\end{equation}
where $D_h$ stands for the distortion caused during the watermark embedding process, and $D_w$ stands for the difference between the decoded watermark and the original watermark, $\lambda\leq0$.

After Eq. \ref{eq:genform} was proposed, two specific EDHVW problems were solved by applying this general formulation. The two problems will both generate and embed a secret pattern (watermark) $W$ into two halftone images, $Y_1$ and $Y_2$, such that when the stego images are overlaid or an XNOR operation is carried out between them, the secret pattern is revealed. In these two problems, one only embeds the secret pattern into one out of two cover images (solved by the general method SEED), while the other embeds the secret pattern into both cover images (solved by the general method DEED). According to the results in [\ref{guo2016hvw}], DEED gives obviously better results compared to SEED, thus only DEED will be briefly introduced here due to limited space.

In DEED, to perform the necessary toggling to change the output halftone value, a distortion $\Delta u_{1,i}$ is added to $x_{1,i}$ and a distortion $\Delta u_{2,i}$ is added to $x_{2,i}$ to change the pixel values before quantization. By letting $\Delta U_1$ and $\Delta U_2$ be the distortions to be added to $X_1$ and $X_2$ during the watermark embedding process and $ED(.)$ be the regular error diffusion process, DEED can be formulated as Eq. \ref{eq:deed1} shows.
\begin{align} \label{eq:deed1}
   \underset{\mathbf{\Delta U_1}, \mathbf{\Delta U_2}}{\operatorname{min}}\text{ } & ||\mathbf{\Delta U_1}||_p^p+||\mathbf{\Delta U_2}||_p^p+\lambda\cdot||\mathbf{W} \nonumber \\
   & -(ED(\mathbf{X_1}+\mathbf{\Delta U_1}) \circ ED(\mathbf{X_2}+\mathbf{\Delta U_2}))||_p^p,
\end{align}
where `$\circ$' stands for the decoding operation which can be either the AND operation ($\otimes$) or XNOR operation ($\odot$).

To our best knowledge, Eq. \ref{eq:deed1} can not be solved directly. Even brute force will have problems since as the image size goes up the possible solution set increases exponentially. However, this optimization problem can be relaxed and then solved accordingly. If we assume the regular error diffusion processing order is from the first pixel $x_{1,1}$($x_{2,1}$) to the last pixel $x_{1,N}$($x_{2,N}$), where $N$ is the number of pixels in $X_1$($X_2$), and $Y_1$ and $Y_2$ are generated simultaneously, Eq. \ref{eq:deed1} can be reformed and relaxed to Eqs. \ref{eq:deed2} and \ref{eq:deed3}.
\begin{align} \label{eq:deed2}
    J_1= & \underset{\Delta u_{1,1}, \Delta u_{2,1}}{\operatorname{min}}\text{ }\{
    |\Delta u_{1,1}|^p+|\Delta u_{2,1}|^p \nonumber \\
    & +\lambda\cdot|ed(x_{2,1}+\Delta u_{2,1})\circ ed(x_{1,1}+\Delta u_{1,1})-w_1|^p\},
\end{align}
\begin{align}\label{eq:deed3}
    J_n= & \underset{\Delta u_{1,n}, \Delta u_{2,n}}{\operatorname{min}}\text{ }\{|\Delta u_{1,n}|^p+|\Delta u_{2,n}|^p \nonumber \\
    & +\lambda\cdot|ed(x_{2,n}+\Delta u_{2,n})\circ ed(x_{1,n}+\Delta u_{1,n})-w_n|^p \nonumber \\
    & +J_{n-1}\},
\end{align}
where $n\in\{2,3,...,N\}$, and $J_1$ and $J_n$ are the minimum cost obtained by the optimization process.

Then the relaxed optimization problem can be solved easily by calculating the best $\Delta u_{1,i}$ and $\Delta u_{2,i}$ for the current pixels $x_{1,i}$ and $x_{2,i}$ and the final stego halftone images $Y_1$ and $Y_2$ can be generated by simply performing regular error diffusion on $X_1+\Delta U_1$ and $X_2+\Delta U_2$ respectively. According to the results in [\ref{guo2016hvw}], when the L-2 Norm is employed in DEED, its performance is superior compared to previous methods. Thus, DEED(L-2), as the latest EDHVW approach, will be compared in the experiments in the latter section.

\section{Performance Analysis of EDHVW Methods}\label{analysis}


Based on our past experience, error diffusion based HVW (EDHVW) methods usually perform differently on different cover images. The performance is limited when certain cover images are employed, which is caused by the limited degree of freedom of halftone images. If users still demand a relatively high correct decoding rate, the embedding distortion will be excessive. In this section, we will introduce the expected performances and limitations of EDHVW methods from a probability perspective by generalizing the analysis of DHSED and DHCED in [\ref{au2003dhced}].

During the analysis, $X_1$ and $X_2$ will always be the original grey-scale images. $Y_1$ and $Y_2$ are the two stego halftone images. In this section, the subscriptions $O$ and $N$ stand for AND operation and XNOR operation respectively.

Although EDHVW methods will embed a secret pattern into two or more halftone images, here the situation of two stego halftone images will be considered and introduced, as the analysis can be easily implied to the situations of more than two stego halftone images.

Before analyzing EDHVW methods, we will discuss the regular error diffusion first. In error diffusion, the halftoning technique tends to preserve the local intensity of the original grey-scale image.
Consider a rectangular region $R_1$ around $(i,j)$ in $X_1$ has a constant intensity $A$ and the co-located region $R_2$ around $(i,j)$ in $X_2$ has a constant intensity $B$, where $X_1$ may or may not be identical to $X_2$. Assume the left half of the region is in the white region $W_w$ of $W$. Similarly, assume the right half of the region is in the black region $W_b$ of $W$. If regular error diffusion is applied to $X_1$ to generate $Y_{1,ED}$, the probability distribution of the halftone pixel $y_{1,ED}(i,j)$ is
\begin{equation} \label{eq:eda1}
    P[y_{1,ED}(i,j)=255]=\frac{A}{255},
\end{equation}
\begin{equation} \label{eq:eda2}
    P[y_{1,ED}(i,j)=0]=\frac{255-A}{255}.
\end{equation}

The expectation of $y_{1,ED}(i,j)$ for $(i,j)\in R_1$ is
\begin{align} \label{eq:eda3}
    E[y_{1,ED}(i,j)] &=0\cdot P[y_{1,ED}(i,j)=0] \nonumber \\
                   & \text{ }\text{ }\text{ }+255\cdot P[y_{1,ED}(i,j)=255] \nonumber \\
                   &=0\cdot \frac{255-A}{255} +255\cdot \frac{A}{255} \nonumber \\
                   &=A,
\end{align}

Therefore, the percentages of black and white pixels in $Y_{1,ED}$ in $R_1$ are $A/255\cdot 100\%$ and $(255-A)/255\cdot 100\%$ respectively and the pixels are distributed evenly in $R_1$.

Assume that when a EDHVW method embeds the secret pattern, the distortion during the embedding process is reasonably small, then the average intensity in $R_1$ and $R_2$ could still be approximately $A$ and $B$ respectively. Then the probability distributions of the stego halftone pixels $y_{1}(i,j)$ and $y_{2}(i,j)$ are
\begin{equation} \label{eq:hvw1}
    P[y_{1}(i,j)=255]\approx\frac{A}{255},
\end{equation}
\begin{equation} \label{eq:hvw2}
    P[y_{1}(i,j)=0]\approx\frac{255-A}{255},
\end{equation}
\begin{equation} \label{eq:hvw3}
    P[y_{2}(i,j)=255]\approx\frac{B}{255},
\end{equation}
\begin{equation} \label{eq:hvw4}
    P[y_{2}(i,j)=0]\approx\frac{255-B}{255}.
\end{equation}

The expectations of $y_{1}(i,j)$ for $(i,j)\in R_1$, $y_{2}(i,j)$ for $(i,j)\in R_2$ are
\begin{align} \label{eq:hvw5}
    E[y_{1}(i,j)] &=0\cdot P[y_{1}(i,j)=0] +255\cdot P[y_{1}(i,j)=255] \nonumber \\
                   &\approx0\cdot \frac{255-A}{255} +255\cdot \frac{A}{255} \nonumber \\
                   &=A,
\end{align}
\begin{align} \label{eq:hvw6}
    E[y_{2}(i,j)] &=0\cdot P[y_{2}(i,j)=0] +255\cdot P[y_{2}(i,j)=255] \nonumber \\
                   &\approx0\cdot \frac{255-B}{255} +255\cdot \frac{B}{255} \nonumber \\
                   &=B.
\end{align}

The percentages of black and white pixels in $R_1$ in $Y_{1}$ are approximately $A/255\cdot 100\%$ and $(255-A)/255\cdot 100\%$. The percentages of black and white pixels in $R_2$ in $Y_{2}$ are approximately $B/255\cdot 100\%$ and $(255-B)/255\cdot 100\%$.


Let $Y_{A}$ be the overlay/AND operation decoded image. Then $y_{A}(i,j)$ is $255$ if and only if $y_1(i,j)=y_2(i,j)=255$. Let $Y_{O}$ be the XNOR operation decoded image. If $y_1(i,j)=y_2(i,j)$, $y_{O}(i,j)$ will be $0$. Otherwise $y_{O}(i,j)=255$.

In the left half of $R_1$ and $R_2$, which are in $W_w$, $y_1(i,j)$ and $y_2(i,j)$ are favored to be identical to each other such that $y_1(i,j)$ and $y_2(i,j)$ tend to be black and white at the same time. To investigate the behavior of $y_{A}(i,j)$ and $y_{O}(i,j)$, we consider four different cases separately: (1)$255\geq A>127$, $255\geq B>127$; (2)$255\geq A>127$, $127\geq B\geq0$; (3)$127\geq A\geq0$, $255\geq B>127$; (4)$127\geq A\geq0$, $127\geq B\geq0$.

From cases (1)-(4) we can conclude that Eq. (\ref{eq:wa1})-(\ref{eq:wa4}) can effectively describe the probability distribution for cases (1)-(4), while Eq. (\ref{eq:wa5}) and (\ref{eq:wa6}) describe the expectations of $Y_{A}$ and $Y_{O}$ where $(i,j)\in W_w$.
\begin{align} \label{eq:wa1}
    P[y_{A}(i,j)=255] &=P[y_1(i,j)=255\cap y_2(i,j)=255] \nonumber\\
                    &\approx min(P[y_1(i,j)=255],P[y_2(i,j)=255]) \nonumber\\
                    &=\frac{min(A,B)}{255}
\end{align}
\begin{equation} \label{eq:wa2}
    P[y_{A}(i,j)=0]\approx \frac{255-min(A,B)}{255}
\end{equation}
\begin{align} \label{eq:wa3}
    P[y_{O}(i,j)=255] & =P[(y_1(i,j) \odot y_2(i,j))=255] \nonumber \\
    & \approx \frac{255-|A-B|}{255}
\end{align}
\begin{equation} \label{eq:wa4}
    P[y_{O}(i,j)=0]\approx\frac{|A-B|}{255}
\end{equation}

Then
\begin{align} \label{eq:wa5}
    E[y_{A}(i,j)] &=0\cdot P[y_{A}(i,j)=0] +255\cdot P[y_{A}(i,j)=255] \nonumber \\
                &\approx0\cdot \frac{min(A,B)}{255} +255\cdot \frac{255-min(A,B)}{255} \nonumber \\
                &=255-min(A,B),
\end{align}
\begin{align} \label{eq:wa6}
    E[y_{O}(i,j)] &=0\cdot P[y_{O}(i,j)=0] +255\cdot P[y_{O}(i,j)=255] \nonumber \\
                &\approx0\cdot \frac{|A-B|}{255} +255\cdot \frac{255-|A-B|}{255} \nonumber \\
                &=255-|A-B|.
\end{align}

In the right half of $R_1$ and $R_2$, which are in $W_b$, $y_1(i,j)$ and $y_2(i,j)$ are favored to be conjugate to each other such that $y_1(i,j)$ and $y_2(i,j)$ tend not to be black at the same time. To investigate the behavior of $y_{A}(i,j)$ and $y_{O}(i,j)$, we also consider four different cases: (1)$255\geq A>127$, $255\geq B>127$; (2)($255\geq A>127$, $127\geq B\geq0$ or $127\geq A\geq0$, $255\geq B>127$) and $A+B\leq255$; (3)($255\geq A>127$, $127\geq B\geq0$ or $127\geq A\geq0$, $255\geq B>127$) and $A+B>255$; (4)$127\geq A\geq0$, $127\geq B\geq0$.
After we divide the problem into cases (1)-(4), we can combine the four cases as two cases: (1)$A+B>255$, (2)$A+B\leq255$.

For case(1), there are more white pixels than black pixels in both $R_1$ and $R_2$. The percentage of black pixels in the right half of $R_1$ of $Y_1$ is about $(255-A)/255\cdot 100\%$. The percentage of black pixels in the right half of $R_2$ of $Y_2$ is about $(255-B)/255\cdot 100\%$. Consider the situation that the black pixels in the right half of $R_1$ and the black pixels in the right half of $R_2$ are at different locations since $y_1(i,j)$ and $y_2(i,j)$ tend not to be black at the same time when $(i,j)\in W_b$. Then
\begin{align} \label{eq:ba1}
    P[y_{A}(i,j)=0] &=P[y_1(i,j)=0]+P[y_2(i,j)=0] \nonumber\\
                  &\approx\frac{255-A}{255}+\frac{255-B}{255} \nonumber\\
                  &=\frac{2\cdot255-A-B}{255},
\end{align}
\begin{equation} \label{eq:ba2}
    P[y_{A}(i,j)=255]\approx \frac{A+B-255}{255},
\end{equation}
\begin{equation} \label{eq:ba3}
    P[y_{O}(i,j)=0]=P[y_{A}(i,j)=0]\approx\frac{2\cdot255-A-B}{255},
\end{equation}
\begin{equation} \label{eq:ba4}
    P[y_{O}(i,j)=255]=P[y_{A}(i,j)=255]\approx\frac{A+B-255}{255}.
\end{equation}

Then
\begin{align} \label{eq:ba5}
    E[y_{A}(i,j)] &=0\cdot P[y_{A}(i,j)=0] +255\cdot P[y_{A}(i,j)=255] \nonumber \\
                &\approx0\cdot \frac{2\cdot255-A-B}{255} +255\cdot \frac{A+B-255}{255} \nonumber \\
                &=A+B-255,
\end{align}
\begin{align} \label{eq:ba6}
    E[y_{O}(i,j)] &=0\cdot P[y_{O}(i,j)=0] +255\cdot P[y_{O}(i,j)=255] \nonumber \\
                &\approx0\cdot \frac{2\cdot255-A-B}{255} +255\cdot \frac{A+B-255}{255} \nonumber \\
                &=A+B-255.
\end{align}

Consequently, the contrasts $Cs$ between the left and right halves of $R$ in $Y_A$ and $Y_O$ can be expressed as
\begin{align}\label{eq:contrast1}
    Cs_{A} &=\frac{E[y_{A}(i,j)|(i,j)\in W_w]-E[y_{A}(i,j)|(i,j)\in W_b]}{E[y_{A}(i,j)|(i,j)\in W_w]} \nonumber \\
               &=\frac{255-min(A,B)-(A+B-255)}{255-min(A,B)},
\end{align}
\begin{align}\label{eq:contrast2}
    Cs_{O} &=\frac{E[y_{O}(i,j)|(i,j)\in W_w]-E[y_{O}(i,j)|(i,j)\in W_b]}{E[y_{O}(i,j)|(i,j)\in W_w]} \nonumber \\
               &=\frac{255-|A-B|-(A+B-255)}{255-|A-B|}.
\end{align}

For case(2), there are more black pixels than white pixels in $R_1$ and $R_2$. Consider the situation that the black pixels in the right half of $R_1$ and the black pixels in the right half of $R_2$ are at different locations, except $\frac{255-A}{255}\cdot 100\%+\frac{255-B}{255}\cdot 100\%-1\cdot 100\%$ percent of pixel pairs which can only both be black. Then
\begin{equation} \label{eq:ba19}
    P[y_{A}(i,j)=0]=P[y_1(i,j)=0\cup y_2(i,j)=0]\approx1,
\end{equation}
\begin{equation} \label{eq:ba20}
    P[y_{A}(i,j)=255]=P[y_1(i,j)=255\cap y_2(i,j)=255]\approx0,
\end{equation}
\begin{align} \label{eq:ba21}
    P[y_{O}(i,j)=0] &=P[(y_1(i,j)\odot y_2(i,j))=0] \nonumber\\
                  &\approx1\cdot 100\%-(\frac{255-A}{255}\cdot 100\% \nonumber \\
                  & \text{ }\text{ }\text{ }+\frac{255-B}{255}\cdot 100\%-1\cdot 100\%) \nonumber\\
                  &=\frac{A+B}{255},
\end{align}
\begin{equation} \label{eq:ba22}
    P[y_{O}(i,j)=255]\approx\frac{255-(A+B)}{255}.
\end{equation}

Then
\begin{align} \label{eq:ba23}
    E[y_{A}(i,j)] &=0\cdot P[y_{A}(i,j)=0] +255\cdot P[y_{A}(i,j)=255] \nonumber \\
                &\approx0\cdot 1 +255\cdot 0 \nonumber \\
                &=0,
\end{align}
\begin{align} \label{eq:ba24}
    E[y_{O}(i,j)] &=0\cdot P[y_{O}(i,j)=0] +255\cdot P[y_{O}(i,j)=255] \nonumber \\
                &\approx0\cdot \frac{A+B}{255} +255\cdot \frac{255-(A+B)}{255} \nonumber \\
                &=255-(A+B).
\end{align}

Consequently, the contrasts $Cs$ between the left and right halves of $R$ in $Y_{A}$ and $Y_{O}$ can be expressed as
\begin{align}\label{eq:contrast7}
    Cs_{A} &=\frac{E[y_{A}(i,j)|(i,j)\in W_w]-E[y_{A}(i,j)|(i,j)\in W_b]}{E[y_{A}(i,j)|(i,j)\in W_w]} \nonumber \\
               &=\frac{255-min(A,B)-0}{255-min(A,B)} \nonumber \\
               &=1,
\end{align}
\begin{align}\label{eq:contrast8}
    Cs_{O} &=\frac{E[y_{O}(i,j)|(i,j)\in W_w]-E[y_{O}(i,j)|(i,j)\in W_b]}{E[y_{O}(i,j)|(i,j)\in W_w]} \nonumber \\
               &=\frac{255-|A-B|-(255-(A+B))}{255-|A-B|}.
\end{align}

To summarize, for $(i,j)\in W_w$, the expected values of $y_A(i,j)$ and $y_O(i,j)$ are
\begin{equation}\label{eq:summarize1}
    E[y_{A}(i,j)|(i,j)\in W_w]\approx 255-min(A,B),
\end{equation}
\begin{equation}\label{eq:summarize2}
    E[y_{O}(i,j)|(i,j)\in W_w]\approx 255-|A-B|.
\end{equation}

For $(i,j)\in W_b$, the expected values of $y_{A}(i,j)$ and $y_{O}(i,j)$ are
\begin{align} \label{eq:summarize3}
E[y_{A}(i,j)|(i,j)\in W_b] \approx \left\{
  \begin{array}{ll}
    0, & \hbox{ $A+B\leq255$ } \\
    A+B-255, & \hbox{ $A+B>255$ }
  \end{array}
\right.
\end{align}
\begin{equation}\label{eq:summarize4}
    E[y_{O}(i,j)|(i,j)\in W_b]\approx |(A+B)-255|.
\end{equation}

The contrasts between the left halves and the right halves of $R$, which are generated by AND operation and XNOR operation, are
\begin{eqnarray} \label{eq:summarize5}
Cs_{A} \approx \left\{
  \begin{array}{ll}
    1, & \hbox{ $A+B\leq255$ } \\
    1-\frac{A+B-255}{255-min(A,B)}, & \hbox{ $A+B>255$ }
  \end{array}
\right.
\end{eqnarray}
\begin{equation}\label{eq:summarize6}
    Cs_{O}\approx \frac{255-|A-B|-|(A+B)-255|}{255-|A-B|}.
\end{equation}

Eqs. (\ref{eq:summarize1})-(\ref{eq:summarize6}) present the expected values and contrast for the AND operation and XNOR operation decoded images in an ideal situation.

The above equations give the expected performances and limitations of EDHVW methods, where the expected performances are highly depending on the content of the cover images and watermark. In practice, the specific decoding results may exceed the theoretical limits if certain embedding distortions are allowed.

\section{Content aware Double-sided Embedding Error Diffusion}\label{method}

According to the analysis in Section \ref{analysis}, the performances of the EDHVW methods are affected by the specific content of the cover images and the secret pattern. To achieve a relatively high correct decoding rate, more embedding distortions, which are less desired, are demanded for the dark/bright cover image regions compared to others. Unfortunately, to our best knowledge, the existing methods, such as DEED in [\ref{guo2016hvw}], DHDCED in [\ref{guo2011dhdced}] and DHCED in [\ref{au2003dhced}], never consider this issue. Besides, the majority of the EDHVW methods including the latest method DEED have not considered the effect of human perception when formulating the embedding distortions and the differences between the reference and decoded secret pattern, though it formulate and solve the specific EDHVW problem decently. Therefore, we improve the problem formulation of DEED and propose a new general method, Content aware Double-sided Embedding Error Diffusion (CaDEED) by considering the content of the cover images and the secret pattern. Note that the definition of $D_w$ in Eq. \ref{eq:genform} is amended here from the differences between the original and decoded watermark to the differences between the reference and decoded watermark, for more generalization.

The existing EDHVW methods only measure the difference between the original and decoded secret pattern. They never consider the performance limitations caused by the content of the cover images and the secret pattern. To achieve a high watermark correct decoding rate, they tend to induce more embedding distortions to the dark/bright image regions, which may further increase the risk of artifact appearances (the traces of the embedded secret pattern) and result in more image quality degradations to the stego images. To better formulate the differences between the reference and decoded secret pattern, by considering the expected performances of specific content of the cover images and the secret pattern, we propose to define $D_w$ in a more general approach as Eq. \ref{eq:cadeedep} shows.
\begin{align} \label{eq:cadeedep}
    D_w=||\alpha*|\mathbf{W}-(\mathbf{Y_1} \circ \mathbf{Y_2})|^p+\beta*|\mathbf{EP}-(\mathbf{Y_1} \circ \mathbf{Y_2})|^p||_1,
\end{align}
where $EP=E[(\mathbf{Y_1} \circ \mathbf{Y_2})]$ stands for the expected decoded secret pattern calculated via Eqs. \ref{eq:summarize1}-\ref{eq:summarize4}, $\mathbf{EP}$, $\mathbf{Y_1}$, $\mathbf{Y_2}$ and $\mathbf{W}$ stand for the vector form of $EP$, $Y_1$, $Y_2$, $W$ and each of them contains $N$ elements, and $\alpha$ and $\beta$ are the assigned weights for controlling the tradeoff between the two calculated similarities. When $\beta/\alpha$ increases, the proposed model favors the decoded secret pattern to be more similar to the expected decoded secret pattern and vice versa.

Different content of the cover images and the secret pattern not only gives different performances, but also affects the human perceived results. Previously, DHCED, DHDCED and DEED all treat the distortions caused during the embedding process as equally important for every pixel in the stego images. In fact, according to the human visual system (HVS), different image gives different perceived content if identical distortions are added. In this paper, since we are embedding the secret pattern by amending both $Y_1$ and $Y_2$, we define the weighted distortions caused during the embedding process $D_h$ as Eq. \ref{eq:cadeeddh} shows.
\begin{align} \label{eq:cadeeddh}
    D_h=D_{Y_1}+D_{Y_2} & =||\mathbf{M_1}.*|\mathbf{X_1}-(\mathbf{X_1}+\mathbf{\Delta U_1})|^p||_1 \nonumber \\
    & \text{ }\text{ }\text{ }+||\mathbf{M_2}.*|\mathbf{X_2}-(\mathbf{X_2}+\mathbf{\Delta U_2})|^p||_1 \nonumber \\
    & =||\mathbf{M_1}.*|\mathbf{\Delta U_1}|^p||_1+||\mathbf{M_2}.*|\mathbf{\Delta U_2}|^p||_1,
\end{align}
where $\mathbf{M_1}$, $\mathbf{M_2}$, $\mathbf{X_1}$, $\mathbf{X_2}$, $\mathbf{\Delta U_1}$ and $\mathbf{\Delta U_2}$ stand for the vector form of $M_1$, $M_2$, $X_1$, $X_2$, $\Delta U_1$ and $\Delta U_2$ and each of them contains $N$ elements. Note that $M_1$ and $M_2$ are masks which assign weights to different distortions at different locations, i.e. different distortion tolerance ability according to different image content. $M_1$ and $M_2$ can be calculated via any specific method.

For the secret pattern, researchers used to treat every pixel equally. From our observation, different pixels in the secret pattern tend to have different levels of importance. The pixels, which contain more structural information, tend to be more important than other pixels, because human eyes are more sensitive to the structural information. Besides, if those pixels can not be correctly decoded, users can only perceive partial or even undistinguishable decoded secret pattern. In the worst case, the secret pattern can hardly be recognized correctly. Therefore, in CaDEED, the weighted differences between the reference and decoded secret pattern are defined as Eq. \ref{eq:cadeeddw} shows.
\begin{align} \label{eq:cadeeddw}
    D_w=& ||\mathbf{\Psi}.*(\alpha*|\mathbf{W}-(\mathbf{Y_1} \circ \mathbf{Y_2})|^p \nonumber \\
    & +\beta*|\mathbf{EP}-(\mathbf{Y_1} \circ \mathbf{Y_2})|^p)||_1,
\end{align}
where $\mathbf{\Psi}$ stand for the vector form of $\Psi$ and contains $N$ elements. Note that $\Psi$ assigns weight to different pixels in the secret pattern and it can also be calculated via any specific method.

The two stego halftone images $Y_1$ and $Y_2$ can be generated by Eqs. \ref{eq:cadeedy1} and \ref{eq:cadeedy2} by letting $ED(.)$ be the regular error diffusion which is described in Fig. \ref{fig:redprocess}.
\begin{align} \label{eq:cadeedy1}
    \mathbf{Y_1}=ED(\mathbf{X_1}+\mathbf{\Delta U_1})
\end{align}
\begin{align} \label{eq:cadeedy2}
    \mathbf{Y_2}=ED(\mathbf{X_2}+\mathbf{\Delta U_2})
\end{align}

By substituting Eq. \ref{eq:cadeeddh}, \ref{eq:cadeeddw}, \ref{eq:cadeedy1} and \ref{eq:cadeedy2} into Eq. \ref{eq:genform}, CaDEED can be described as Eq. \ref{eq:cadeed} shows.
\begin{align} \label{eq:cadeed}
   \underset{\mathbf{\Delta U_1}, \mathbf{\Delta U_2}}{\operatorname{min}}\text{ } & ||\mathbf{M_1}.*|\mathbf{\Delta U_1}|^p||_1+||\mathbf{M_2}.*|\mathbf{\Delta U_2}|^p||_1 \nonumber \\
   & +\lambda*||\mathbf{\Psi}.*(\alpha*|\mathbf{W}-(\mathbf{Y_1} \circ \mathbf{Y_2})|^p \nonumber \\
   & +\beta*|\mathbf{EP}-(\mathbf{Y_1} \circ \mathbf{Y_2})|^p)||_1
\end{align}

The solution to Eq. \ref{eq:cadeed} is essentially the global optimum solution of CaDEED's problem. However, to our best knowledge, there is no closed-form solution to Eq. \ref{eq:cadeed} due to the unique property of halftone images and the error diffusion process. Since the size of the possible solutions set increases exponentially as the image size increases, brute force also cannot be applied. Fortunately, by applying a similar optimization method as in [\ref{guo2016hvw}], we can manage to find a relaxed optimal solution to Eq. \ref{eq:cadeed} as follows.

In error diffusion process, there exists a feedback loop. Then, latter generated halftone pixels are depending on former pixels. Thus the distortions to be added to the latter pixels are also depending on previous distortions as shown in Eq. \ref{eq:cadeed2} which is obtained by rewriting Eq. \ref{eq:cadeed} using the elements instead of vectors and considering the dependencies among the elements.
\begin{align} \label{eq:cadeed2}
    & \underset{\Delta u_{1,1},\Delta u_{1,2},...,\Delta u_{1,N},\Delta u_{2,1},\Delta u_{2,2},...,\Delta u_{2,N}}{\operatorname{min}}\text{ }\{|m_{1,1}*|\Delta u_{1,1}|^p| \nonumber \\
    & +|m_{1,2}*|\Delta u_{1,2}|^p|+|m_{1,3}*|\Delta u_{1,3}|^p|+... \nonumber \\
    & +|m_{1,N}*|\Delta u_{1,N}|^p|+|m_{2,1}*|\Delta u_{2,1}|^p|+|m_{2,2}*|\Delta u_{2,2}|^p| \nonumber \\
    & +|m_{2,3}*|\Delta u_{2,3}|^p|+...+|m_{2,N}*|\Delta u_{2,N}|^p| \nonumber \\
    & +\lambda*(|\psi_1*(\alpha*|ed(x_{2,1}+\Delta u_{2,1})\circ ed(x_{1,1}+\Delta u_{1,1}) \nonumber \\
    & -w_1|^p+\beta*|ed(x_{2,1}+\Delta u_{2,1})\circ ed(x_{1,1}+\Delta u_{1,1})-ep_1|^p|) \nonumber \\
    & +|\psi_2*(\alpha*|ed(x_{2,2}+\Delta u_{2,2})\circ ed(x_{1,2}+\Delta u_{1,2})-w_2|^p \nonumber \\
    & +\beta*|ed(x_{2,2}+\Delta u_{2,2})\circ ed(x_{1,2}+\Delta u_{1,2})-ep_2|^p)| \nonumber \\
    & +|\psi_3*(\alpha*|ed(x_{2,3}+\Delta u_{2,3})\circ ed(x_{1,3}+\Delta u_{1,3})-w_3|^p \nonumber \\
    & +\beta*|ed(x_{2,3}+\Delta u_{2,3})\circ ed(x_{1,3}+\Delta u_{1,3})-ep_3|^p)|+... \nonumber \\
    & +|\psi_N*(\alpha*|ed(x_{2,N}+\Delta u_{2,N})\circ ed(x_{1,N}+\Delta u_{1,N})-w_N|^p \nonumber \\
    & +\beta*|ed(x_{2,N}+\Delta u_{2,N})\circ ed(x_{1,N}+\Delta u_{1,N})-ep_N|^p)|)\},
\end{align}
where $\forall i \in \{2, 3, ..., N\}$, $\Delta u_{1,i}$ depends on $\{\Delta u_{1,1}, ..., \Delta u_{1,i-1}\}$, $\forall i \in \{2, 3, ..., N\}$, $\Delta u_{2,i}$ depends on $\{\Delta u_{2,1}, ..., \Delta u_{2,i-1}\}$ and $ed(.)$ stands for carrying out regular error diffusion, as described in Eqs. \ref{eq:ed1}-\ref{eq:ed3}, on the current pixel and diffusing its quantization error to the latter pixels.

If we assume the regular error diffusion processing order is from $1$ to $N$, the joint optimization problem in Eq. \ref{eq:cadeed2} can be reformed into ordered separate optimization problems in Eq. \ref{eq:cadeed2r}.
\begin{align} \label{eq:cadeed2r}
    & \underset{\Delta u_{1,1},\Delta u_{2,1}}{\operatorname{min}}\text{ }\{\underset{\Delta u_{1,2},\Delta u_{2,2}}{\operatorname{min}}\text{ }\{...\text{ }\underset{\Delta u_{1,N},\Delta u_{2,N}}{\operatorname{min}}\text{ }\{|m_{1,1}*|\Delta u_{1,1}|^p| \nonumber \\
    & +|m_{1,2}*|\Delta u_{1,2}|^p|+|m_{1,3}*|\Delta u_{1,3}|^p|+... \nonumber \\
    & +|m_{1,N}*|\Delta u_{1,N}|^p|+|m_{2,1}*|\Delta u_{2,1}|^p|+|m_{2,2}*|\Delta u_{2,2}|^p| \nonumber \\
    & +|m_{2,3}*|\Delta u_{2,3}|^p|+...+|m_{2,N}*|\Delta u_{2,N}|^p| \nonumber \\
    & +\lambda*(|\psi_1*(\alpha*|ed(x_{2,1}+\Delta u_{2,1})\circ ed(x_{1,1}+\Delta u_{1,1}) \nonumber \\
    & -w_1|^p+\beta*|ed(x_{2,1}+\Delta u_{2,1})\circ ed(x_{1,1}+\Delta u_{1,1})-ep_1|^p)| \nonumber \\
    & +|\psi_2*(\alpha*|ed(x_{2,2}+\Delta u_{2,2})\circ ed(x_{1,2}+\Delta u_{1,2})-w_2|^p \nonumber \\
    & +\beta*|ed(x_{2,2}+\Delta u_{2,2})\circ ed(x_{1,2}+\Delta u_{1,2})-ep_2|^p)| \nonumber \\
    & +|\psi_3*(\alpha*|ed(x_{2,3}+\Delta u_{2,3})\circ ed(x_{1,3}+\Delta u_{1,3})-w_3|^p \nonumber \\
    & +\beta*|ed(x_{2,3}+\Delta u_{2,3})\circ ed(x_{1,3}+\Delta u_{1,3})-ep_3|^p)|+... \nonumber \\
    & +|\psi_N*(\alpha*|ed(x_{2,N}+\Delta u_{2,N})\circ ed(x_{1,N}+\Delta u_{1,N})-w_N|^p \nonumber \\
    & +\beta*|ed(x_{2,N}+\Delta u_{2,N})\circ ed(x_{1,N}+\Delta u_{1,N})-ep_N|^p)|) \nonumber \\
    & \}...\}\}
\end{align}

For convenience, $S_i$ in Eq. \ref{eq:costi} is employed to represent the cost calculated when processing the $i$-th pixels of $X_1$ and $X_2$.
\begin{align} \label{eq:costi}
    S_i= & |m_{1,i}*|\Delta u_{1,i}|^p|+|m_{2,i}*|\Delta u_{2,i}|^p| \nonumber \\
    & +\lambda*|\psi_i*(\alpha*|ed(x_{2,i}+\Delta u_{2,i})\circ ed(x_{1,i}+\Delta u_{1,i}) \nonumber \\
    & -w_i|^p+\beta*|ed(x_{2,i}+\Delta u_{2,i})\circ ed(x_{1,i}+\Delta u_{1,i}) \nonumber \\
    &-ep_i|^p)|, \forall i \in \{1, 2, 3, ..., N\},
\end{align}
where $\forall i \in \{2,3,...,N\}, S_i$ depends on $\{S_1,...,S_{i-1}\}$ since $\forall i \in \{2,3,...,N\}, \Delta u_i$ depends on $\{\Delta u_1,..., \Delta u_{i-1}\}$.

Then Eq. \ref{eq:cadeed2r} can be reformed to Eq. \ref{eq:cadeed3} with Eq. \ref{eq:costi}.
\begin{align} \label{eq:cadeed3}
    & \underset{\Delta u_{1,1},\Delta u_{2,1}}{\operatorname{min}}\text{ }\{S_1+ \underset{\Delta u_{1,2},\Delta u_{2,2}}{\operatorname{min}}\text{ }\{
    S_2+\text{...} \nonumber \\
    & +\underset{\Delta u_{1,N-2},\Delta u_{2,N-2}}{\operatorname{min}}\text{ }\{S_{N-2}+\underset{\Delta u_{1,N-1},\Delta u_{2,N-1}}{\operatorname{min}}\text{ }\{S_{N-1} \nonumber \\
    & +\underset{\Delta u_{1,N},\Delta u_{2,N}}{\operatorname{min}}\text{ }\{S_N\}\}\}\text{...}\}\}
\end{align}

Currently, these separate problems' optimization order in Eq. \ref{eq:cadeed3} is from $N$ to $1$, which is inconsistent with the error diffusion processing order. To our best knowledge, due to the existing dependencies among all the variables, Eq. \ref{eq:cadeed3} can only be solved if the optimization order is relaxed to be consistent with the error diffusion processing order.

Since $\Delta u_{1,N}$ and $\Delta u_{2,N}$ depend on all previous variable distortions, Eq. \ref{eq:cadeed3} can be relaxed to Eq. \ref{eq:cadeed4} by relaxing the optimization order between the variable pairs $(\Delta u_{1,N}$, $\Delta u_{2,N})$ and $(\Delta u_{1,N}$, $\Delta u_{2,N})$.
\begin{align} \label{eq:cadeed4}
    & \underset{\Delta u_{1,1},\Delta u_{2,1}}{\operatorname{min}}\text{ }\{S_1+ \underset{\Delta u_{1,2},\Delta u_{2,2}}{\operatorname{min}}\text{ }\{
    S_2+\text{...} \nonumber \\
    & +\underset{\Delta u_{1,N-2},\Delta u_{2,N-2}}{\operatorname{min}}\text{ }\{S_{N-2}+\underset{\Delta u_{1,N-1},\Delta u_{2,N-1}}{\operatorname{min}}\text{ }\{S_{N-1}\} \nonumber \\
    & +\underset{\Delta u_{1,N},\Delta u_{2,N}}{\operatorname{min}}\text{ }\{S_N\}\}\text{...}\}\}
\end{align}

Then similar relaxations can be performed on the remaining joint optimization problem. If we continue to carry out relaxations, Eq. (\ref{eq:cadeed5}) can be finally obtained.
\begin{align} \label{eq:cadeed5}
    & \underset{\Delta u_{1,1},\Delta u_{2,1}}{\operatorname{min}}\text{ }\{S_1\}+ \underset{\Delta u_{1,2},\Delta u_{2,2}}{\operatorname{min}}\text{ }\{
    S_2\}+\text{...} \nonumber \\
    & +\underset{\Delta u_{1,N-2},\Delta u_{2,N-2}}{\operatorname{min}}\text{ }\{S_{N-2}\}+\underset{\Delta u_{1,N-1},\Delta u_{2,N-1}}{\operatorname{min}}\text{ }\{S_{N-1}\} \nonumber \\
    & +\underset{\Delta u_{1,N},\Delta u_{2,N}}{\operatorname{min}}\text{ }\{S_N\}
\end{align}

Then Eq. \ref{eq:cadeed5} can be reformed to Eq. \ref{eq:cadeed6} based on the processing order of regular error diffusion (from $1$ to $N$).
\begin{align} \label{eq:cadeed6}
    & \underset{\Delta u_{1,N},\Delta u_{2,N}}{\operatorname{min}}\text{ }\{\text{... }\underset{\Delta u_{1,2},\Delta u_{2,2}}{\operatorname{min}}\text{ }\{\underset{\Delta u_{1,1},\Delta u_{2,1}}{\operatorname{min}}\text{ }\{
    S_1\}+S_2\}+... \nonumber \\
    & +S_N\}
\end{align}

Then Eq. \ref{eq:cadeedfinal1} and \ref{eq:cadeedfinal2} can be derived from Eqs. \ref{eq:cadeed6} and \ref{eq:costi}, and the variables can be easily solved from $(x_{1,1}$, $x_{2,1})$ to $(x_{1,N}$, $x_{2,N})$ once a pair.
\begin{align}\label{eq:cadeedfinal1}
    C_1= & \underset{\Delta u_{1,1},\Delta u_{2,1}}{\operatorname{min}}\text{ }\{
    |m_{1,1}*|\Delta u_{1,1}|^p|+|m_{2,1}*|\Delta u_{2,1}|^p| \nonumber \\
    & +\lambda*|\psi_1*(\alpha*|ed(x_{2,1}+\Delta u_{2,1})\circ ed(x_{1,1}+\Delta u_{1,1}) \nonumber \\
    & -w_1|^p+\beta*|ed(x_{2,1}+\Delta u_{2,1})\circ ed(x_{1,1}+\Delta u_{1,1}) \nonumber \\
    & -ep_1|^p)|\},
\end{align}
\begin{align}\label{eq:cadeedfinal2}
    C_n= & \underset{\Delta u_{1,n},\Delta u_{2,n}}{\operatorname{min}}\text{ }\{|m_{1,n}*|\Delta u_{1,n}|^p|+|m_{2,n}*|\Delta u_{2,n}|^p| \nonumber \\
    & +\lambda*|\psi_n*(\alpha*|ed(x_{2,n}+\Delta u_{2,n})\circ ed(x_{1,n}+\Delta u_{1,n}) \nonumber \\
    & -w_n|^p+\beta*|ed(x_{2,n}+\Delta u_{2,n})\circ ed(x_{1,n}+\Delta u_{1,n}) \nonumber \\
    & -ep_n|^p)|+C_{n-1}\},
\end{align}
where $n\in\{2,3,...,N\}$ and $C_1$ and $C_n$ are the minimum cost obtained during the optimization process.

With $\Delta U_1$ and $\Delta U_2$ solved according to Eq. \ref{eq:cadeedfinal1} and \ref{eq:cadeedfinal2}, $Y_1$ and $Y_2$ can be obtained by Eq. \ref{eq:cadeedy1} and \ref{eq:cadeedy2}, respectively.

In practice, CaDEED will calculate the optimal $\Delta u_{1,i}$ and $\Delta u_{2,i}$ for the current pixels $x_{1,i}$ and $x_{2,i}$ first. Then CaDEED will perform regular error diffusion on $x_{1,i}+\Delta u_{1,i}$ and $x_{2,i}+\Delta u_{2,i}$ separately. When calculating the optimal $\Delta u_{1,i}$ and $\Delta u_{2,i}$, the possible solutions set can be reduced to a much smaller set with only four possible solutions: (a) $\Delta u_{1,i}=\Delta u_{2,i}=0$; (b) $\Delta u_{1,i}=0$, $\Delta u_{2,i}\neq0$; (c) $\Delta u_{2,i}=0$, $\Delta u_{1,i}\neq0$; (d) $\Delta u_{1,i}\neq0$ and $\Delta u_{2,i}\neq0$. Since the cost in option (d) is indeed larger or equal to the other three options, the final possible solution set contains three possible solutions. The CaDEED algorithm is shown in Algorithm \ref{alg:cadeed}.

\begin{algorithm}[ht]
\caption{CaDEED}
\begin{algorithmic}[1]
\Require ~~\ 
$\mathbf{X_1}$, $\mathbf{X_2}$, $\mathbf{W}$, $p$, $\lambda$, $\alpha$, $\beta$, $\circ$, the error diffusion kernel, the specific methods to calculate $M_1$, $M_2$, $\Psi$
\Ensure ~~\ 
$\mathbf{Y_1}$, $\mathbf{Y_2}$
\State Obtain $EP$, $M_1$, $M_2$, $\Psi$.
\For{$i=1$ to $N$}
\State Obtain $[\Delta u_{1,i}\text{ }\Delta u_{2,i}] =\underset{\Delta u_{1,i},\Delta u_{2,i}}{\operatorname{argmin}}{S_i}$ \Comment refer to (\ref{eq:costi})
\State $y_{1,i}=ed(x_{1,i}+\Delta u_{1,i})$
\State $y_{2,i}=ed(x_{2,i}+\Delta u_{2,i})$
\EndFor
\State \Return $\mathbf{Y_1}$, $\mathbf{Y_2}$
\end{algorithmic}
\label{alg:cadeed}
\end{algorithm}

In CaDEED, different parameters, $M_1$, $M_2$ and $\Psi$ give different specific EDHVW method. Note that if $\alpha=1$, $\beta=0$, $M_1=M_2=1$ and $\Psi=1$, CaDEED is essentially equivalent to DEED. Besides, if a $\Delta U_1=0$ constraint is added to the formulation, CaDEED becomes a single-sided embedding algorithm. For example, if $\alpha=1$, $\beta=0$, $M_1=M_2=1$, $\Psi=1$ and a $\Delta U_1=0$ constraint is added to CaDEED, CaDEED is equivalent to SEED.

%
%
%
%


To demonstrate the effectiveness of CaDEED, CaDEED generates CaDEED with expectation constraint (CaDEED-EC), by selecting $\alpha=\beta=1$, $M_1=M_2=1$ and $\Psi=1$. Besides of CaDEED-EC, CaDEED-NVF\&IF (CaDEED-N\&I) is also proposed with $\alpha=\beta=1$. In this scheme, the Noise Visibility Function (NVF) [\ref{volo2000nvf}] is selected to calculate $M_1$ and $M_2$, and a simple yet effective method, which calculates different importance factors (IF) for different pixels, is proposed. According to DEED(L-2)'s experience [\ref{guo2016hvw}], $p$ is set to be $2$.

The Noise Visibility Function [\ref{volo2000nvf}] in CaDEED-N\&I is described in Eqs. \ref{eq:nvf1} and \ref{eq:nvf2}.
\begin{equation} \label{eq:nvf1}
    m_1(i,j)=v_1(i,j)=\frac{1}{1+\theta_1\sigma^2_{Rx_1}(i,j)},
\end{equation}
\begin{equation} \label{eq:nvf2}
    m_2(i,j)=v_2(i,j)=\frac{1}{1+\theta_2\sigma^2_{Rx_2}(i,j)},
\end{equation}
where $Rx_1$ and $Rx_2$ are the local neighborhoods centered at $x_1(i,j)$ and $x_2(i,j)$ respectively. $\sigma^2_{Rx_1}(i,j)$ and $\sigma^2_{Rx_2}(i,j)$ are the local variances for $Rx_1$ and $Rx_2$ respectively. $\theta_1$, $\theta_2$ are calculated as Eq. (\ref{eq:nvf3}) and (\ref{eq:nvf4}) show.
\begin{equation} \label{eq:nvf3}
    \theta_1=\frac{D}{\sigma^2_{Rx_{1,max}}},
\end{equation}
\begin{equation} \label{eq:nvf4}
    \theta_2=\frac{D}{\sigma^2_{Rx_{2,max}}},
\end{equation}
where $\sigma^2_{Rx_{1,max}}$ and $\sigma^2_{Rx_{2,max}}$ are the maximum local variances in $X_1$ and $X_2$ respectively. $D=75$ is an experimentally determined parameter.

\begin{figure}
\begin{center}
\includegraphics[width=60mm]{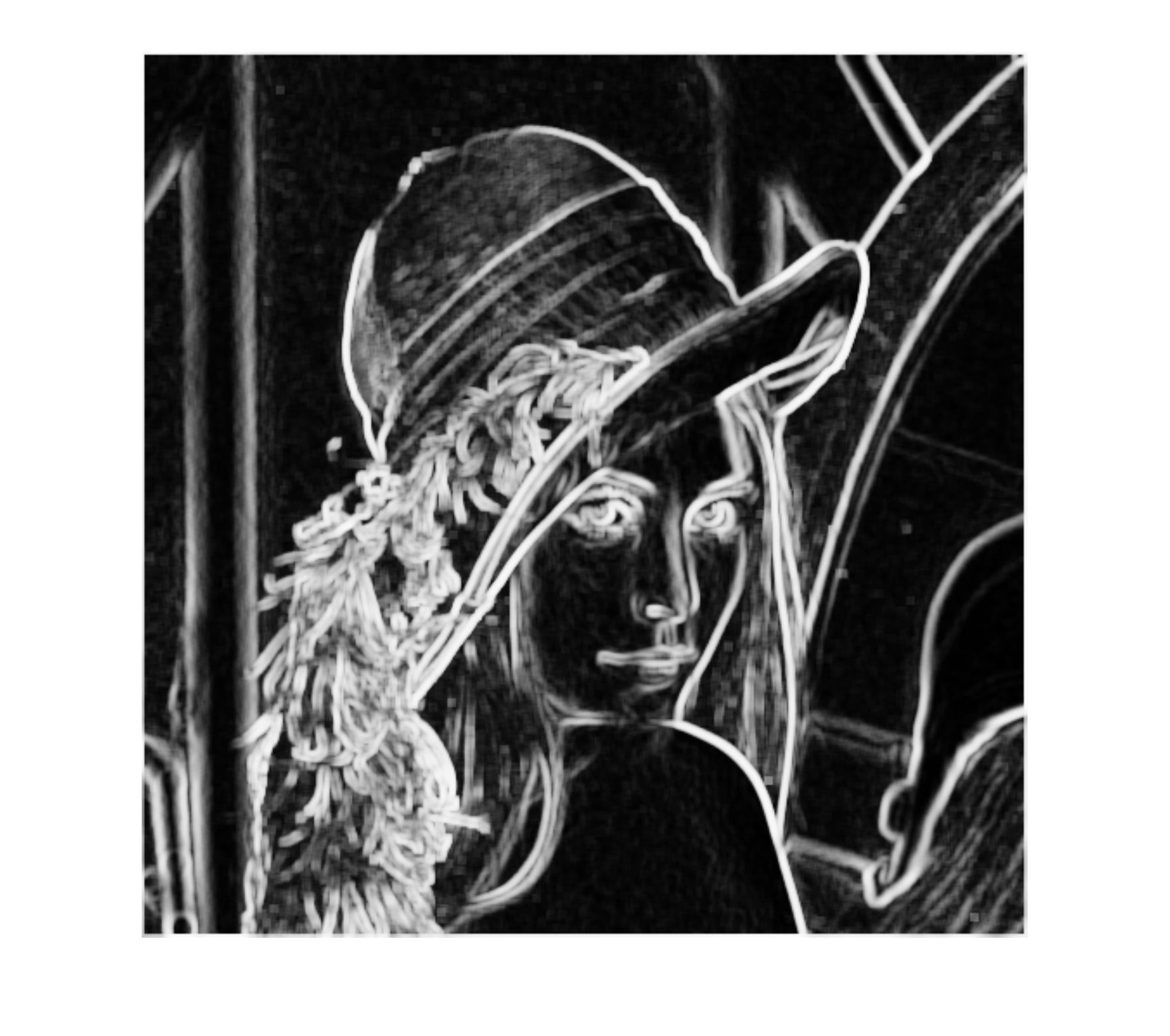}
\caption{Lena's NVF masking image. To highlight the pixels which can tolerate more noise, here darker pixels represent the original pixels that possess weak noise tolerance ability, while brighter pixels represent the original pixels that possess high noise tolerance ability. \label{fig:lenanvf}}
\end{center}
\vspace{-0.5cm}
\end{figure}

NVF generates a masking image where the smaller a masking value is, the more noise the original pixel can tolerate compared to other pixels. An example is shown in Fig. \ref{fig:lenanvf}. As we can observe, the texture regions can tolerate more noise compared to the smooth regions.

To calculate $\Psi$ in CaDEED-N\&I, we propose a simple but effective method to calculate a Importance Factor (IF) $\gamma$ for each pixel in the secret pattern as $\Psi$.
The calculation of $\gamma$ is shown in Eq. (\ref{eq:im}).
\begin{equation} \label{eq:im}
    \psi(i,j)=\gamma(i,j)=1+\frac{\sigma^2_{Rw}}{\sigma^2_{Rw_max}},
\end{equation}
where $Rw$ is the local neighborhood centered at $w(i,j)$, $\sigma^2_{Rw}(i,j)$ is the local variance for $Rw$ and $\sigma^2_{Rw_max}$ is the maximum local variances in $W$.

\section{Experimental Results}\label{results}



\subsection{Experimental Setups}

\begin{figure*}
  \centering
  \subfigure[]{
    \includegraphics[width=2.0cm]{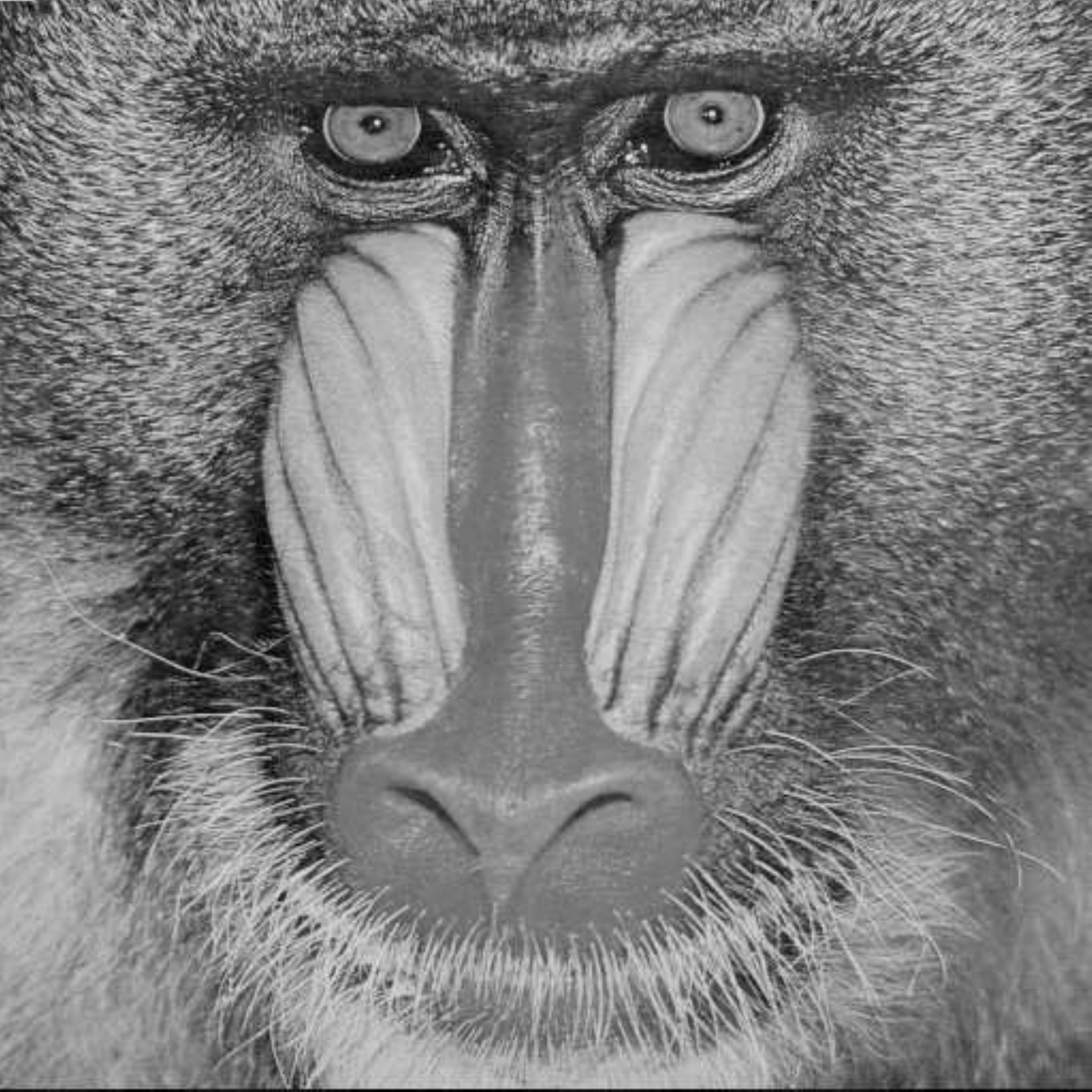}\label{fig:baboon}}
  \subfigure[]{
    \includegraphics[width=2.0cm]{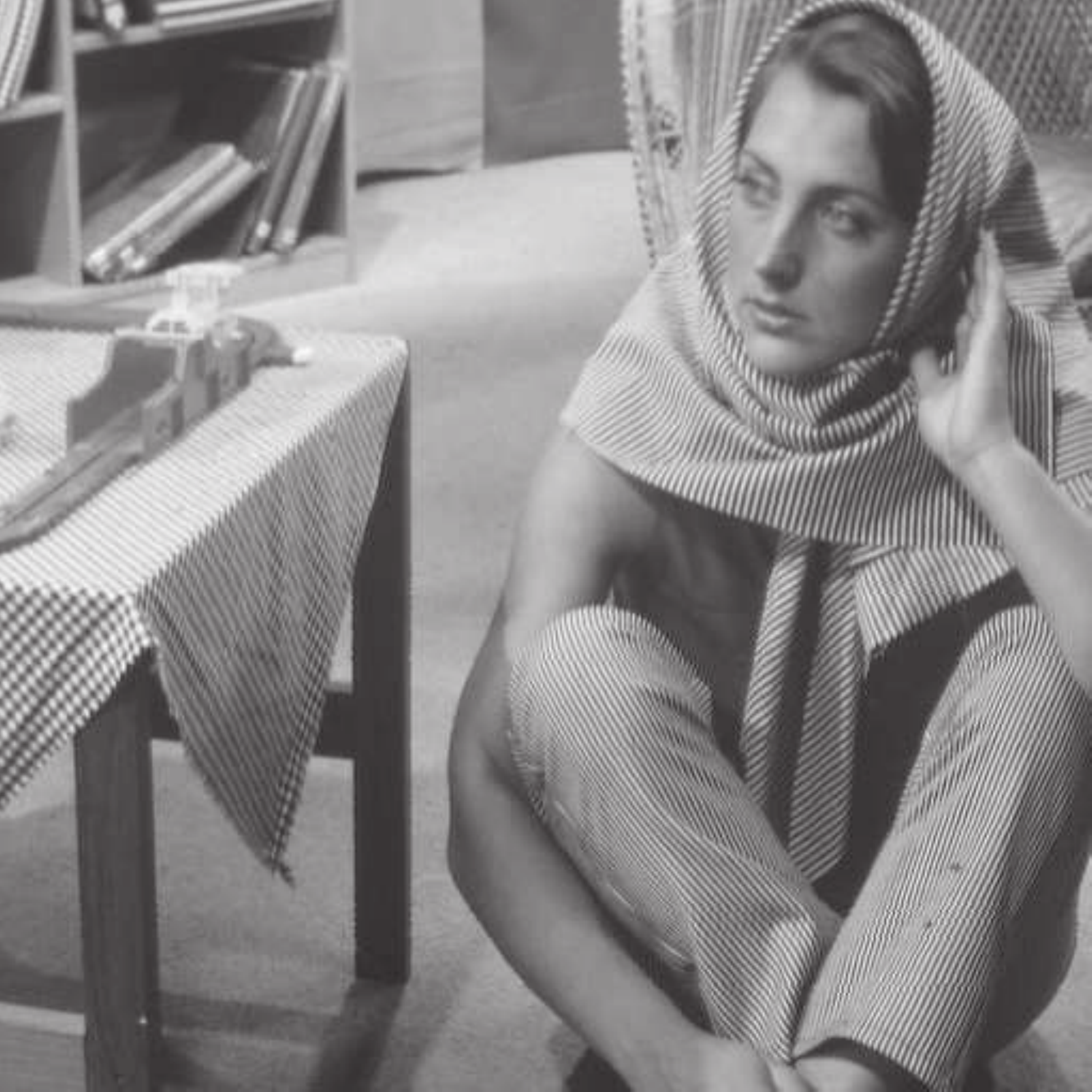}\label{fig:barbara}}
  \subfigure[]{
    \includegraphics[width=2.0cm]{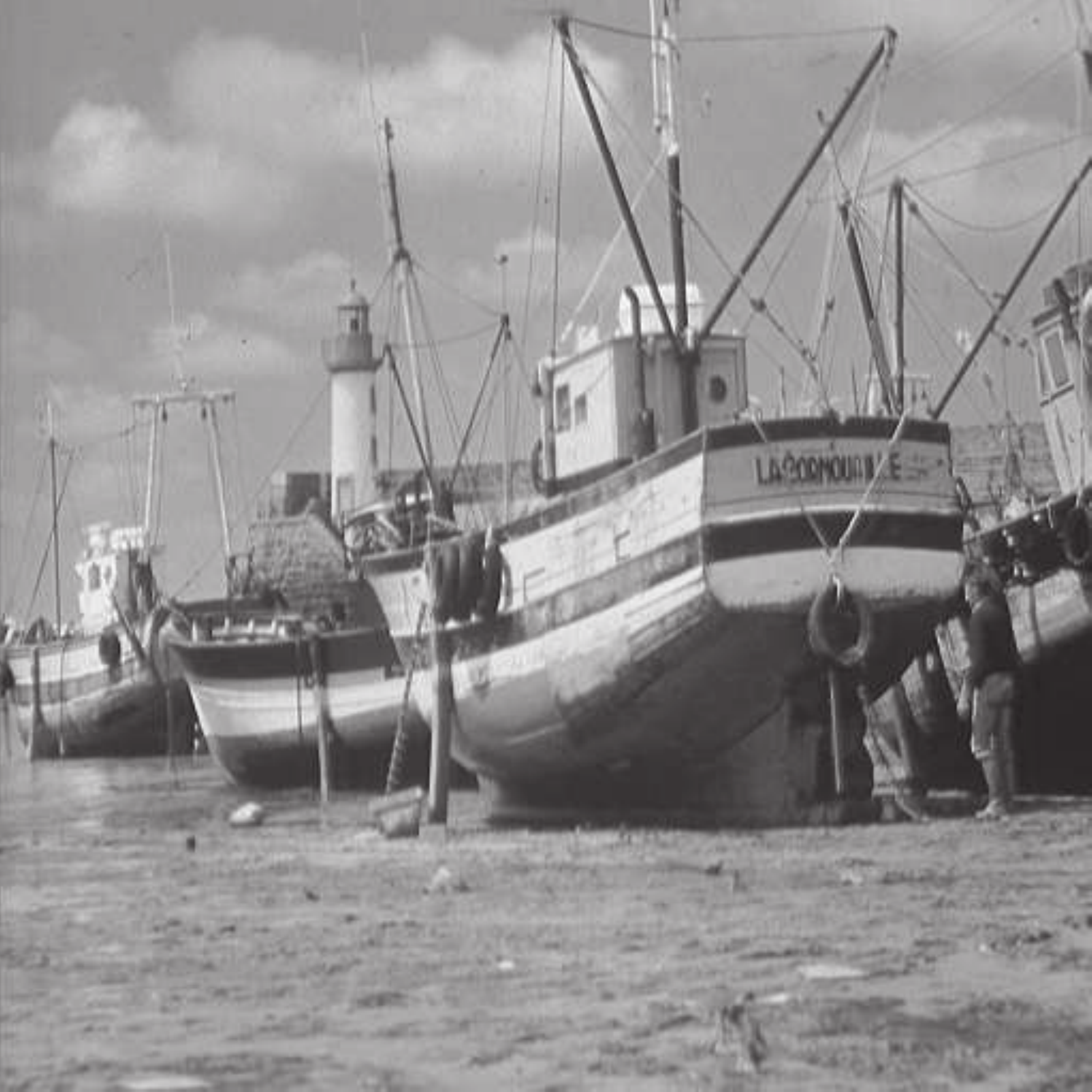}\label{fig:boat}}
  \subfigure[]{
    \includegraphics[width=2.0cm]{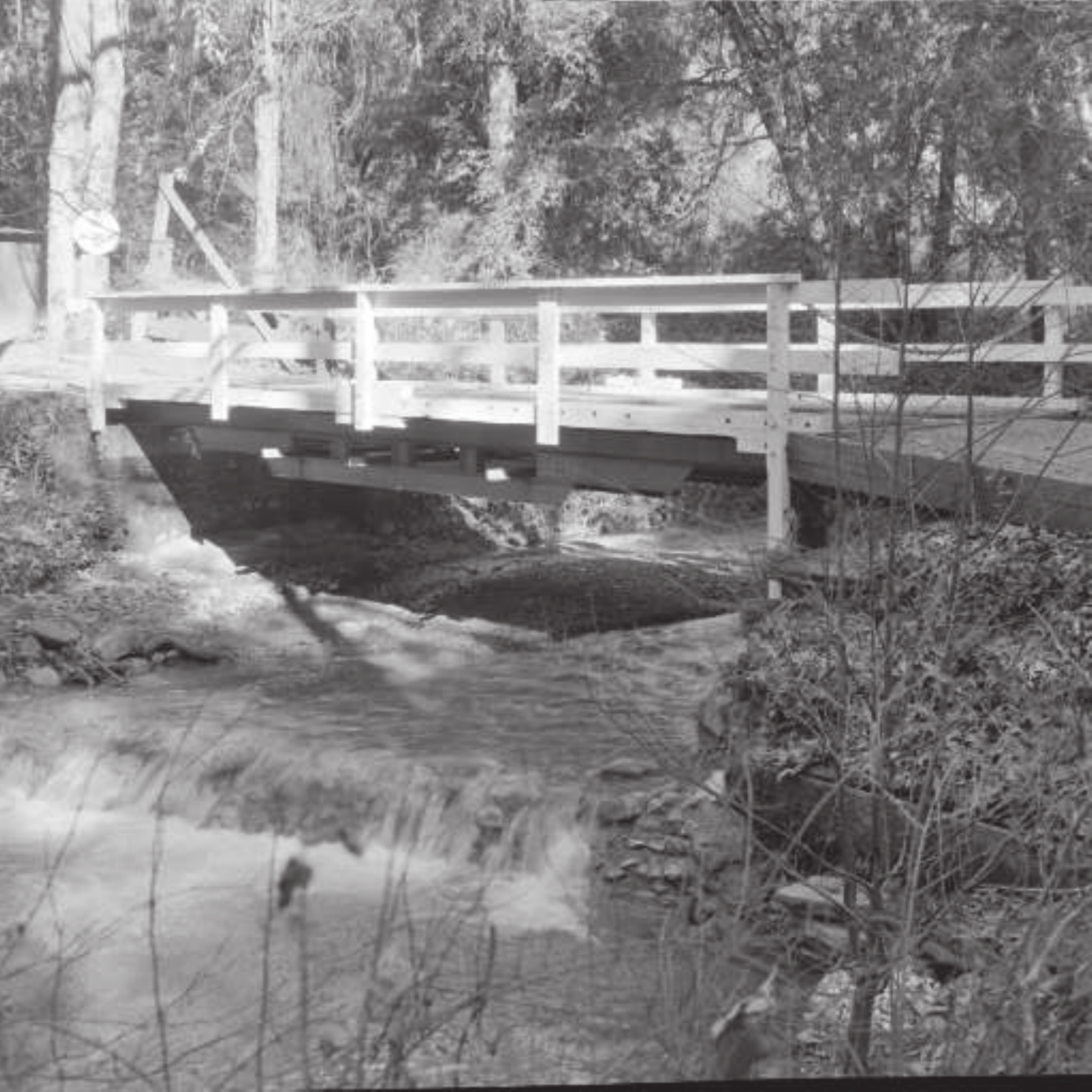}\label{fig:bridge}}
  \subfigure[]{
    \includegraphics[width=2.0cm]{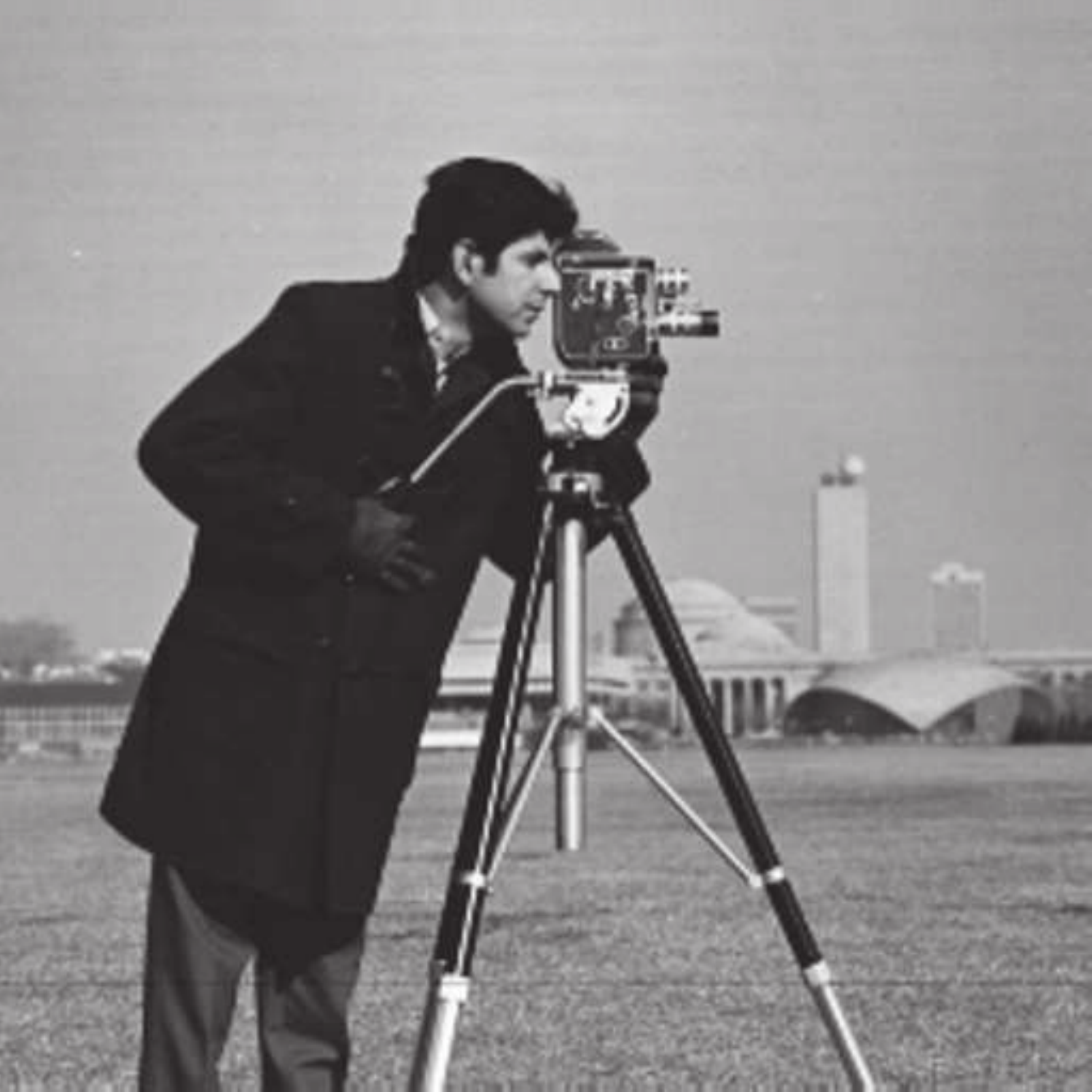}\label{fig:cameraman}}
  \subfigure[]{
    \includegraphics[width=2.0cm]{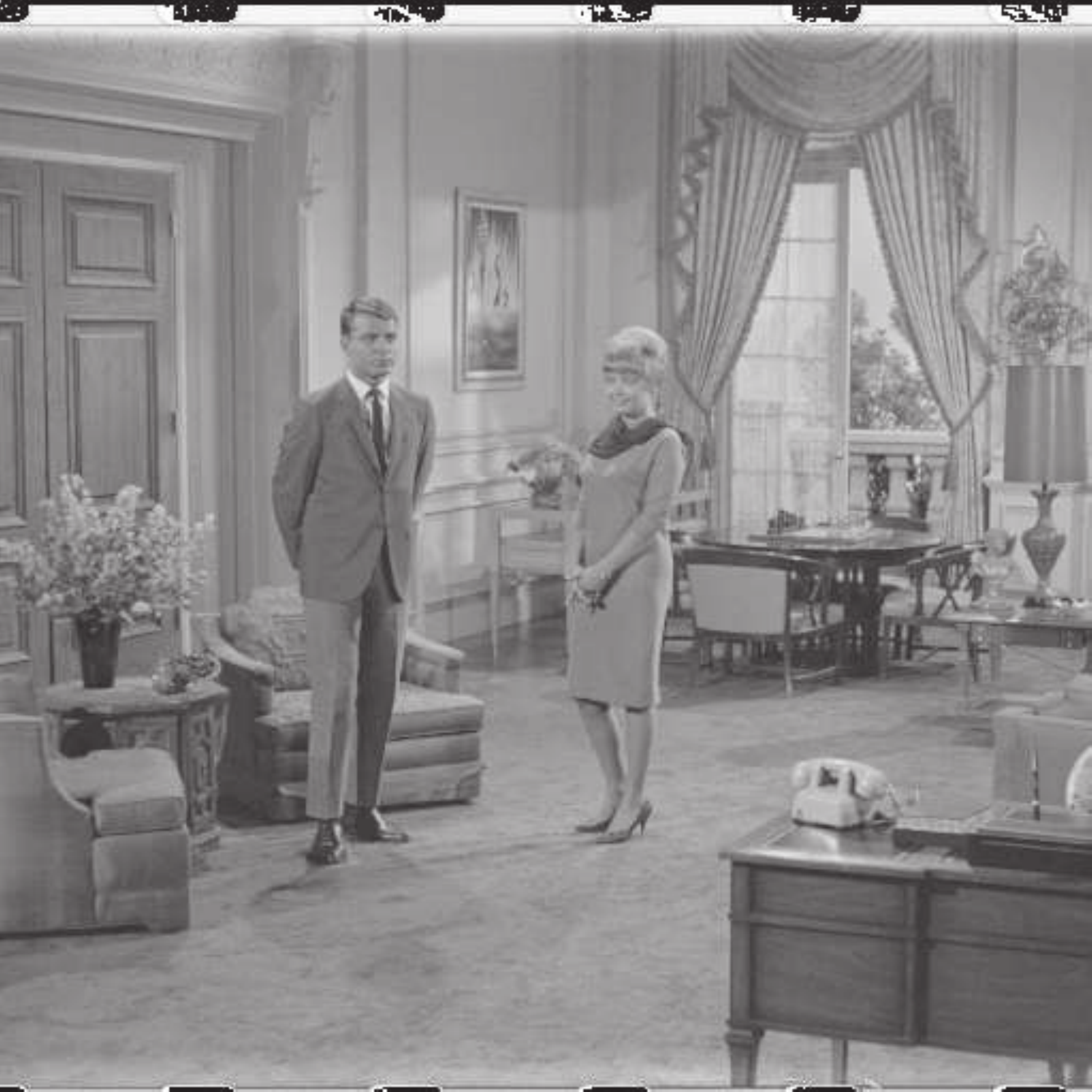}\label{fig:couple}}
  \\
  \subfigure[]{
    \includegraphics[width=2.0cm]{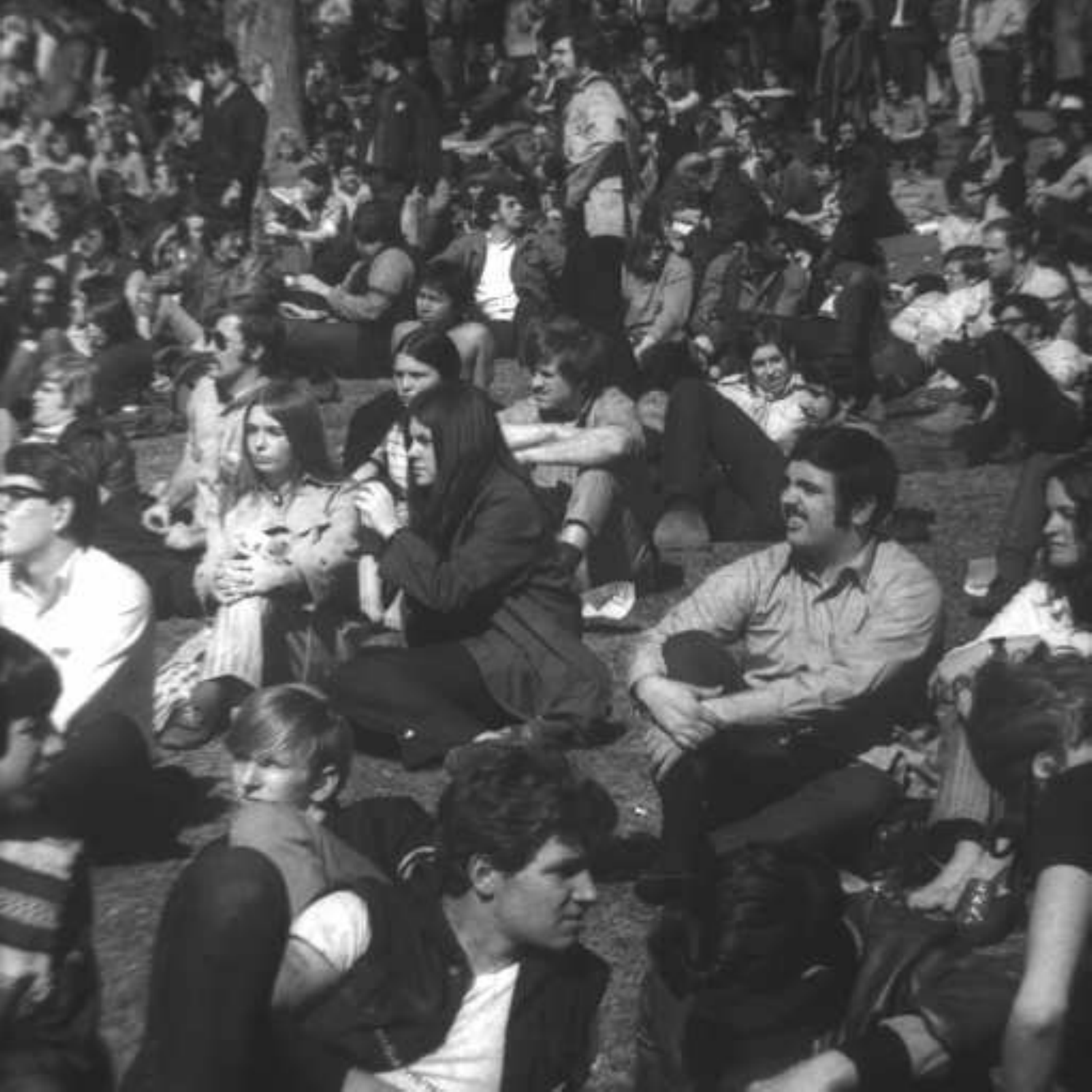}\label{fig:crowd}}
  \subfigure[]{
    \includegraphics[width=2.0cm]{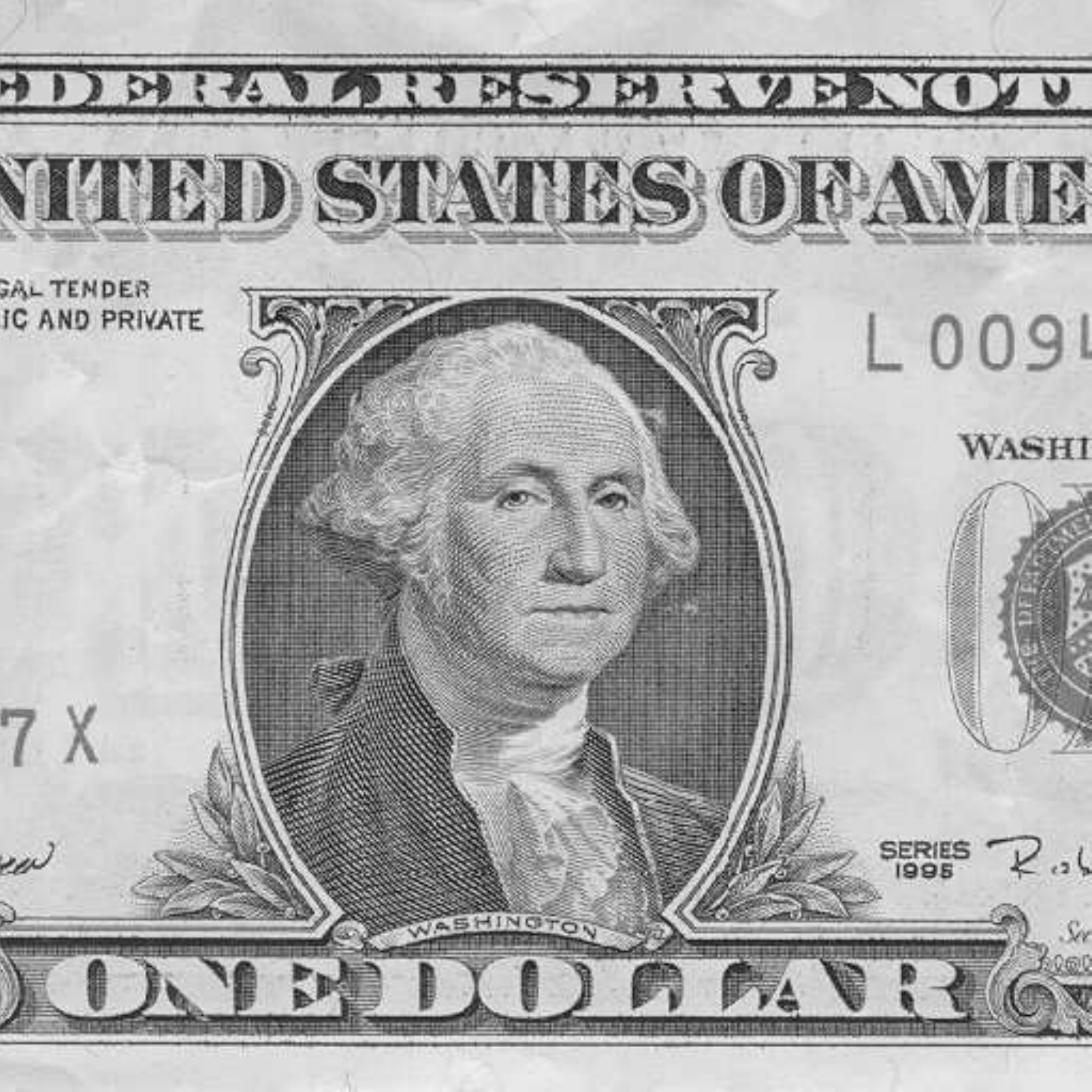}\label{fig:dollar}}
  \subfigure[]{
    \includegraphics[width=2.0cm]{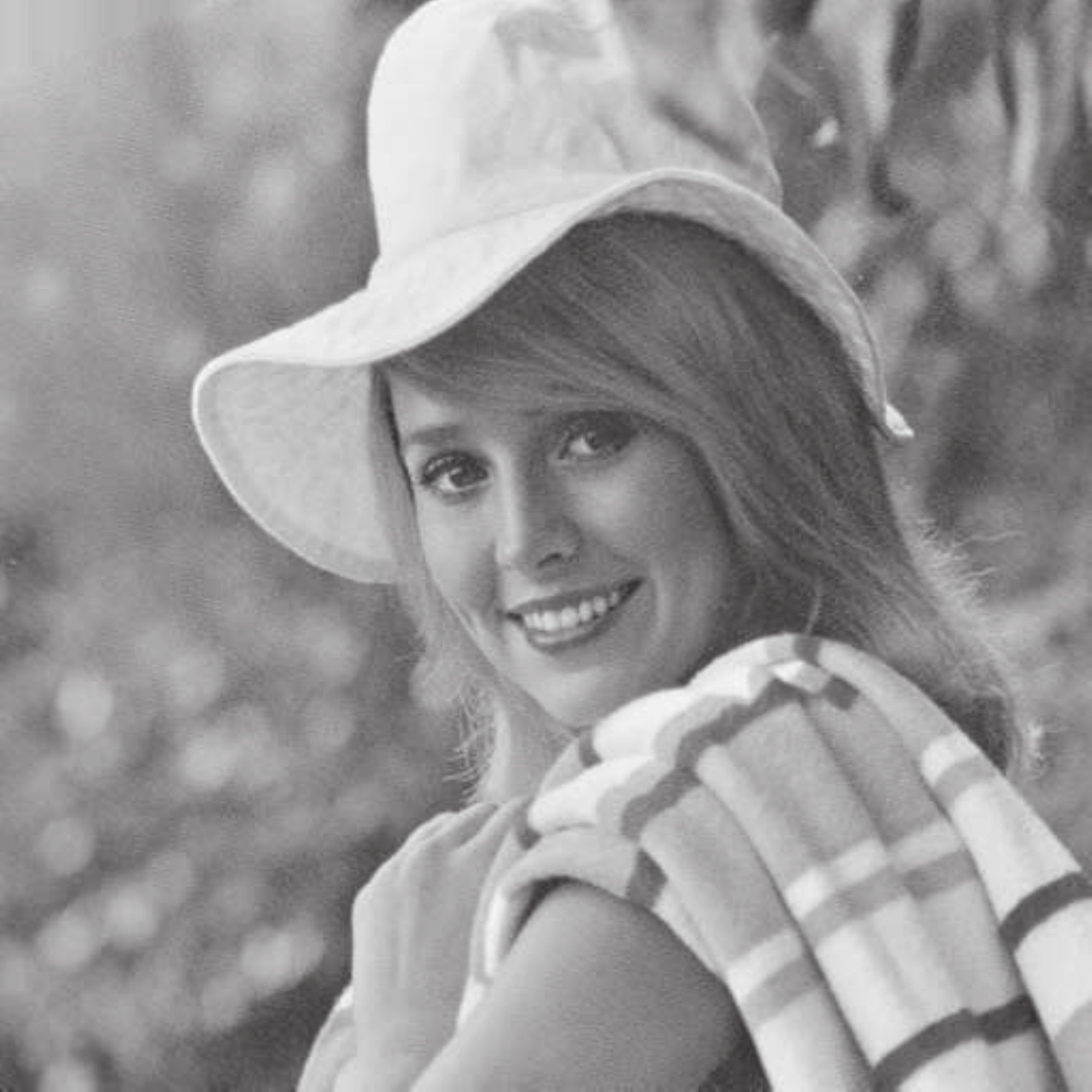}\label{fig:elaine}}
  \subfigure[]{
    \includegraphics[width=2.0cm]{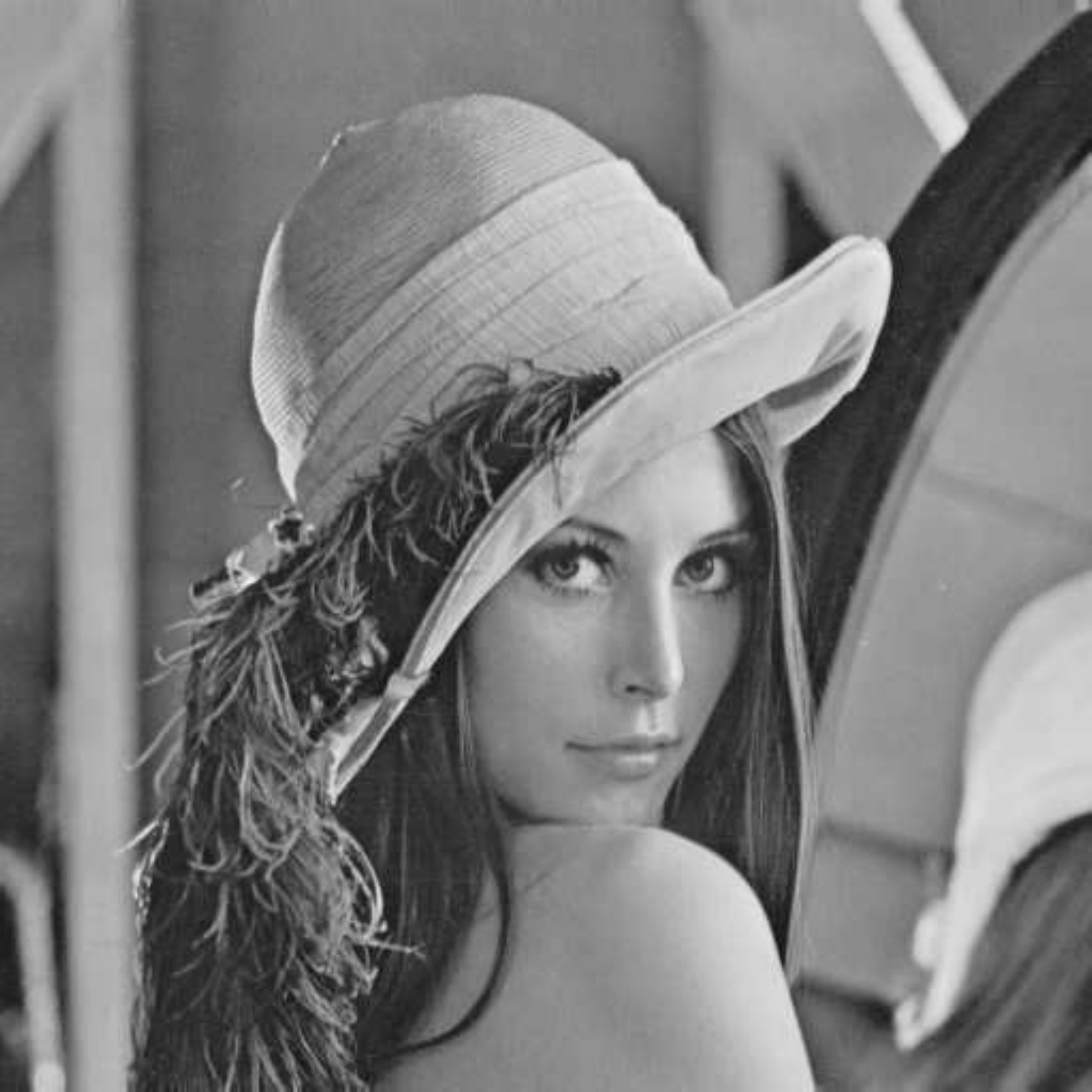}\label{fig:lena}}
  \subfigure[]{
    \includegraphics[width=2.0cm]{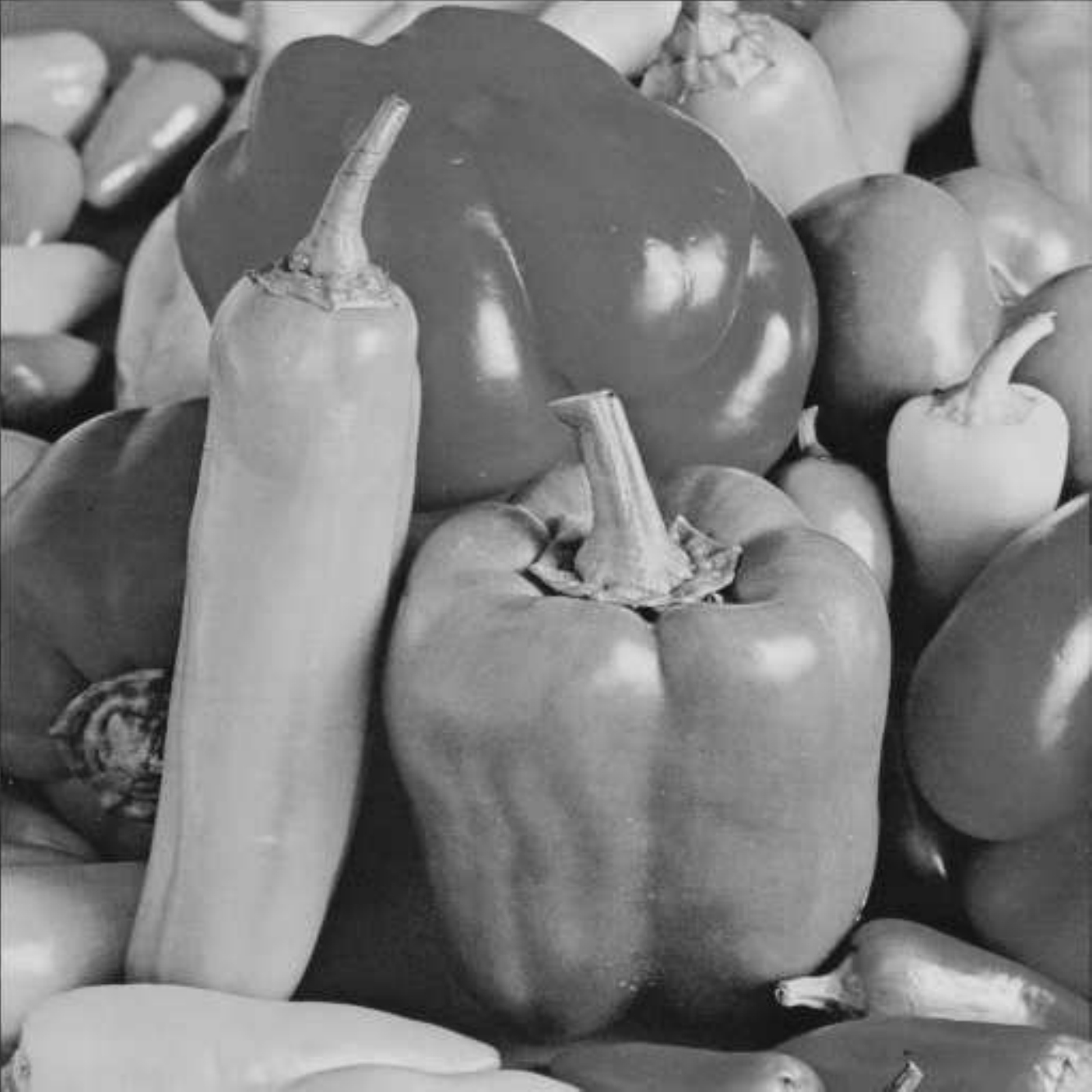}\label{fig:pepper}}
  \subfigure[]{
    \includegraphics[width=2.0cm]{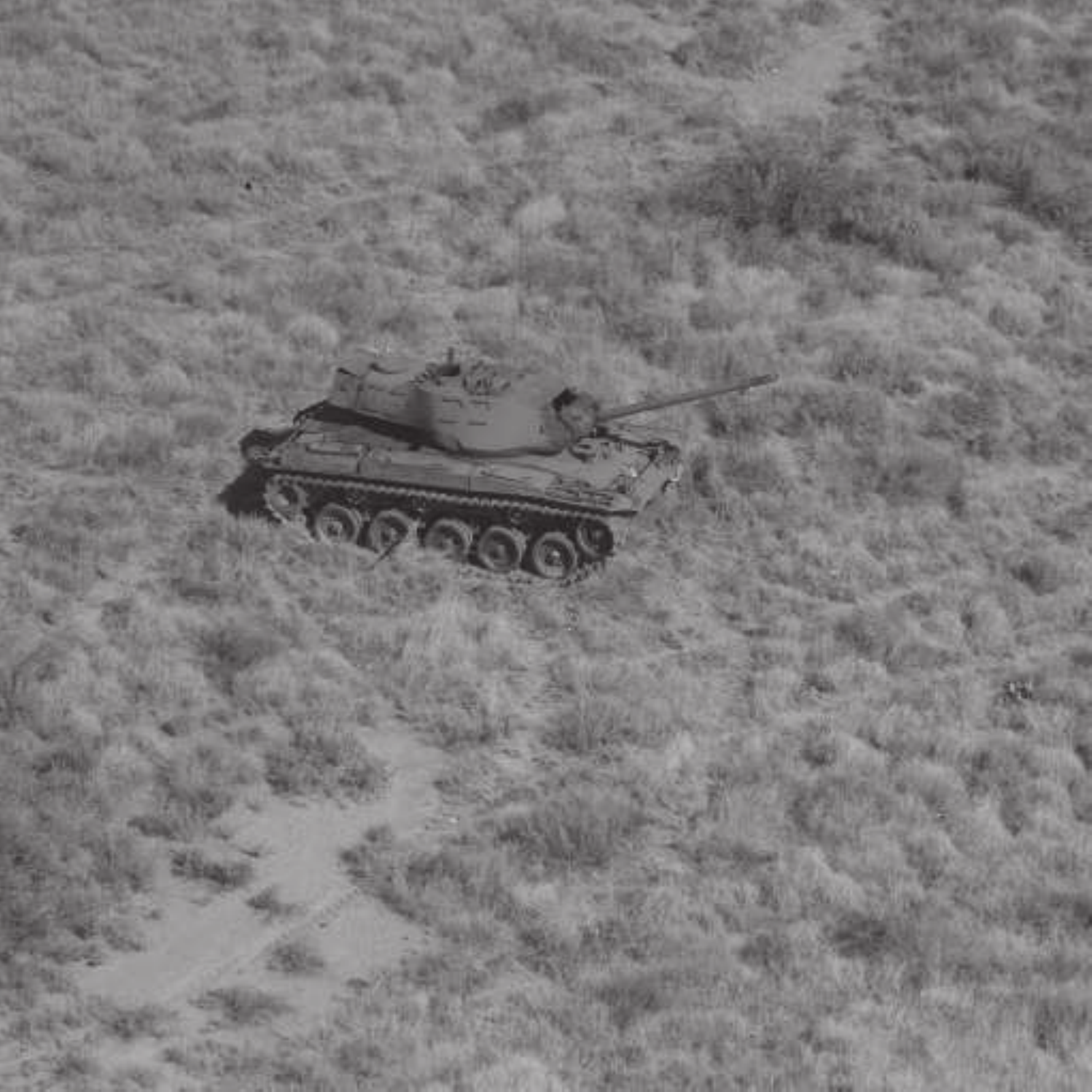}\label{fig:tank}}
  \subfigure[]{
    \setlength{\fboxrule}{1pt} \setlength{\fboxsep}{0cm} \fbox{\includegraphics[width=2.0cm]{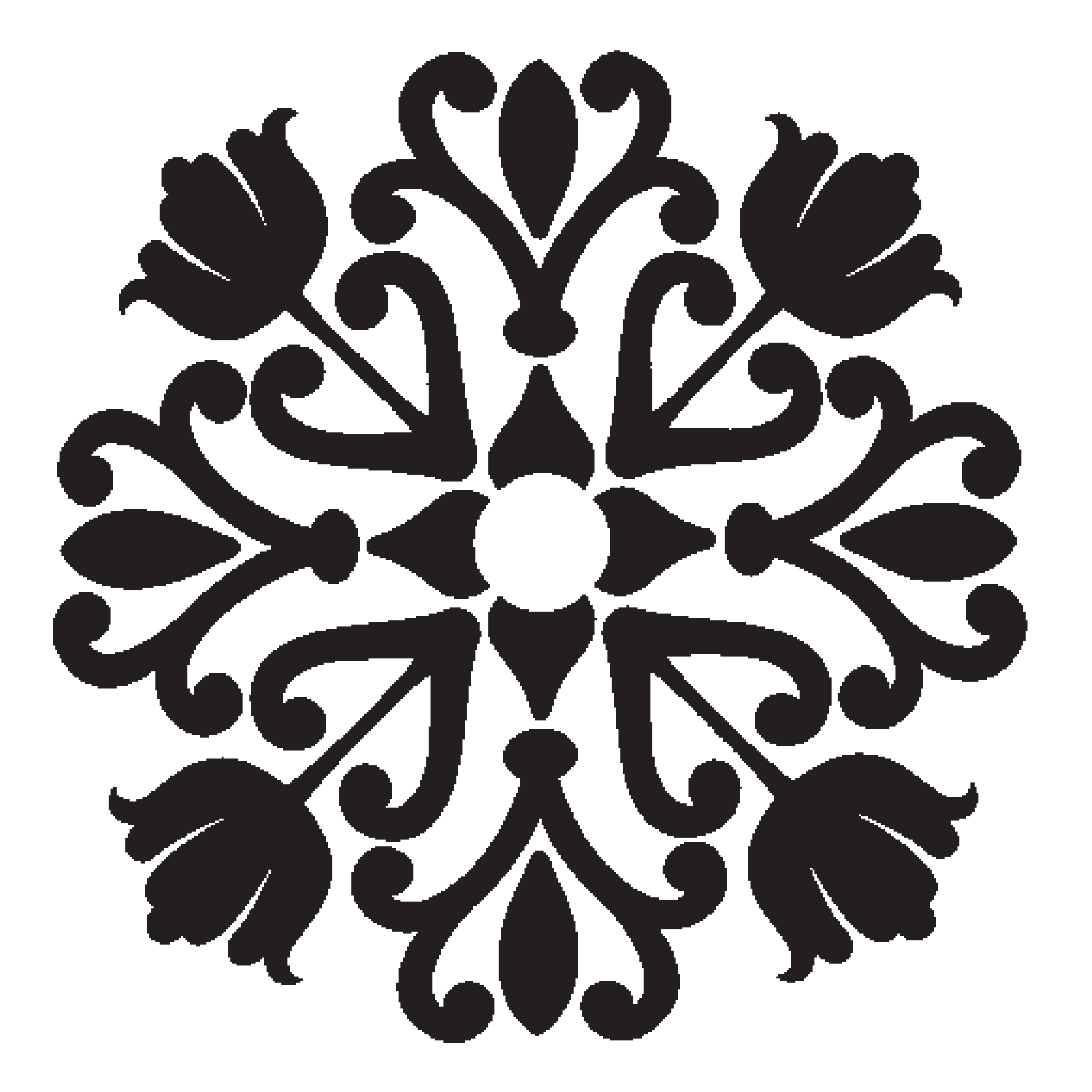}}\label{fig:secret}}
\vspace{-0.2cm}
\caption{Test images: (a)`Baboon'; (b)`Barbara'; (c)`Boat'; (d)`Bridge'; (e)`Cameraman'; (f)`Couple'; (g)`Crowd'; (h)`Dollar'; (i)`Elaine'; (j)`Lena'; (k)`Pepper'; (l)`Tank'; (m)Secret pattern to be embedded. \label{fig:testimage}}
\vspace{-0.3cm}
\end{figure*}

During the experiments, the $512\times 512$ test images in Fig. \ref{fig:testimage} are employed. Fig. \ref{fig:secret} shows the $512\times 512$ secret pattern to be embedded. For convenience, Steinberg kernel [\ref{steinberg1976}] and $\circ=\odot=\text{XNOR}$ will be employed in this paper. Note that in real implementations, CaDEED-N\&I will force $y_1(i,j)$ and $y_2(i,j)$ to be identical for $(i,j)\in W_w$ when $X_1=X_2$.

\subsection{Validation Test}

\begin{figure*}
  \centering
  \subfigure[]{
    \includegraphics[width=4.0cm]{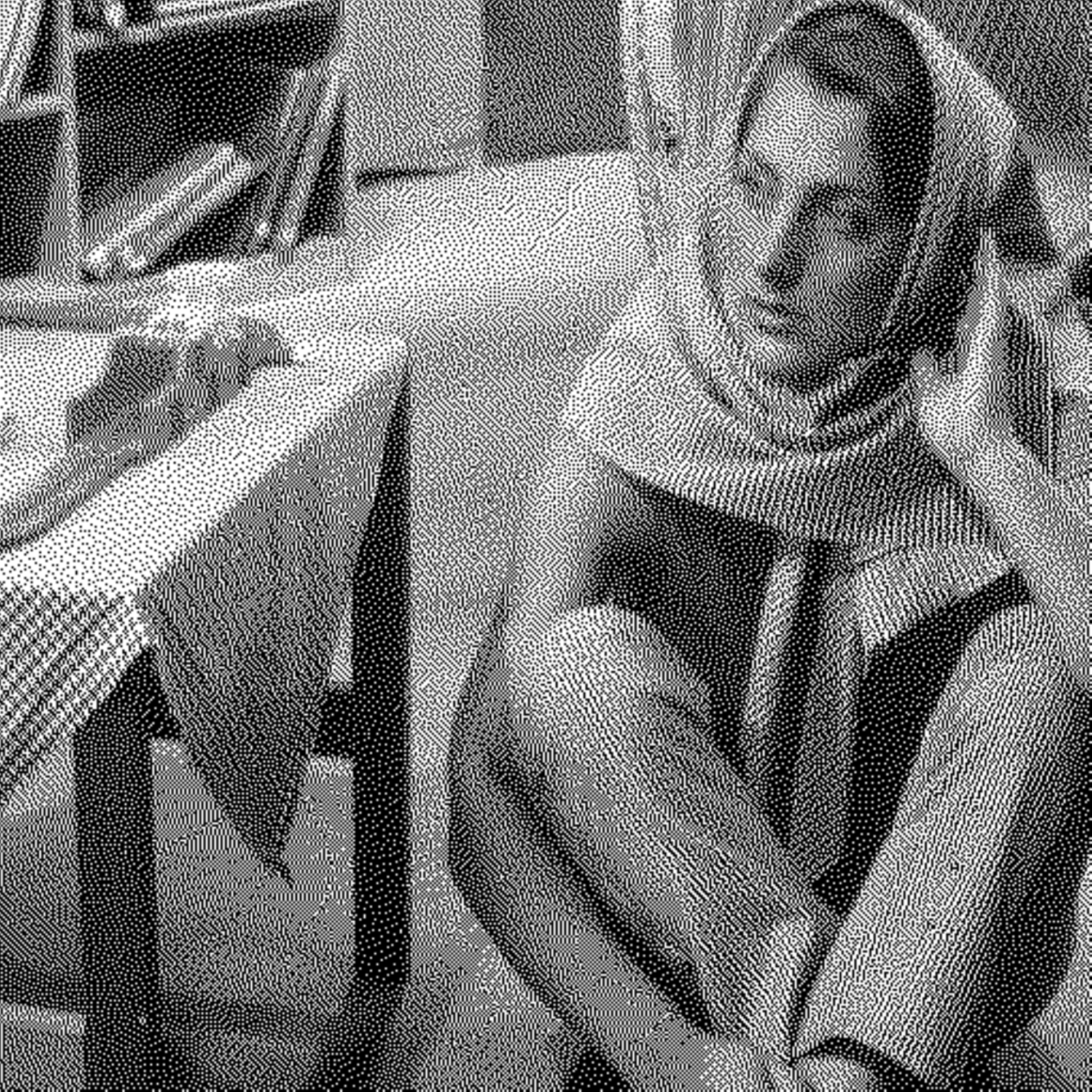}\label{fig:cbdcedy1}}
  \subfigure[]{
    \includegraphics[width=4.0cm]{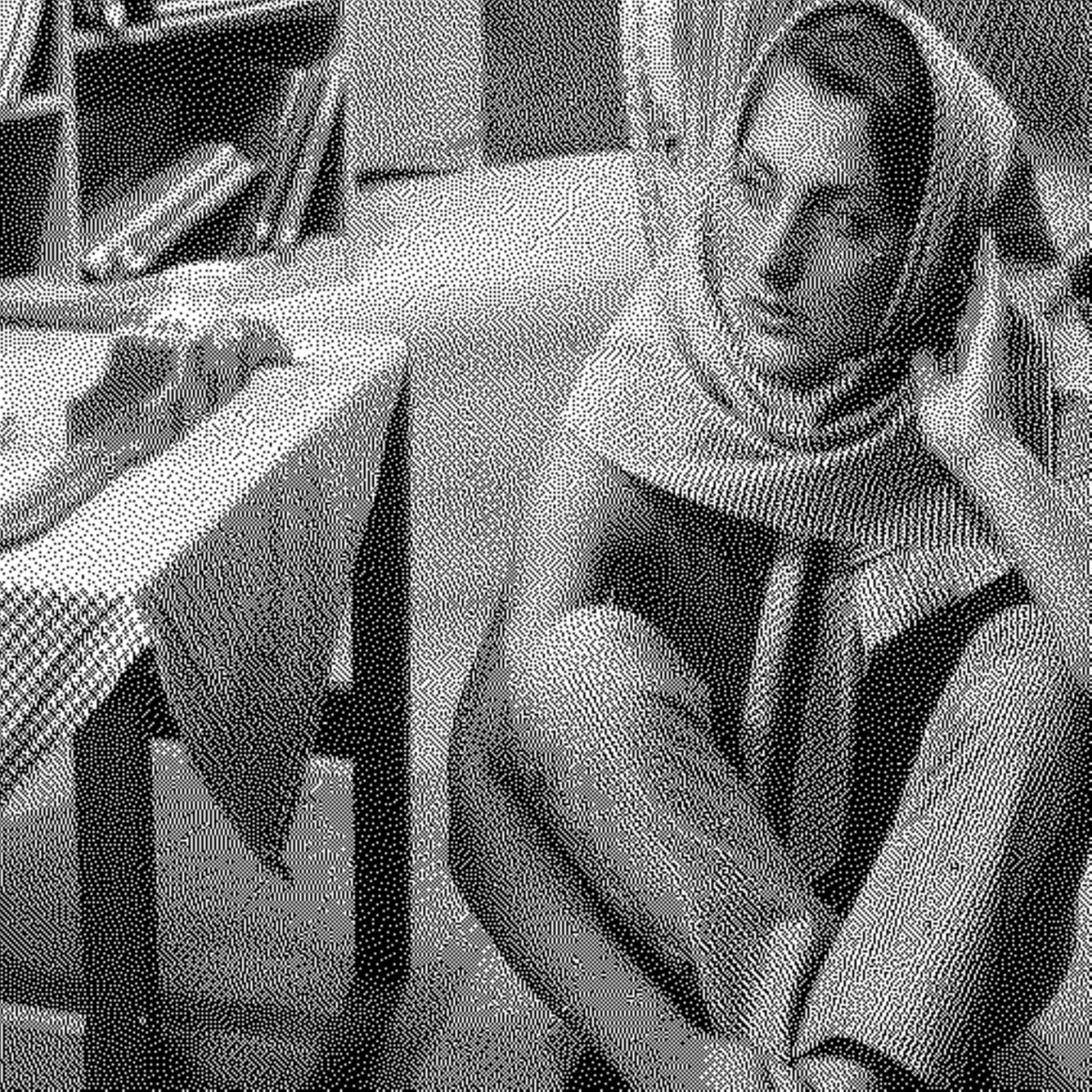}\label{fig:cbdcedy2}}
  \subfigure[]{
    \includegraphics[width=4.0cm]{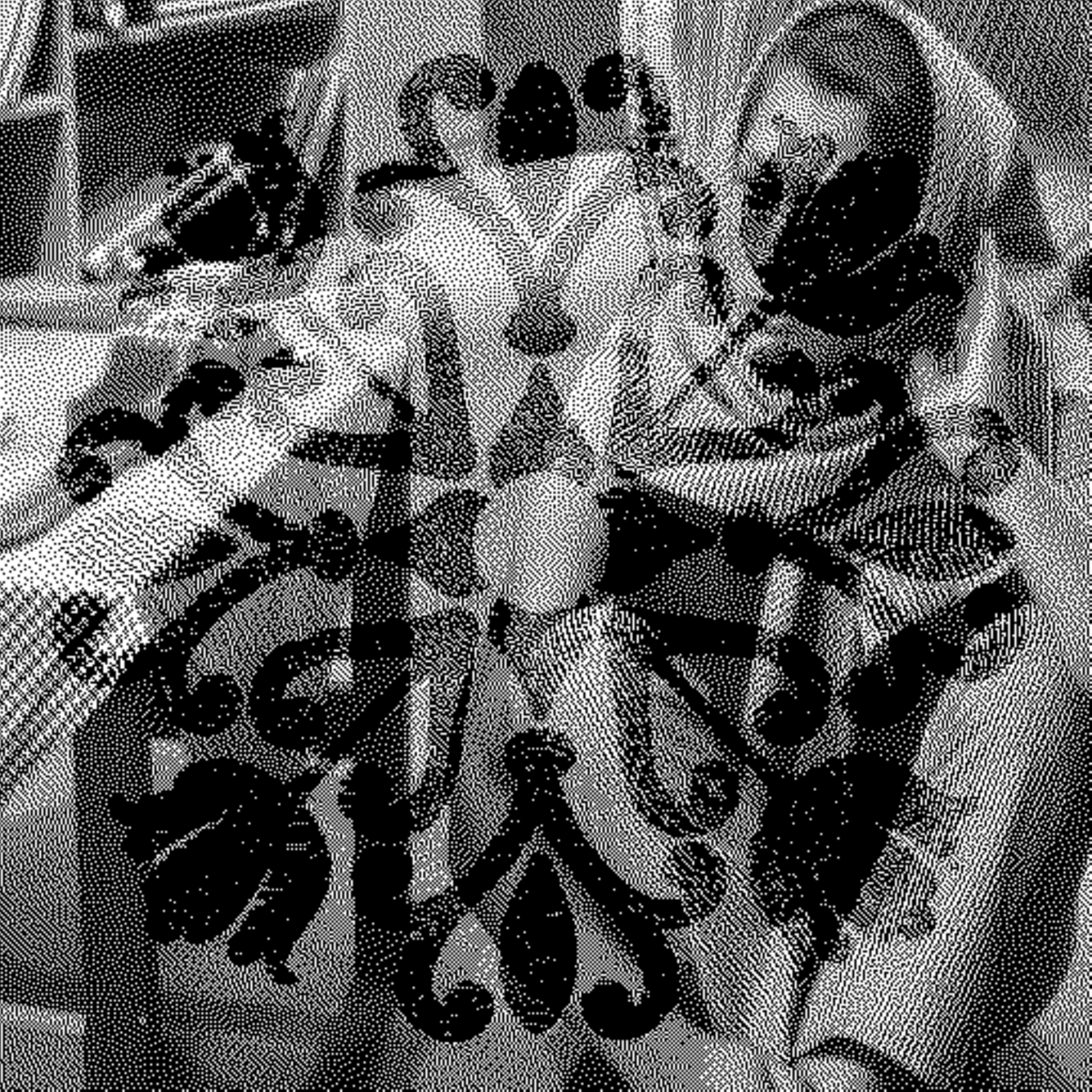}\label{fig:cbdcedoverlay}}
  \subfigure[]{
    \setlength{\fboxrule}{1pt} \setlength{\fboxsep}{0cm}
    \fbox{\includegraphics[width=4.0cm]{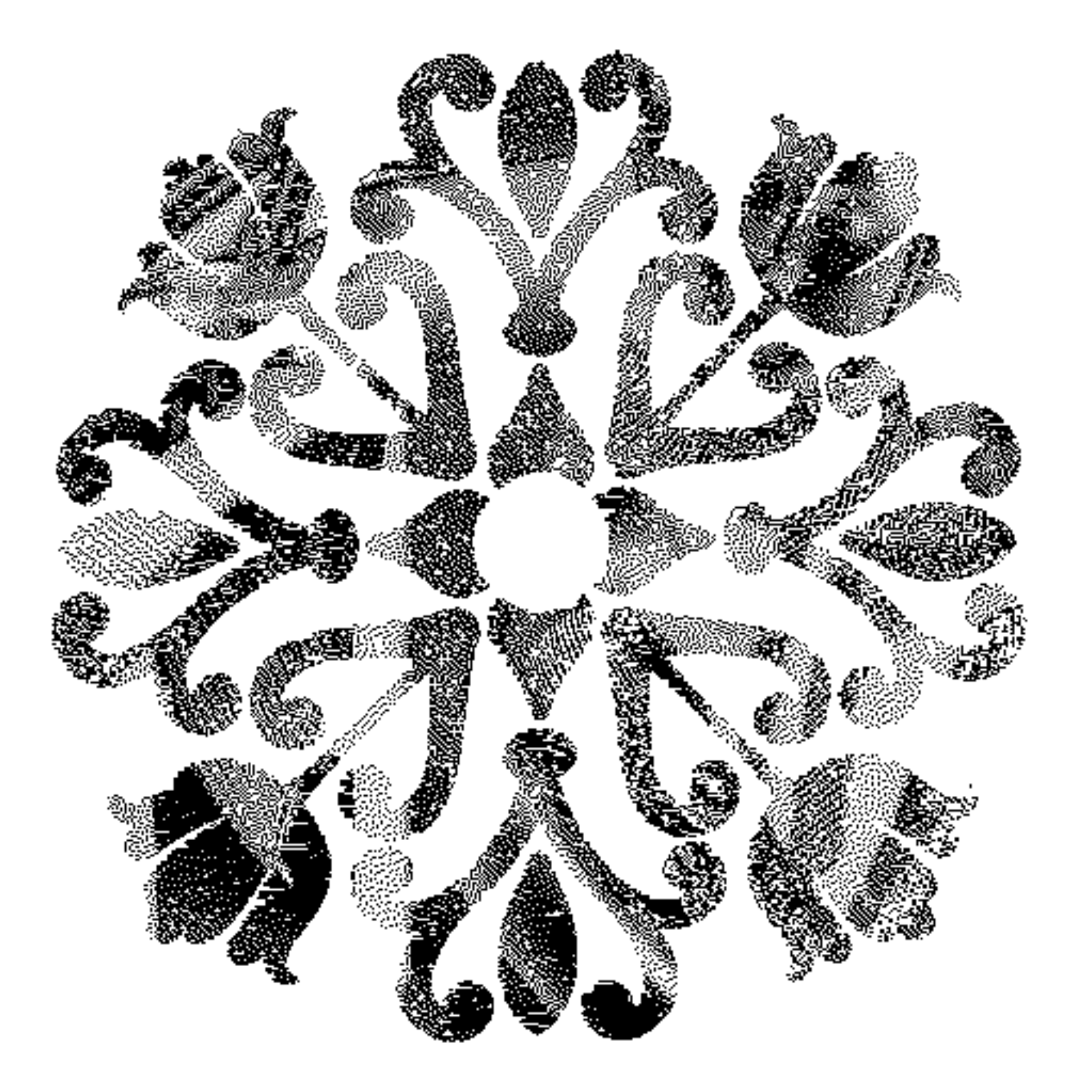}\label{fig:cbdcedxnor}}}
\vspace{-0.2cm}
\caption{(a)CaDEED-EC $Y_1$. $X_1=X_2=`Barbara'$, Steinberg kernel.
(b)CaDEED-EC $Y_2$. $X_1=X_2=`Barbara'$, Steinberg kernel. (c)CaDEED-EC AND operation decoded image. $X_1=X_2=`Barbara'$, Steinberg kernel.
(d)CaDEED-EC XNOR operation decoded image. $X_1=X_2=`Barbara'$, Steinberg kernel.
\label{fig:cbdced}}
\vspace{-0.3cm}
\end{figure*}

\begin{figure*}
  \centering
  \subfigure[]{
    \includegraphics[width=4.0cm]{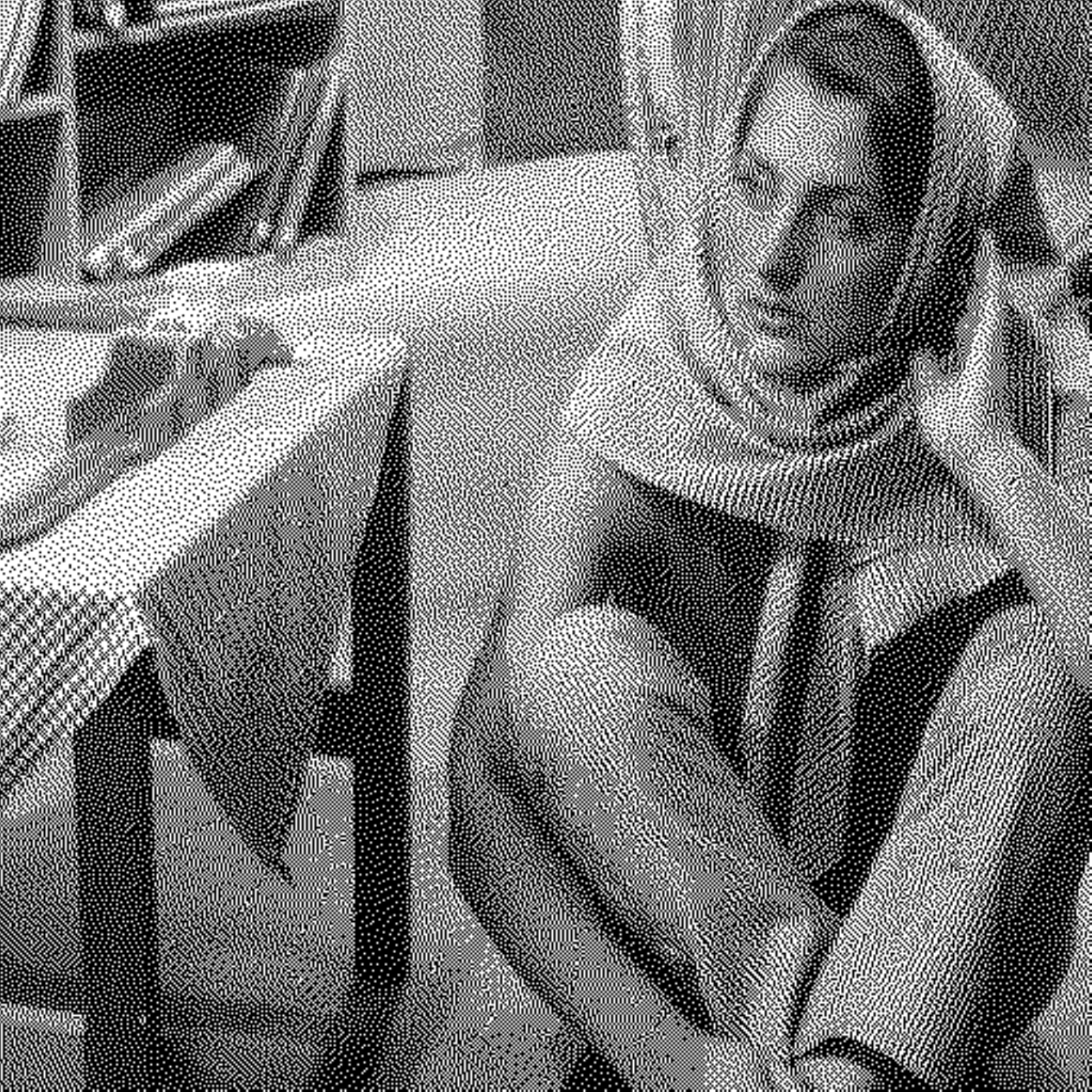}\label{fig:cbdcedy12}}
  \subfigure[]{
    \includegraphics[width=4.0cm]{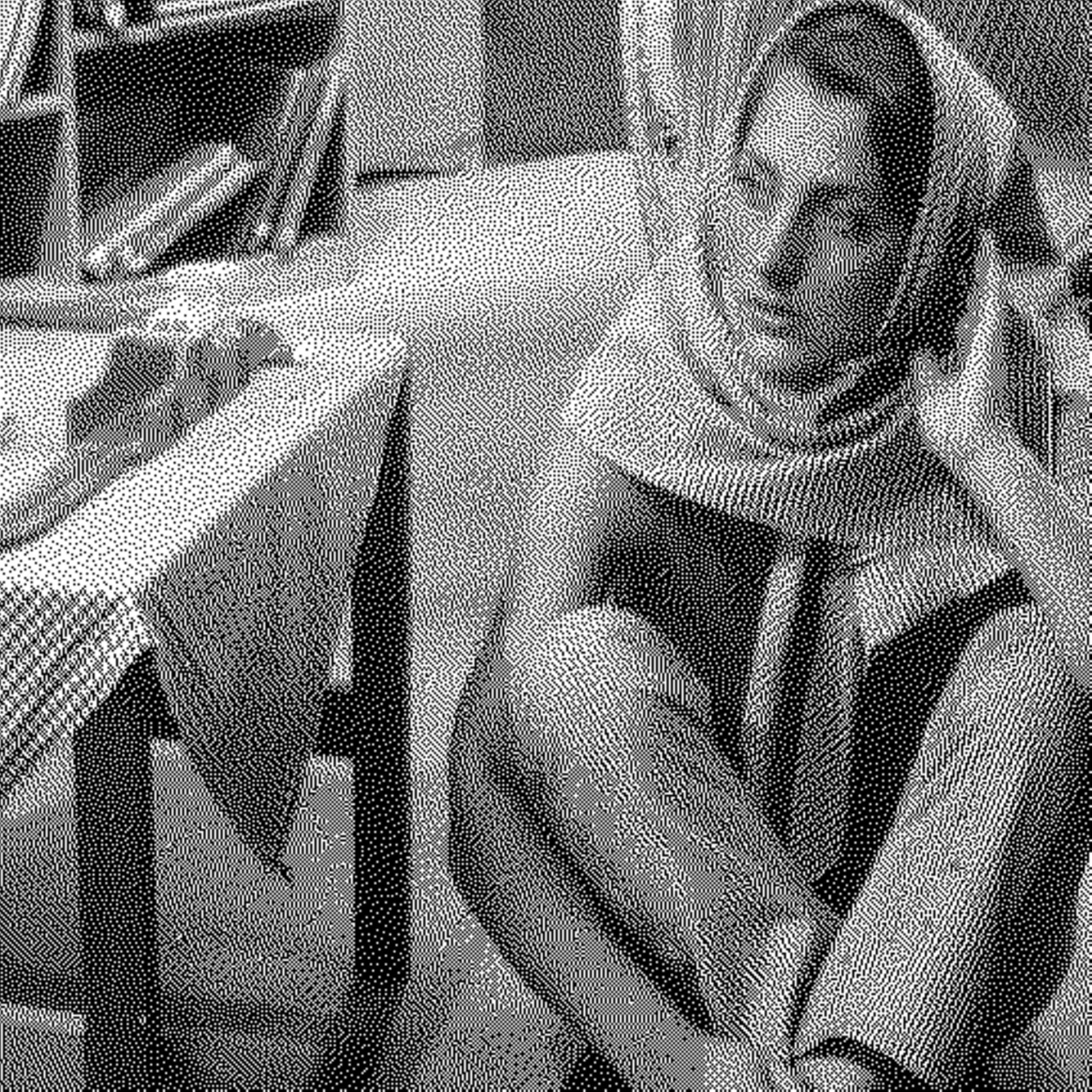}\label{fig:cbdcedy22}}
  \subfigure[]{
    \includegraphics[width=4.0cm]{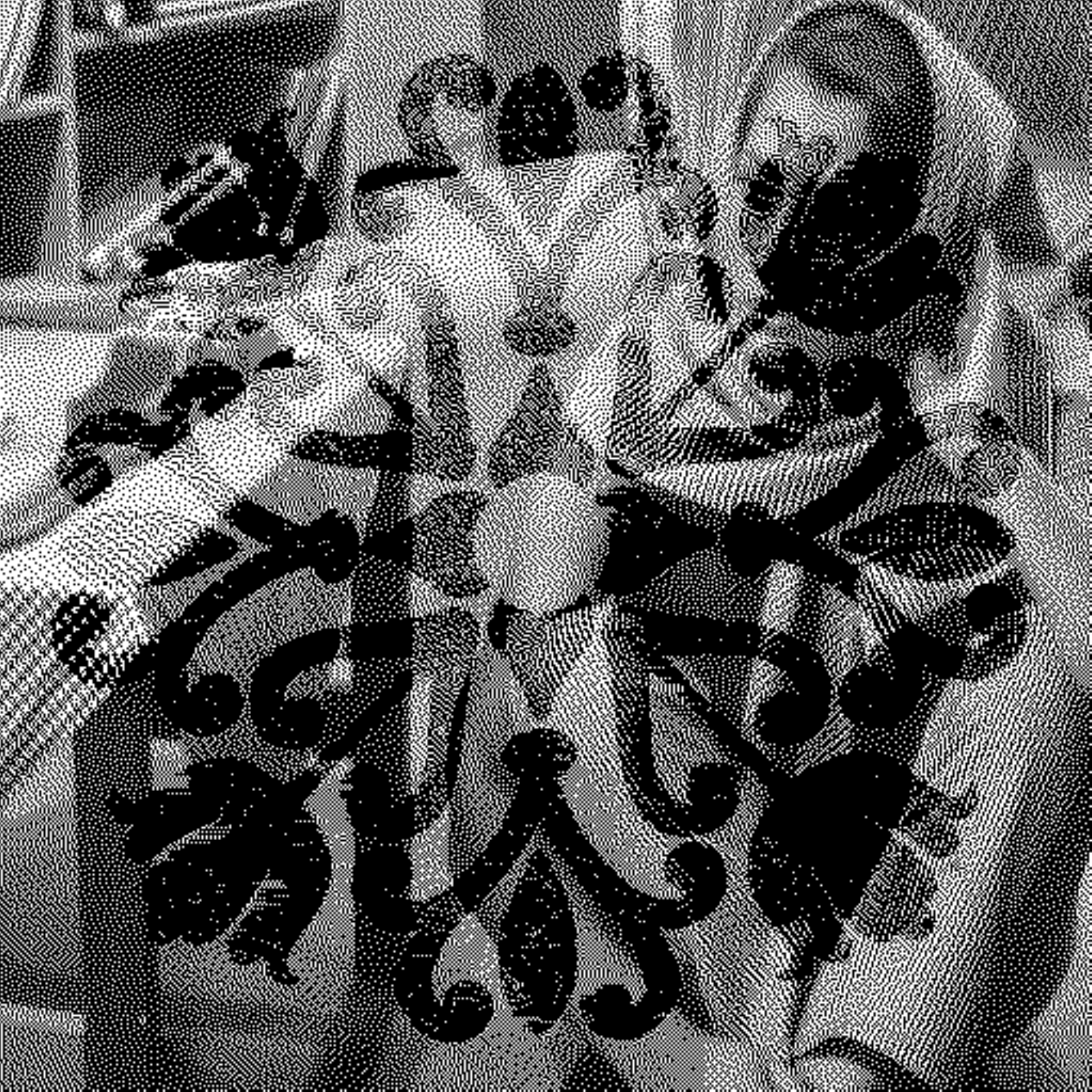}\label{fig:cbdcedoverlay2}}
  \subfigure[]{
    \setlength{\fboxrule}{1pt} \setlength{\fboxsep}{0cm}
    \fbox{\includegraphics[width=4.0cm]{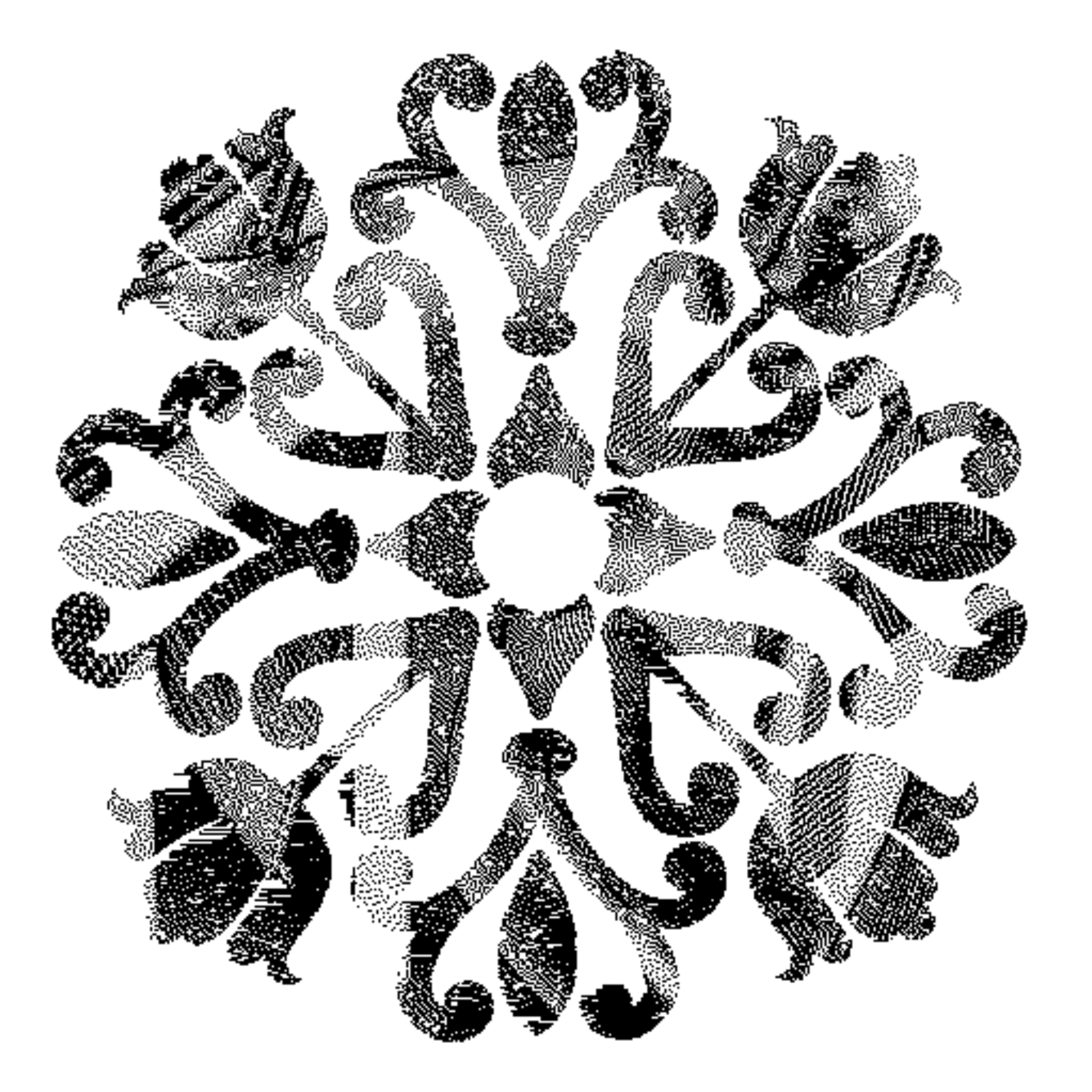}\label{fig:cbdcedxnor2}}}
\vspace{-0.2cm}
\caption{(a)CaDEED-N\&I $Y_1$. $X_1=X_2=`Barbara'$, Steinberg kernel.
(b)CaDEED-N\&I $Y_2$. $X_1=X_2=`Barbara'$, Steinberg kernel. (c)CaDEED-N\&I AND operation decoded image. $X_1=X_2=`Barbara'$, Steinberg kernel.
(d)CaDEED-N\&I XNOR operation decoded image. $X_1=X_2=`Barbara'$, Steinberg kernel.
\label{fig:cbdced2}}
\vspace{-0.3cm}
\end{figure*}

Figs. \ref{fig:cbdced}-\ref{fig:cbdced2} give the validations of CaDEED-EC and CaDEED-N\&I. `Baboon' in Fig. \ref{fig:baboon} is employed as the original grey-scale image and $X_1=X_2$. In Fig. \ref{fig:cbdced}, Fig. \ref{fig:cbdcedy1} and Fig. \ref{fig:cbdcedy2} are the generated $Y_1$ and $Y_2$ respectively. The AND operation decoded image generated from $Y_1$ and $Y_2$ is in Fig. \ref{fig:cbdcedoverlay} and the XNOR operation decoded image is in Fig. \ref{fig:cbdcedxnor}. Similarly, in Fig. \ref{fig:cbdced2}, Fig. \ref{fig:cbdcedy12} and Fig. \ref{fig:cbdcedy22} are $Y_1$ and $Y_2$ respectively. The AND operation decoded image is in Fig. \ref{fig:cbdcedoverlay2}. The XNOR operation decoded image is in Fig. \ref{fig:cbdcedxnor2}.

As we can observe from Figs. \ref{fig:cbdced}-\ref{fig:cbdced2}, both CaDEED-EC and CaDEED-N\&I perform well.

\subsection{Parameter Selection}


After the validation tests, the local region sizes for the calculation of NVF and IF will be discussed and suggested. The measurement employed in the parameter selection is the Sum of Squared Error (SSE) and the correct decoding rate (CDR).

In CaDEED (DHCED, DHDCED, DEED(L-2) also), the generation process of $Y_1$ and $Y_2$ can also be treated equivalently to adding specific noises, which are bounded by the thresholds, to the original images and carrying out the regular error diffusion process on the noisy images. Therefore, we can calculate SSE as Eq. \ref{eq:mse} between the original images and the equivalent noisy images to measure the embedding distortion. Compared to previous modified PSNR, which firstly let the halftone images pass through a low pass filter and then calculate the PSNR between the low-passed halftone images and the original images, this SSE measurement doesn't possess the low-pass filter selection problem and its result will not be affected by the imperfection of the halftoning process, where some embedding distortions may actually improve the measured quality of the stego halftone image.
\begin{equation} \label{eq:mse}
    \textrm{SSE}=\sum(\Delta u^2_1(i,j))+\sum(\Delta u^2_2(i,j))
\end{equation}

Traditionally, the correct decoding rate (CDR) is ultilized to measure the differences between the reference and decoded secret pattern. The CDR is defined as in Eq. (\ref{eq:cdr}), where we let $D$ be the decoded image in this paper.
\begin{equation} \label{eq:cdr}
    \textrm{CDR}=\frac{\sum{|w(i,j) \odot d(i,j)|}}{IW\times IH}
\end{equation}
where $IW$ and $IH$ are the image width and height respectively.

\begin{figure*}
  \centering
  \subfigure[]{
    \includegraphics[width=5.8cm]{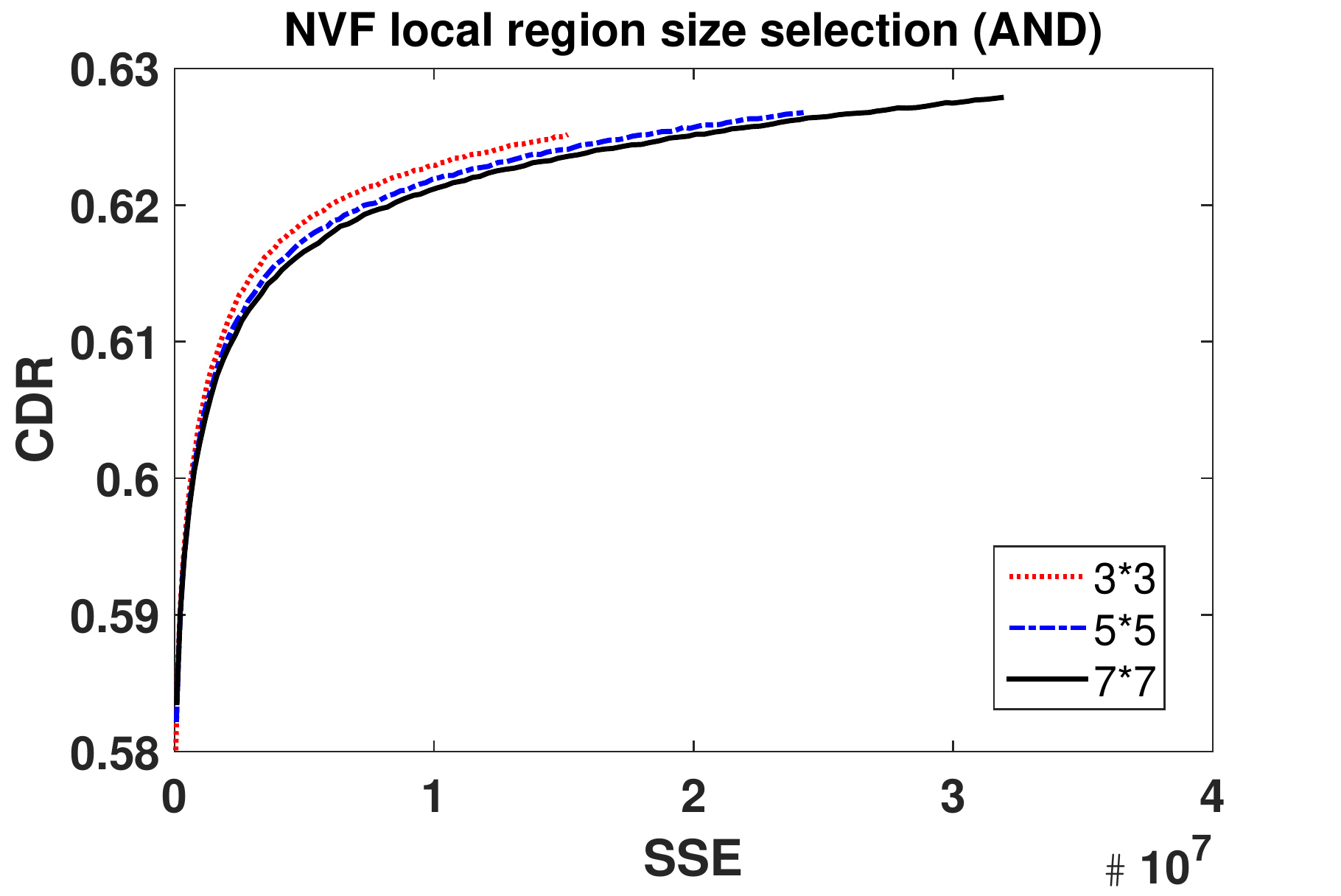}\label{fig:nvfsizeandcompare}}
  \subfigure[]{
    \includegraphics[width=5.8cm]{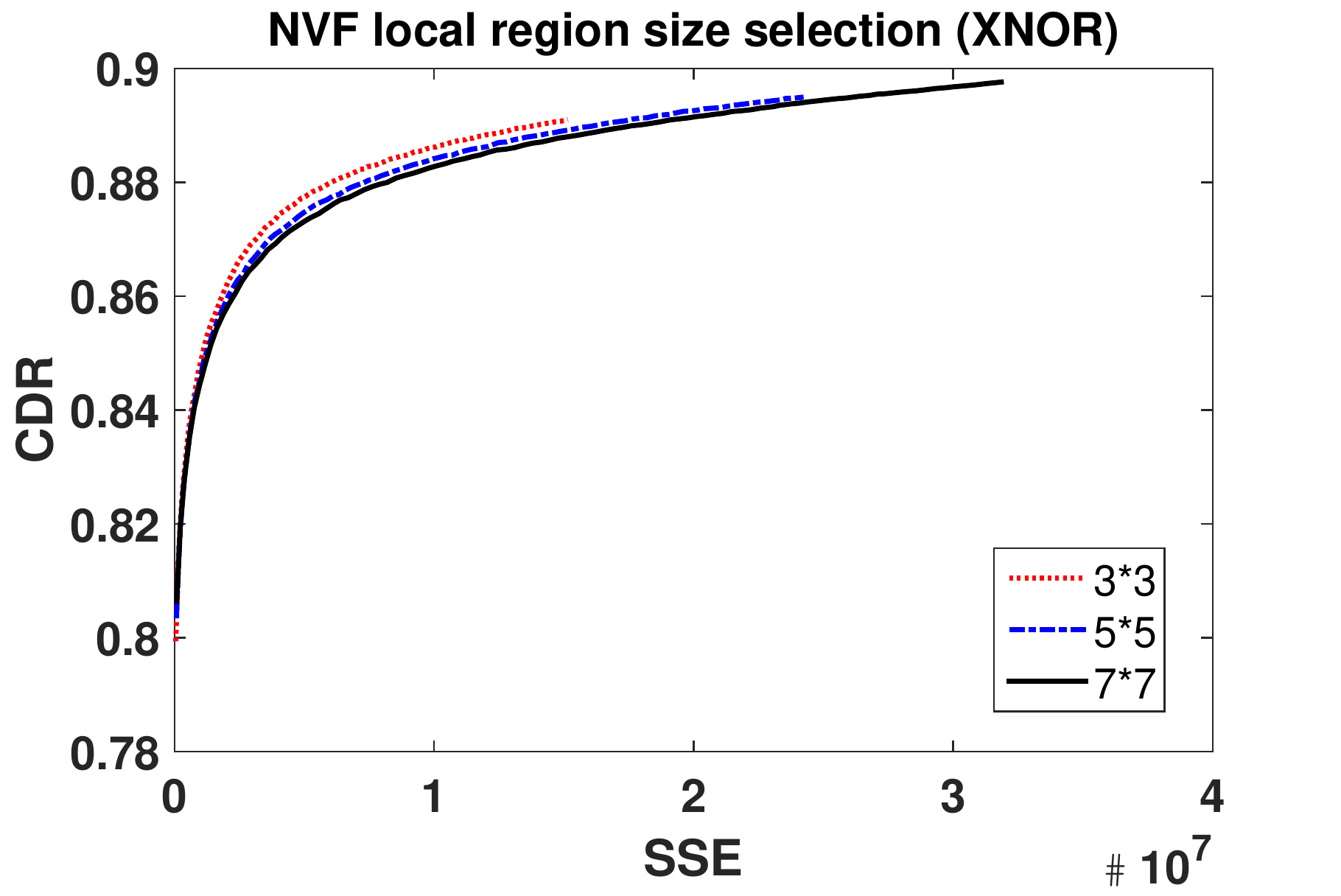}\label{fig:nvfsizexnorcompare}}
\vspace{-0.2cm}
\caption{NVF local region size selection. (a) AND operation decoded results. Steinberg kernel.
(b) XNOR operation decoded results. Steinberg kernel.
\label{fig:nvfsizecompare}}
\vspace{-0.3cm}
\end{figure*}

First, the local region size of NVF calculation $SR_x$ is explored. Here, IF is disabled. $3\times3$, $5\times5$ and $7\times7$ local region sizes are tested and compared. According to the results, $SR_x=3$ is recommended since it gives the best performance and it possesses the lowest computational complexity. The average results are introduced in Fig. \ref{fig:nvfsizecompare}.

\begin{figure*}
  \centering
  \subfigure[]{
    \includegraphics[width=5.8cm]{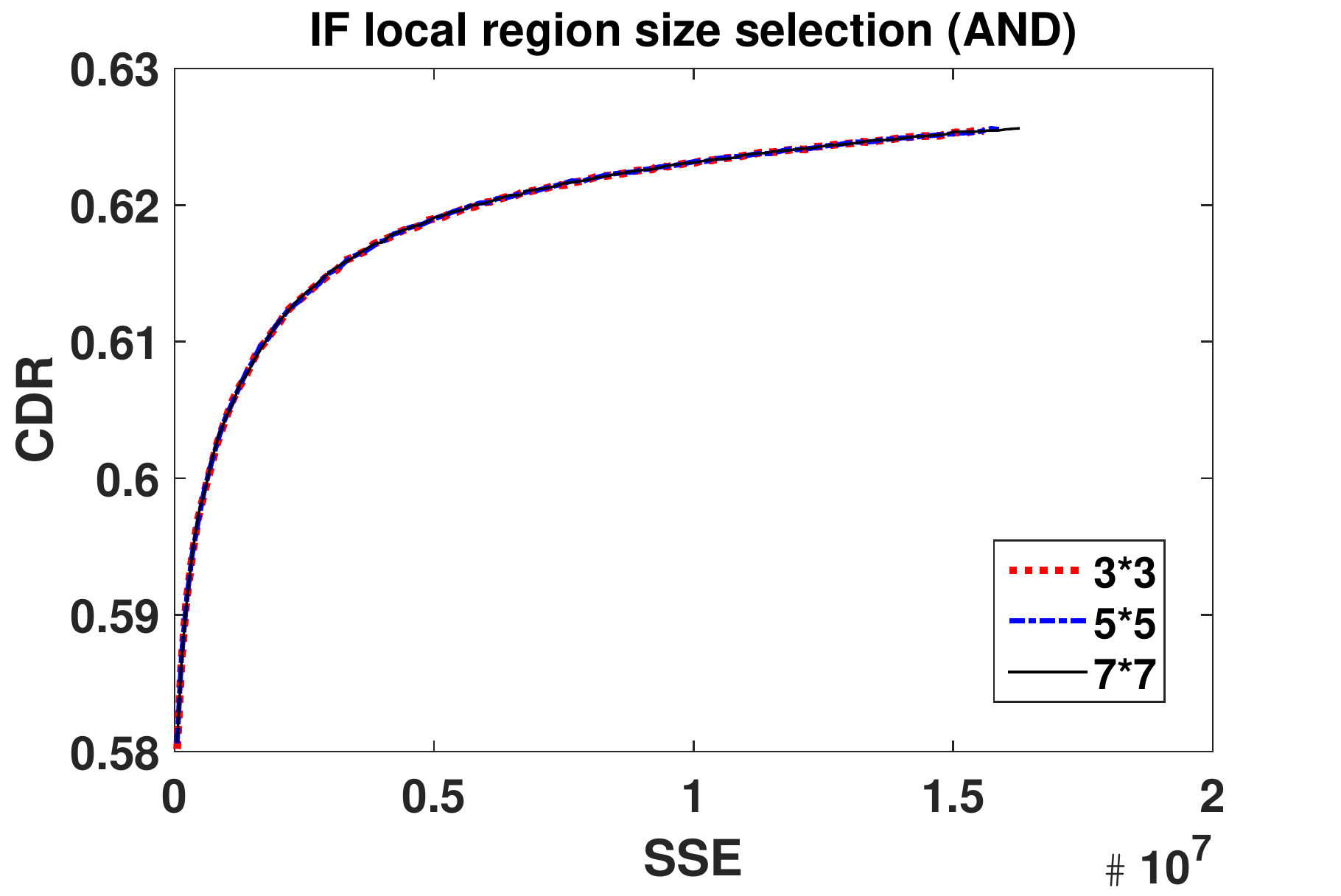}\label{fig:ifsizeandcompare}}
  \subfigure[]{
    \includegraphics[width=5.8cm]{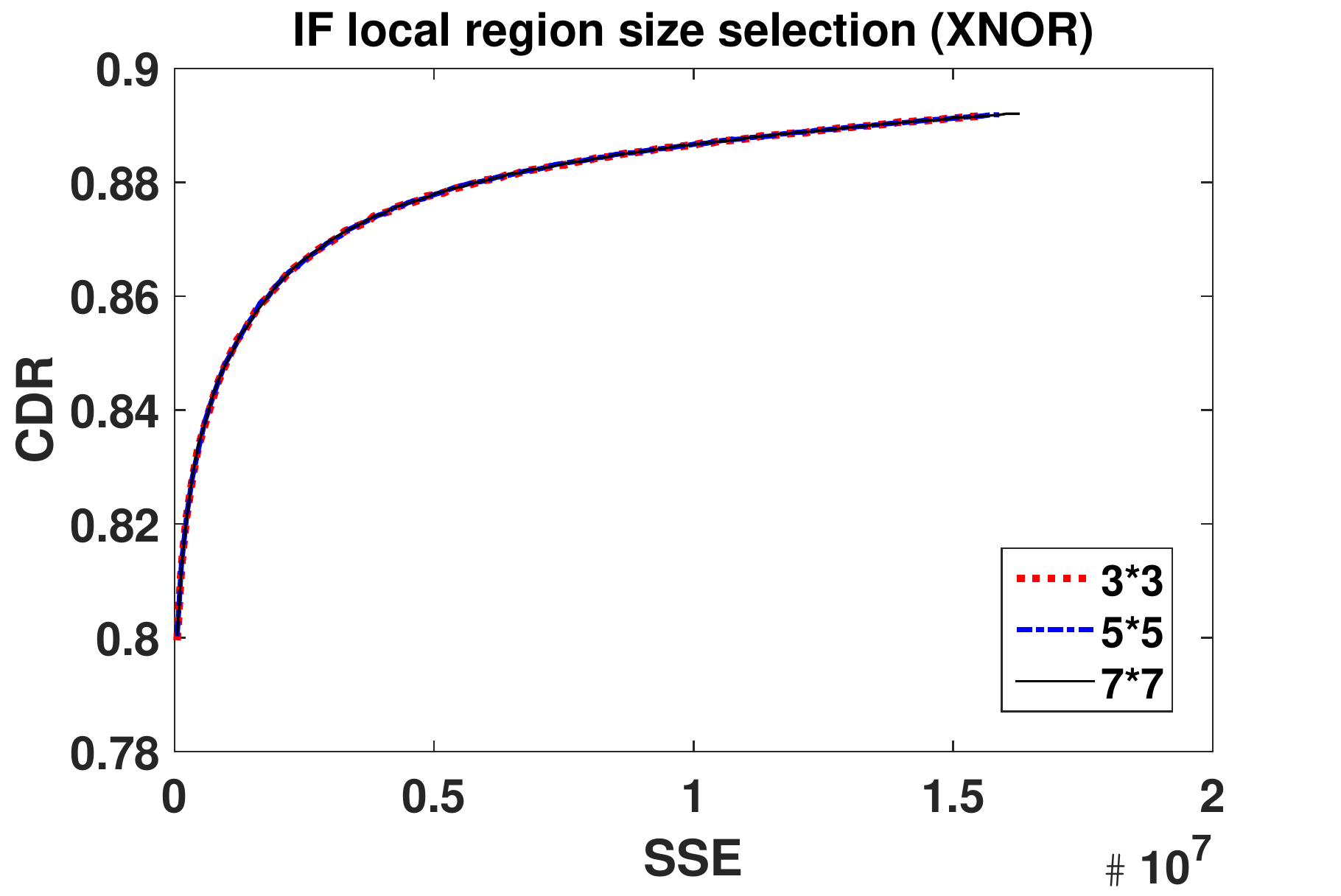}\label{fig:ifsizexnorcompare}}
\vspace{-0.2cm}
\caption{IF local region size selection. (a) AND operation decoded results. Steinberg kernel.
(b) XNOR operation decoded results. Steinberg kernel.
\label{fig:ifsizecompare}}
\vspace{-0.3cm}
\end{figure*}

Then, the local region size of IF $SR_w$ is explored. In this test, the local region size of NVF is set to be $3\times3$. Three different sizes $3\times3$, $5\times5$ and $7\times7$ are tested and compared. According to the experiments, though the performances of the three tested local region sizes are similar, $SR_w=3$ is recommended because it possesses the lowest computational complexity. The average results are displayed in Fig. \ref{fig:ifsizecompare}.

\subsection{Comparison Test}

With CaDEED-N\&I's parameters determined, CaDEED-EC and CaDEED-N\&I are compared to the classical and latest previous EDHVW methods, DHCED [\ref{au2003dhced}], DHDCED [\ref{guo2011dhdced}] and DEED(L-2) [\ref{guo2016hvw}].

\begin{figure*}
  \centering
  \subfigure[]{
    \includegraphics[width=5.8cm]{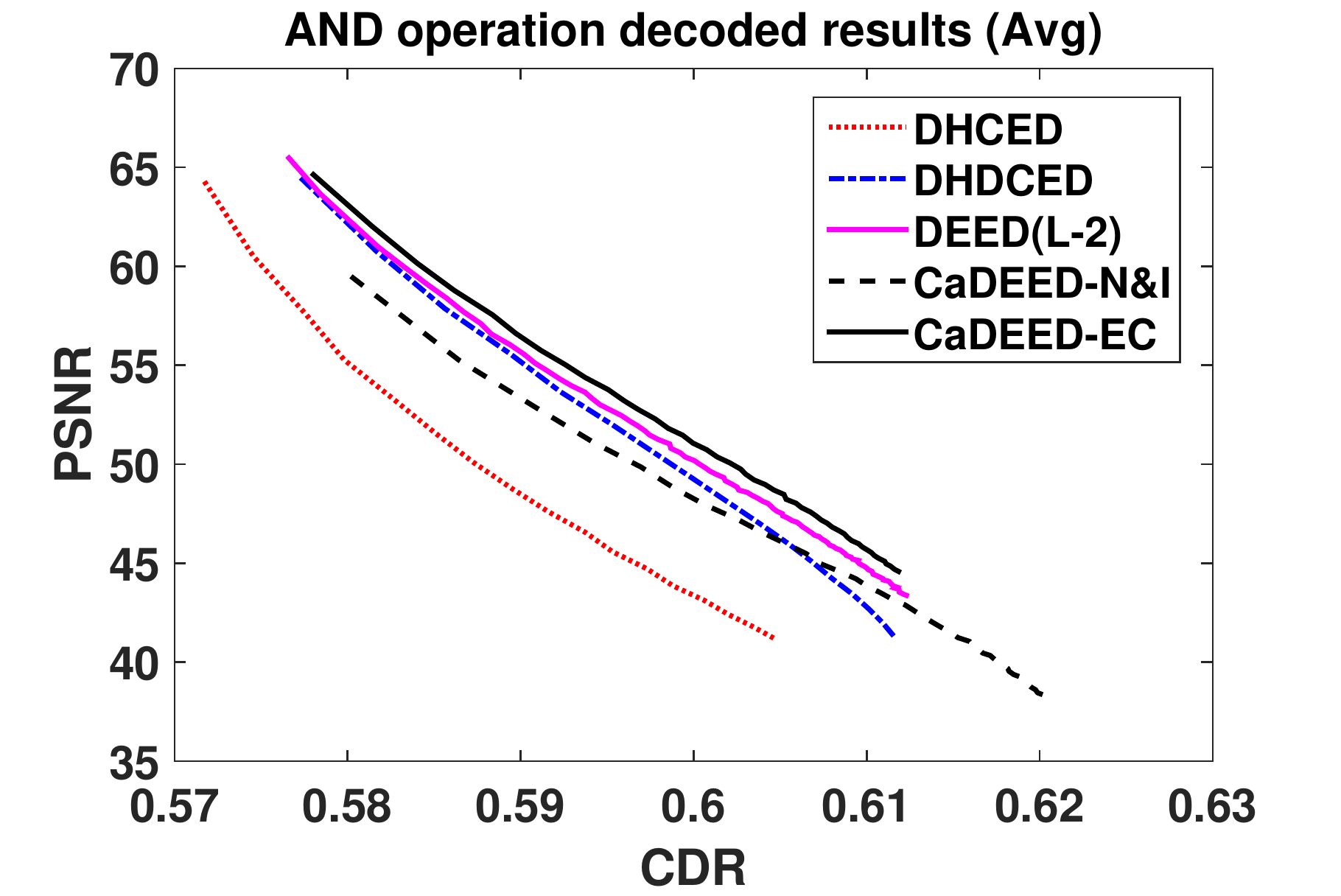}\label{fig:avganddecodedpsnr}}
  \subfigure[]{
    \includegraphics[width=5.8cm]{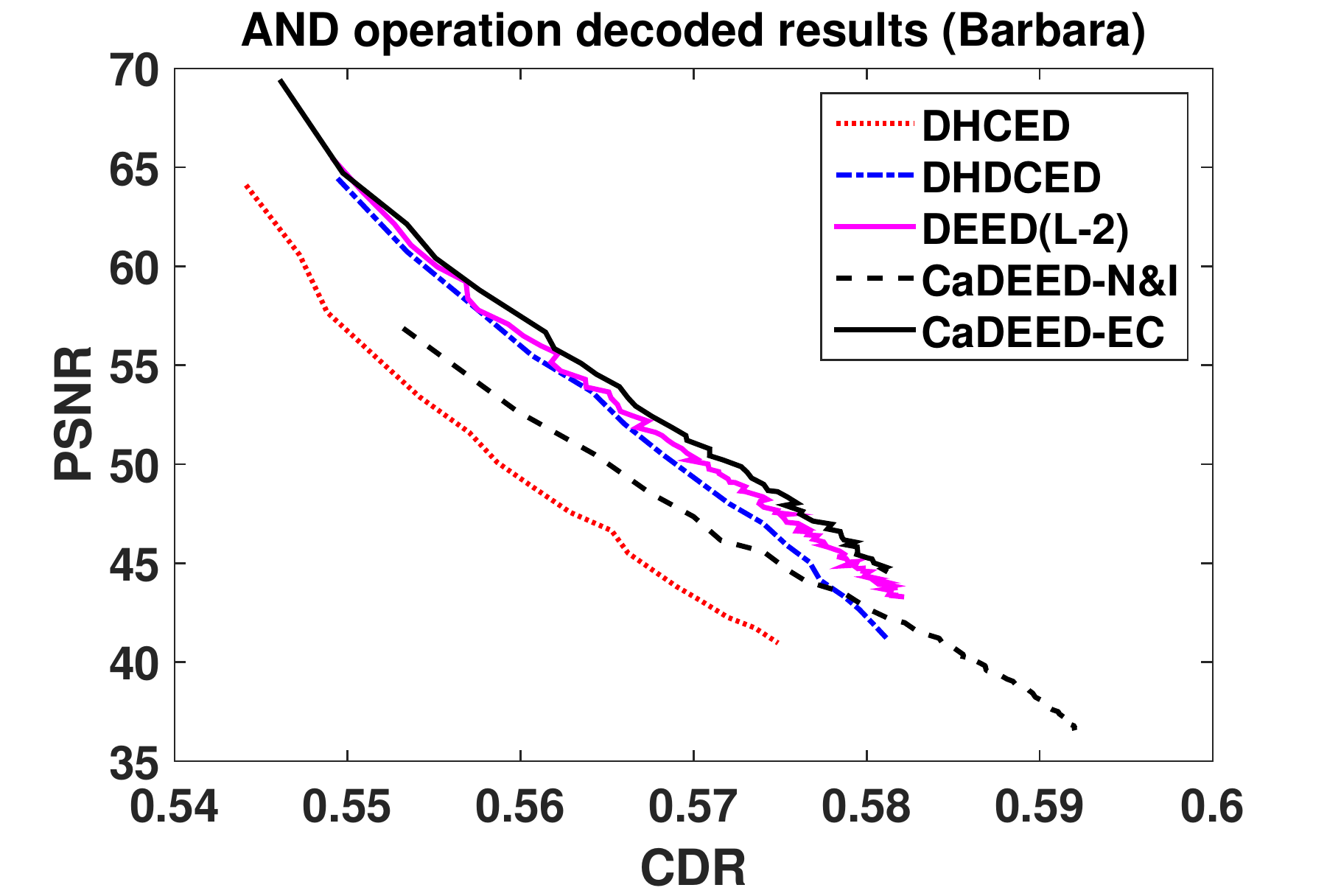}\label{fig:barbaraanddecodedpsnr}}
  \subfigure[]{
    \includegraphics[width=5.8cm]{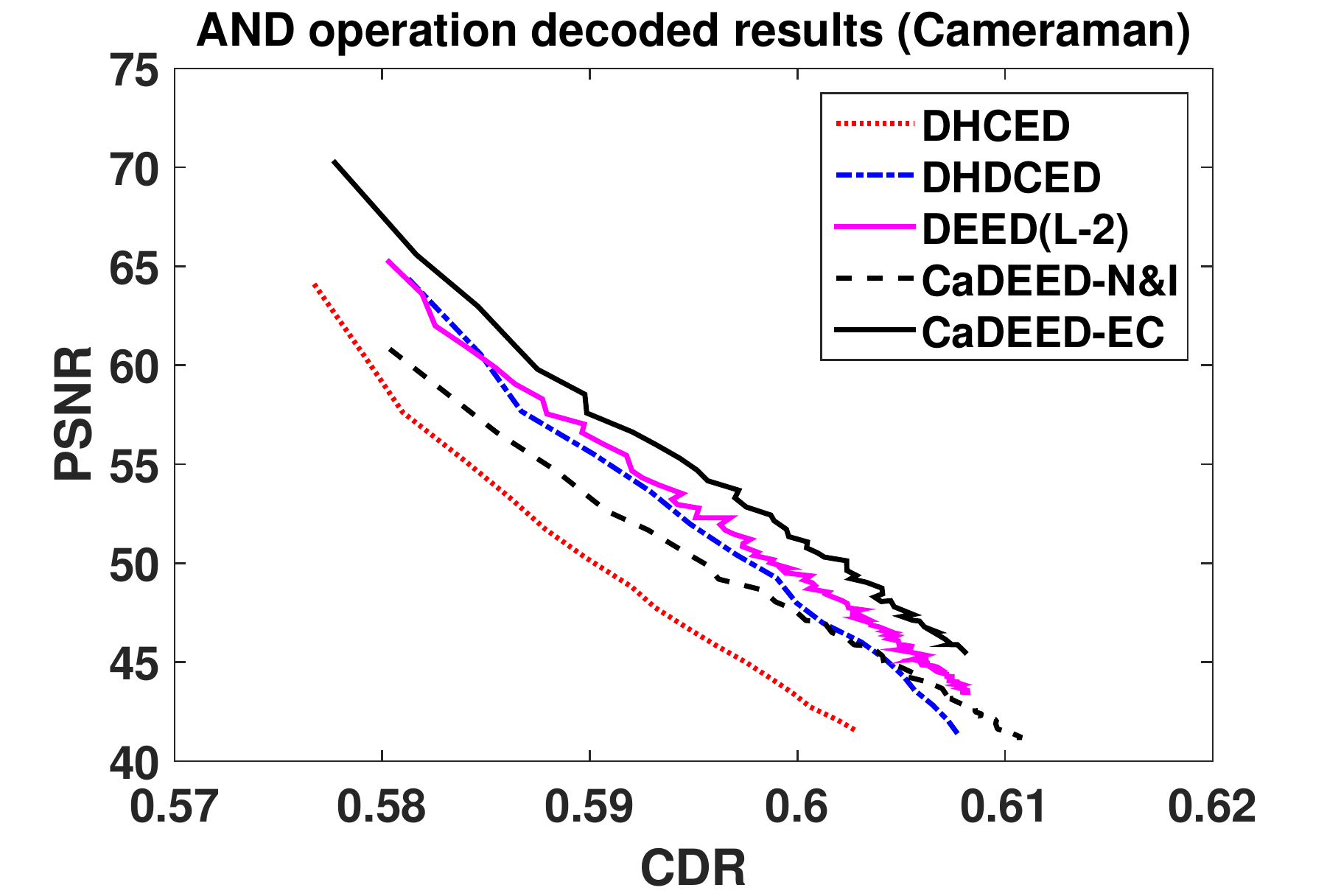}\label{fig:cameramananddecodedpsnr}}
\vspace{-0.2cm}
\caption{(a) Average PSNRs of the AND operation decoded results. Steinberg kernel.
(b) PSNRs of the AND operation decoded results. $X_1=X_2=`Barbara'$, Steinberg kernel. (c) PSNRs of the AND operation decoded results. $X_1=X_2=`Cameraman'$, Steinberg kernel.
\label{fig:psnranddecodedresults}}
\vspace{-0.3cm}
\end{figure*}

\begin{figure*}
  \centering
  \subfigure[]{
    \includegraphics[width=5.8cm]{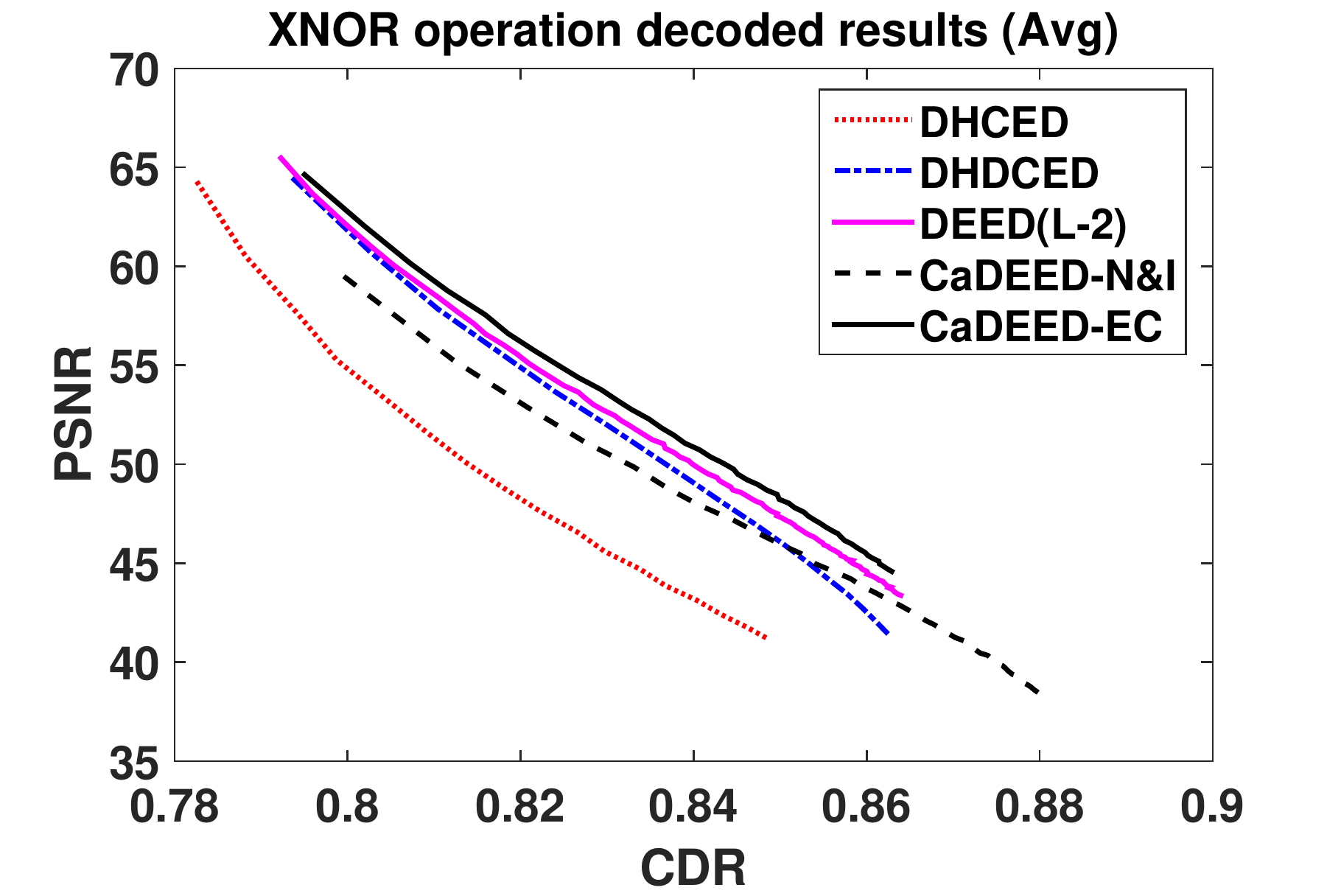}\label{fig:avgxnordecodedpsnr}}
  \subfigure[]{
    \includegraphics[width=5.8cm]{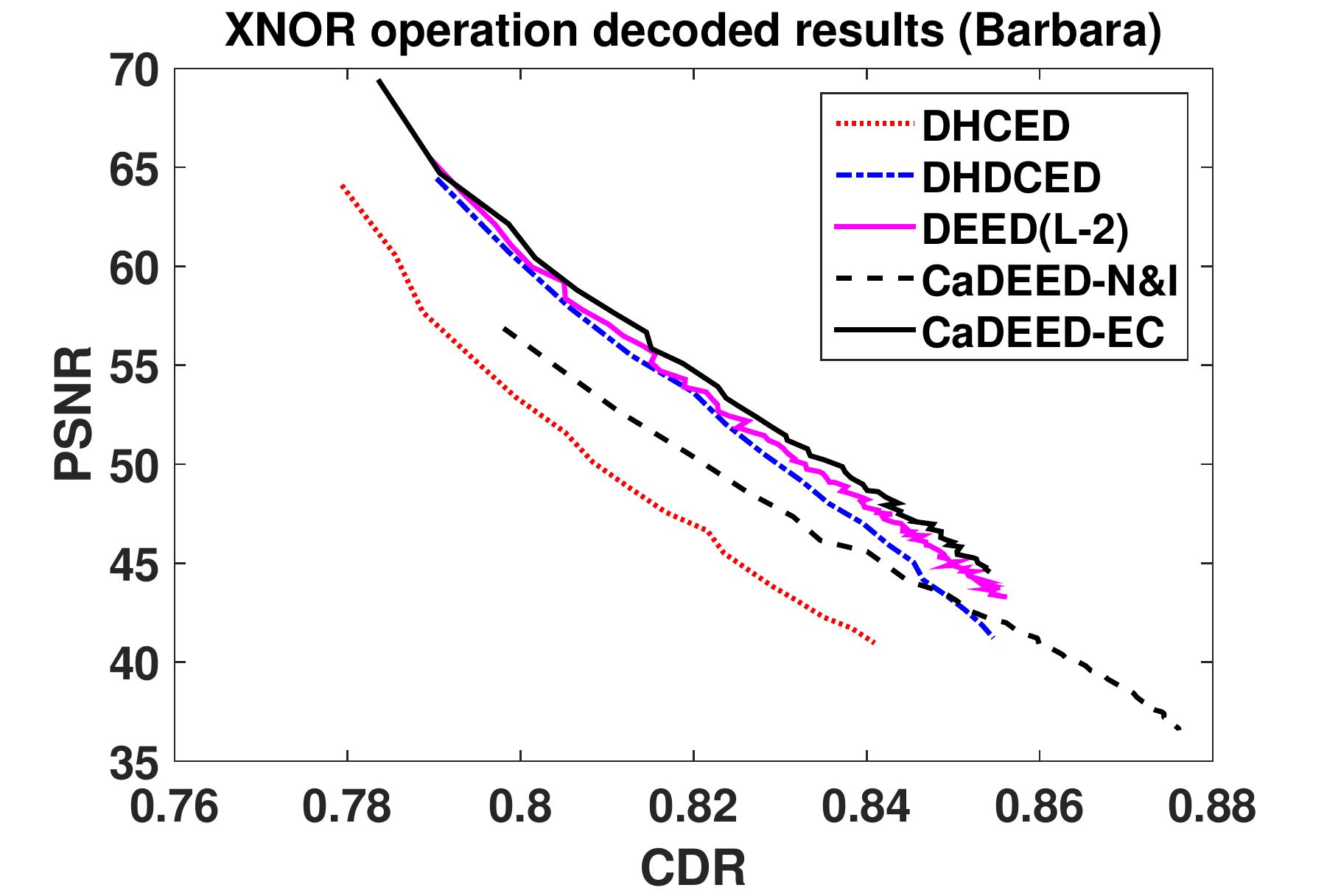}\label{fig:barbaraxnordecodedpsnr}}
  \subfigure[]{
    \includegraphics[width=5.8cm]{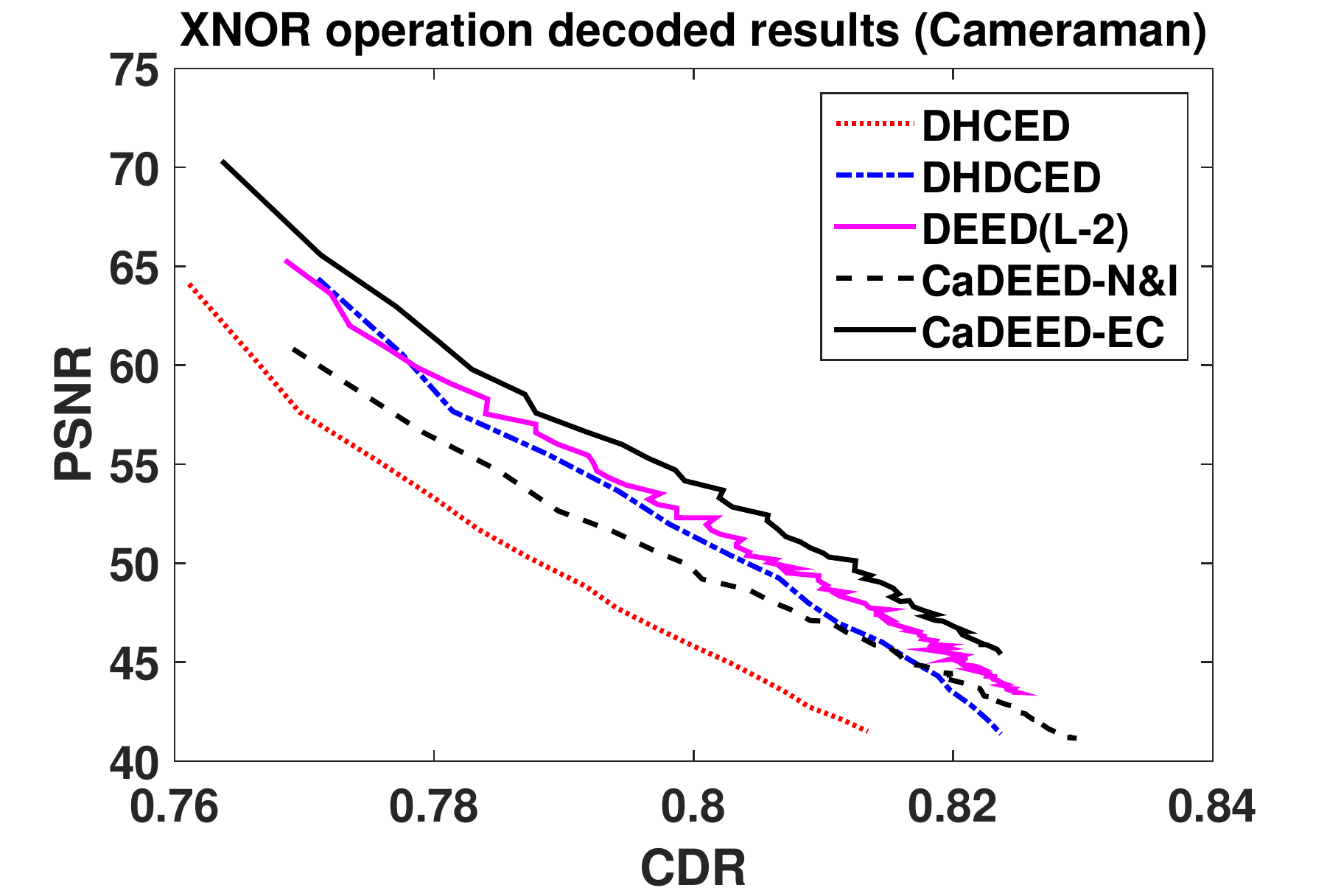}\label{fig:cameramanxnordecodedpsnr}}
\vspace{-0.2cm}
\caption{(a) Average PSNRs of the XNOR operation decoded results. Steinberg kernel.
(b) PSNRs of the XNOR operation decoded results. $X_1=X_2=`Barbara'$, Steinberg kernel. (c) PSNRs of the XNOR operation decoded results. $X_1=X_2=`Cameraman'$, Steinberg kernel.
\label{fig:psnrxnordecodedresults}}
\vspace{-0.3cm}
\end{figure*}

Similar to [\ref{guo2016hvw}], the traditional measurements Peak Signal-to-Noise Ratio (PSNR) and CDR are employed to measure the performances of the EDHVW methods. Figs. \ref{fig:psnranddecodedresults} and \ref{fig:psnrxnordecodedresults} gives the results of DHCED, DHDCED, DEED(L-2), CaDEED-EC and CaDEED-N\&I. Figs. \ref{fig:avganddecodedpsnr} and \ref{fig:avgxnordecodedpsnr} presents the average results, while Figs. \ref{fig:barbaraanddecodedpsnr}-\ref{fig:cameramananddecodedpsnr} and \ref{fig:barbaraxnordecodedpsnr}-\ref{fig:cameramanxnordecodedpsnr} introduces the results for specific test images, `Barbara' and `Cameraman'. Note that for fair comparison, when we plot these figures, the PSNRs results of DHDCED, DEED(L-2) and CaDEED are the average PSNRs of the two stego images. As we can observe, CaDEED-EC performs obviously better than the previous methods with an approximately 1dB gain when $\textrm{CB-CDR}=0.6$ (in Fig. \ref{fig:avganddecodedpsnr}), $0.84$ (in Fig. \ref{fig:avgxnordecodedpsnr}). Since CaDEED-N\&I allows larger distortions for the image regions which contain complex content and PSNR penalizes large distortions more, it is not surprising that its results are not impressive with the traditional measurements.

To better illustrate the superiority of CaDEED-N\&I, since complex regions in a image can tolerate more noise than the flat regions, we modify the PSNR measurement in [\ref{guo2016hvw}] to noise tolerant PSNR (NT-PSNR), which is defined in Eq. (\ref{eq:ntmse}).
\begin{equation} \label{eq:ntmse}
    \textrm{NT-PSNR}=10*log_{10}(\frac{255^2}{\frac{\sum(\Delta u^2_k(i,j)*v_k(i,j))}{IW_k*IH_k}}), k \in \{1, 2\},
\end{equation}

Also, the traditional CDR will process every pixel in the secret pattern with equal weights. However, the edge/texture pixels in the secret pattern tend to play a bigger role in maintaining its structure. Therefore, by assigning more weights to those edge/texture pixels, here we propose our new measurement for assessing the decoded secret pattern, which is called the content based correct decoding rate (CB-CDR).
\begin{equation} \label{eq:cbcdr}
    \textrm{CB-CDR}=\frac{\sum{(\gamma(i,j)\times |w(i,j) \odot d(i,j)|)}}{\sum{\gamma(i,j)}}
\end{equation}

\begin{figure*}
  \centering
  \subfigure[]{
    \includegraphics[width=5.8cm]{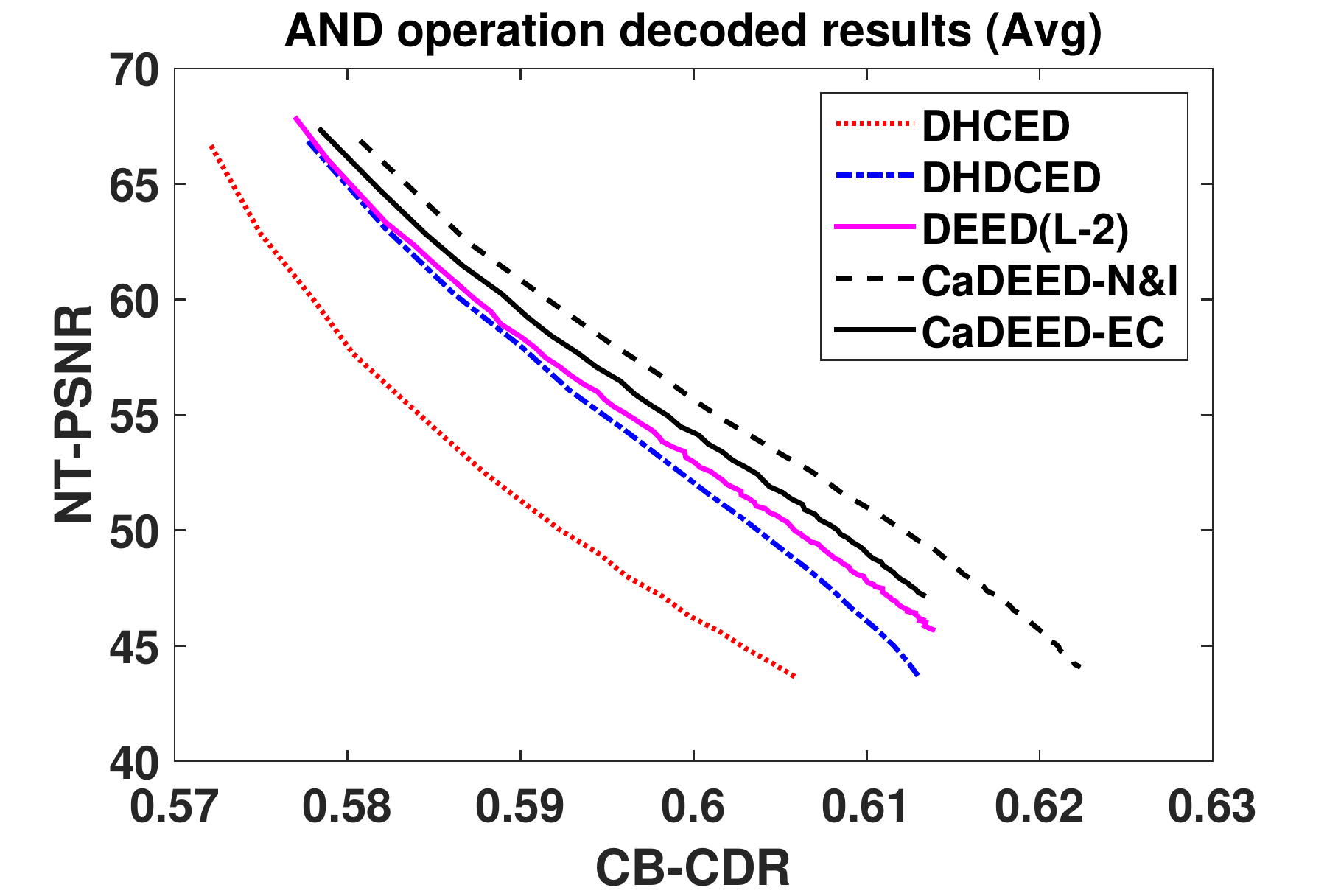}\label{fig:avganddecodedntpsnr}}
  \subfigure[]{
    \includegraphics[width=5.8cm]{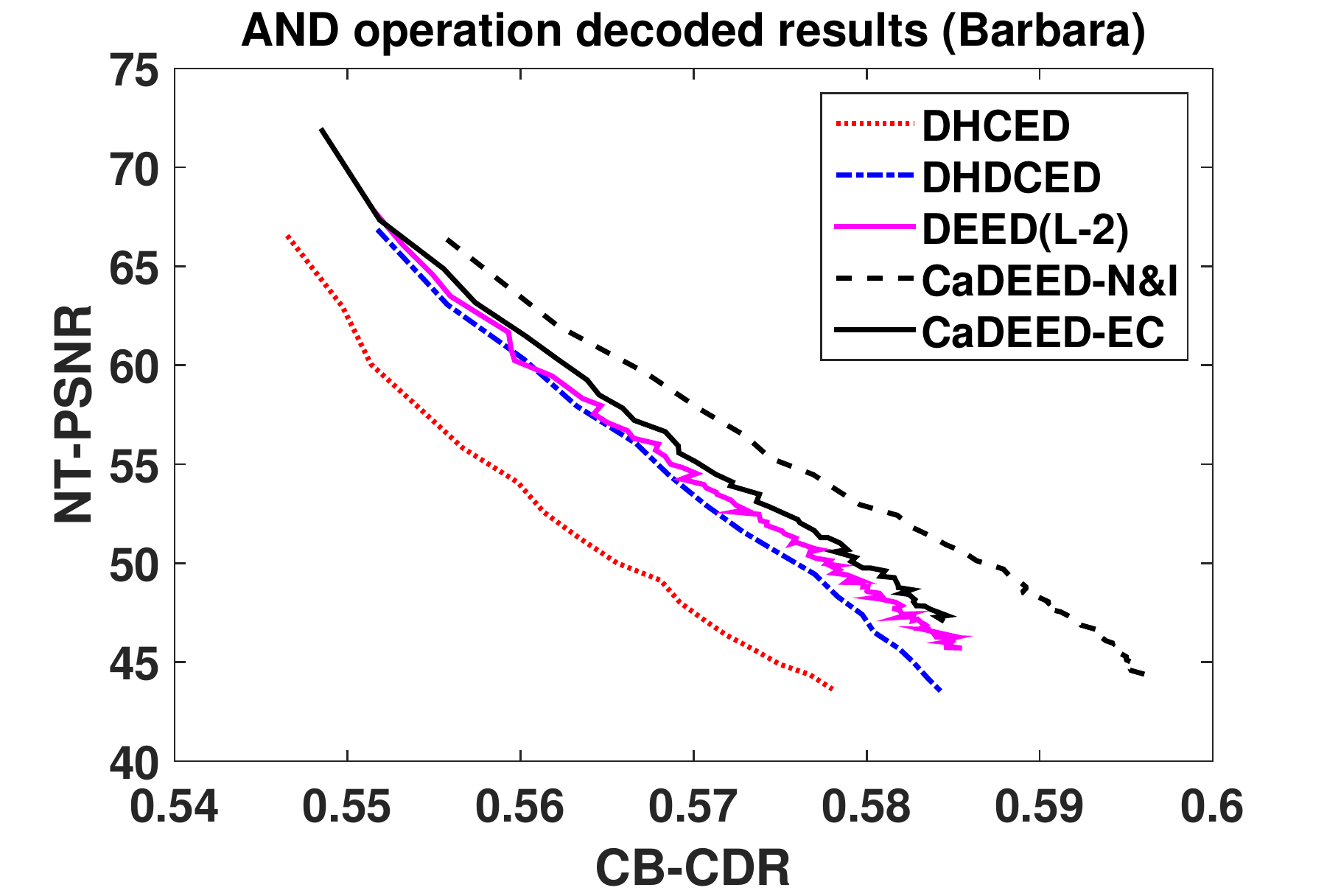}\label{fig:barbaraanddecodedntpsnr}}
  \subfigure[]{
    \includegraphics[width=5.8cm]{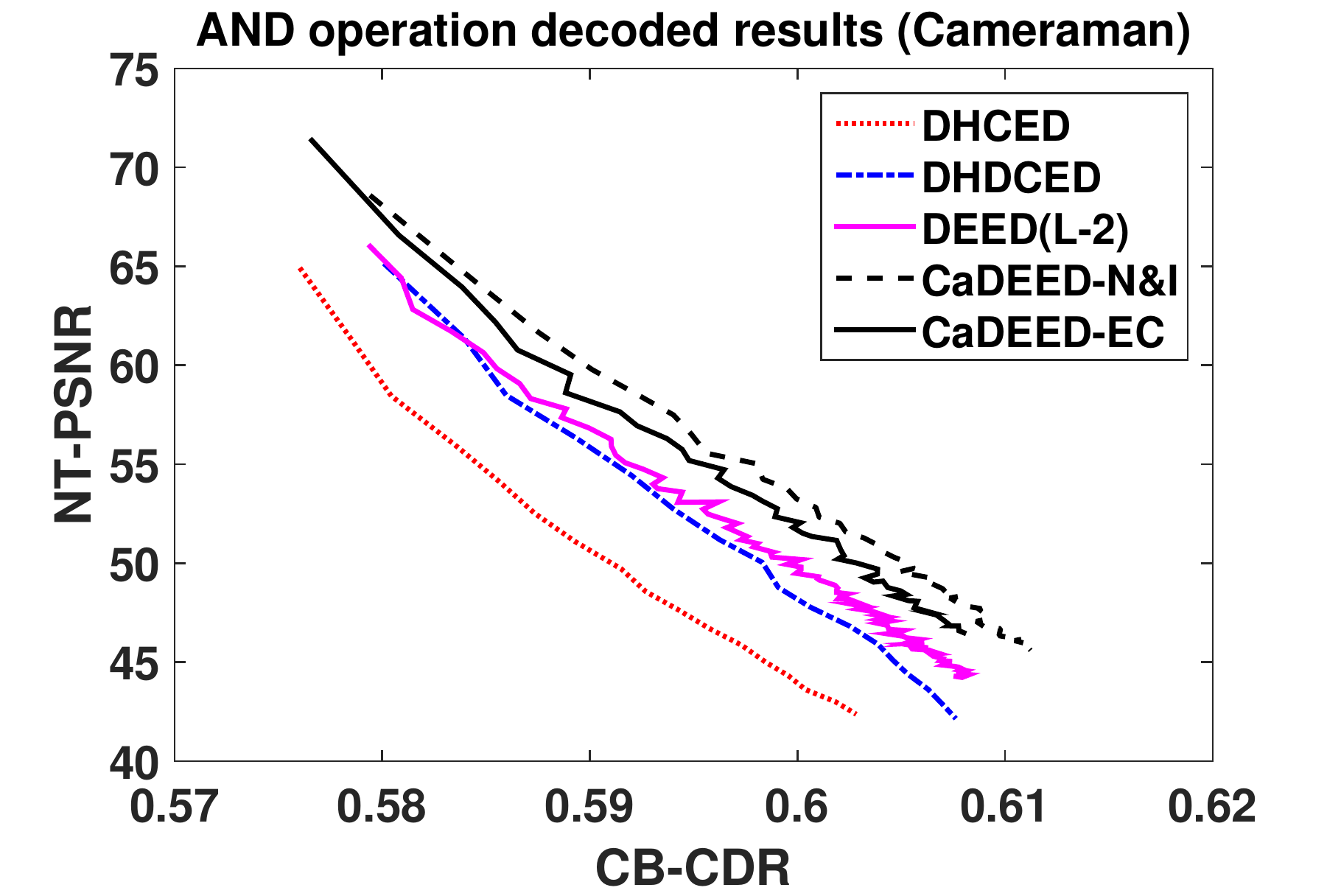}\label{fig:cameramananddecodedntpsnr}}
\vspace{-0.2cm}
\caption{(a) Average NT-PSNRs of the AND operation decoded results. Steinberg kernel.
(b) NT-PSNRs of the AND operation decoded results. $X_1=X_2=`Barbara'$, Steinberg kernel. (c) NT-PSNRs of the AND operation decoded results. $X_1=X_2=`Cameraman'$, Steinberg kernel.
\label{fig:ntpsnranddecodedresults}}
\vspace{-0.3cm}
\end{figure*}

\begin{figure*}
  \centering
  \subfigure[]{
    \includegraphics[width=5.8cm]{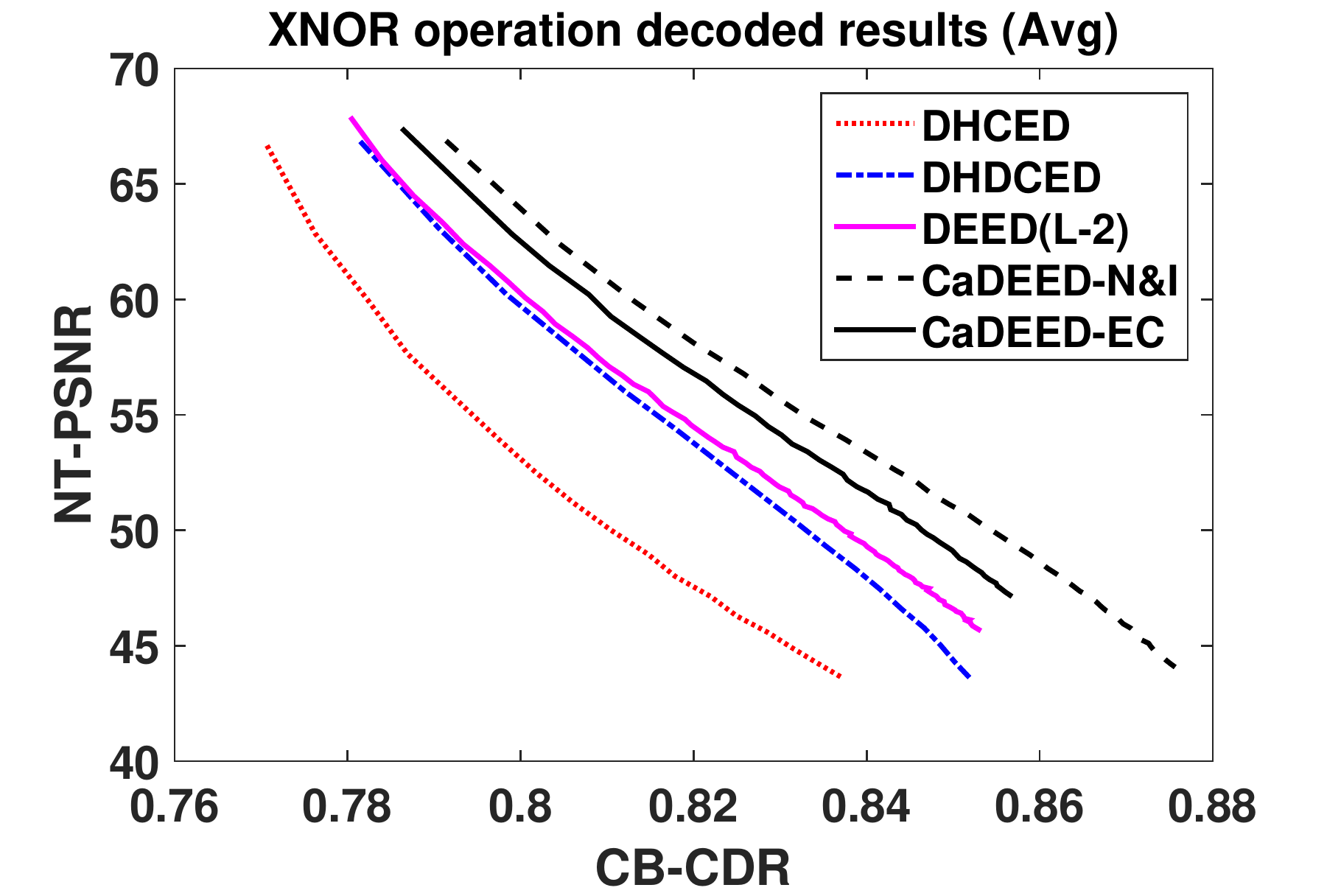}\label{fig:avgxnordecodedntpsnr}}
  \subfigure[]{
    \includegraphics[width=5.8cm]{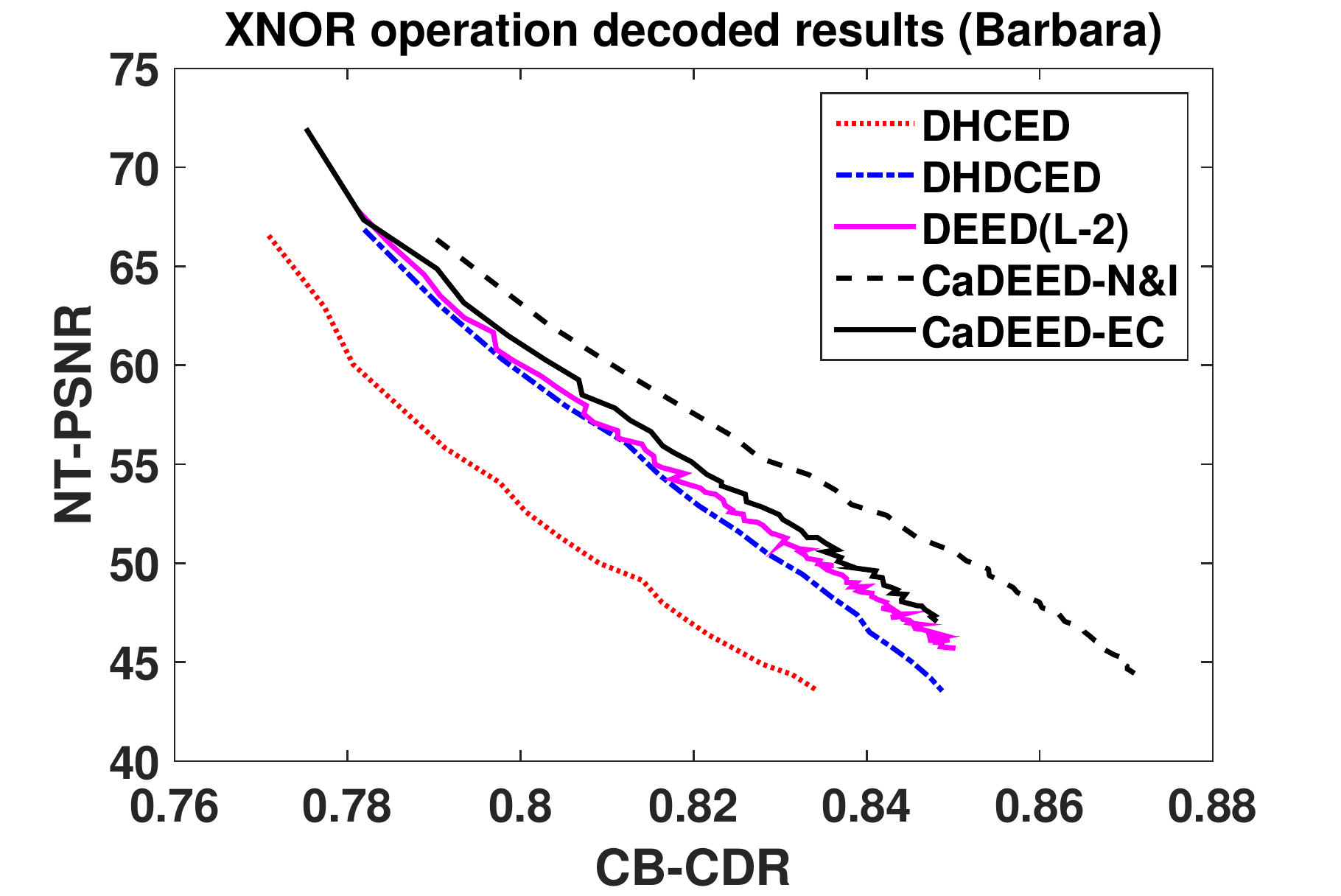}\label{fig:barbaraxnordecodedntpsnr}}
  \subfigure[]{
    \includegraphics[width=5.8cm]{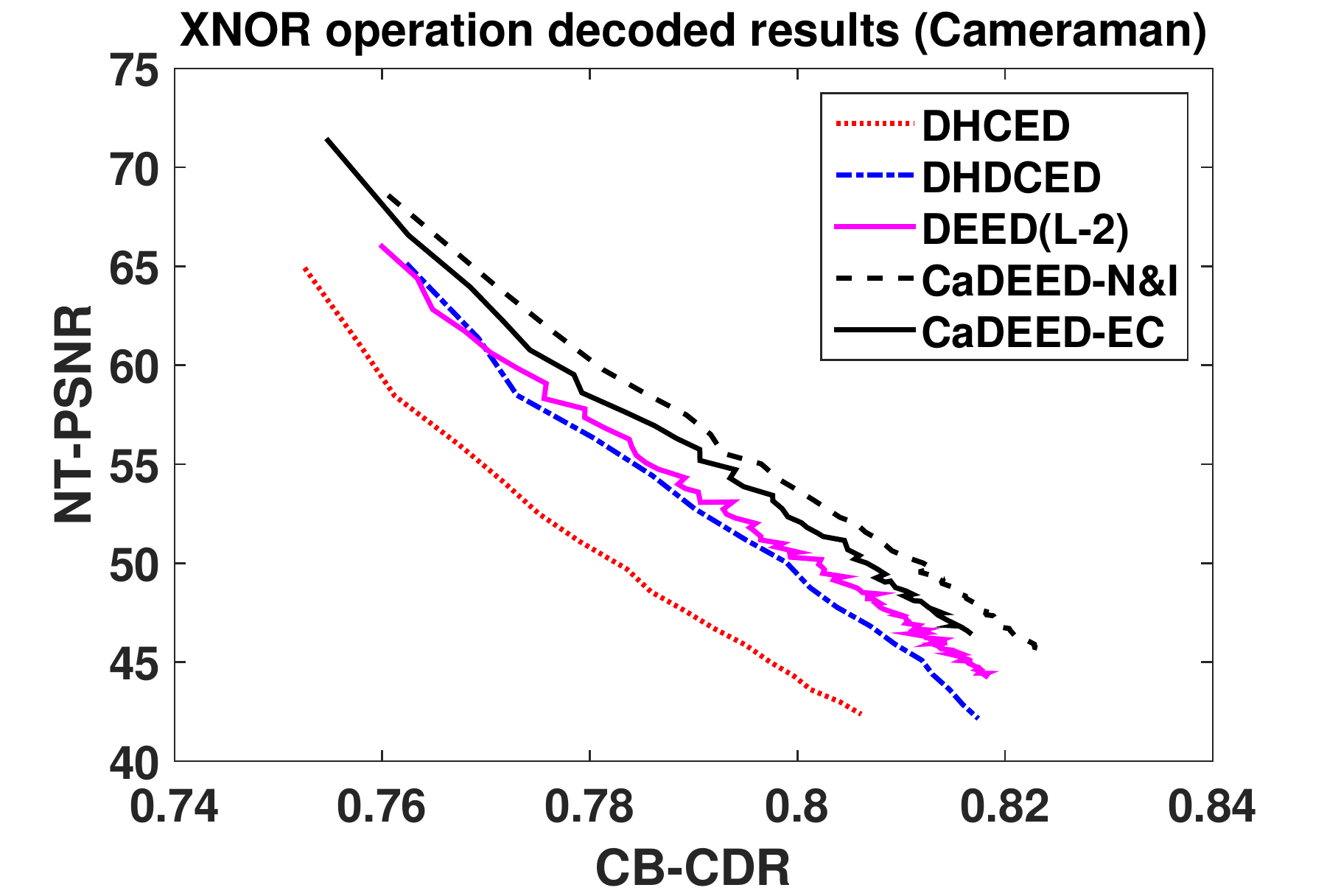}\label{fig:cameramanxnordecodedntpsnr}}
\vspace{-0.2cm}
\caption{(a) Average NT-PSNRs of the XNOR operation decoded results. Steinberg kernel.
(b) NT-PSNRs of the XNOR operation decoded results. $X_1=X_2=`Barbara'$, Steinberg kernel. (c) NT-PSNRs of the XNOR operation decoded results. $X_1=X_2=`Cameraman'$, Steinberg kernel.
\label{fig:ntpsnrxnordecodedresults}}
\vspace{-0.3cm}
\end{figure*}

Similar to Figs. \ref{fig:psnranddecodedresults} and \ref{fig:psnrxnordecodedresults}, Figs. \ref{fig:avganddecodedntpsnr} and \ref{fig:avgxnordecodedntpsnr} presents the average NT-PSNR results, while Figs. \ref{fig:barbaraanddecodedntpsnr}-\ref{fig:cameramananddecodedntpsnr} and \ref{fig:barbaraxnordecodedntpsnr}-\ref{fig:cameramanxnordecodedntpsnr} introduces the results for specific test images, `Barbara' and `Cameraman'. Note that for fair comparison, when we plot these figures, the NT-PSNRs results of DHDCED, DEED(L-2) and CaDEED are the average NT-PSNRs of the two stego images. As we can observe, CaDEED-EC and CaDEED-N\&I both perform obviously better than the previous methods. For $\textrm{CB-CDR}=0.6$ (in Fig. \ref{fig:avganddecodedntpsnr}), CaDEED-EC gives an approximately 1.2dB gain while CaDEED-N\&I presents an approximately 2.8dB gain compared to DEED(L-2). For $CB-CDR=0.84$ (in Fig. \ref{fig:avgxnordecodedntpsnr}), CaDEED-EC presents an approximately 2.3dB gain and CaDEED-N\&I gives an approximately 4dB gain compared to DEED(L-2).

\begin{figure*}
  \centering
  \subfigure[]{
    \includegraphics[width=3.0cm]{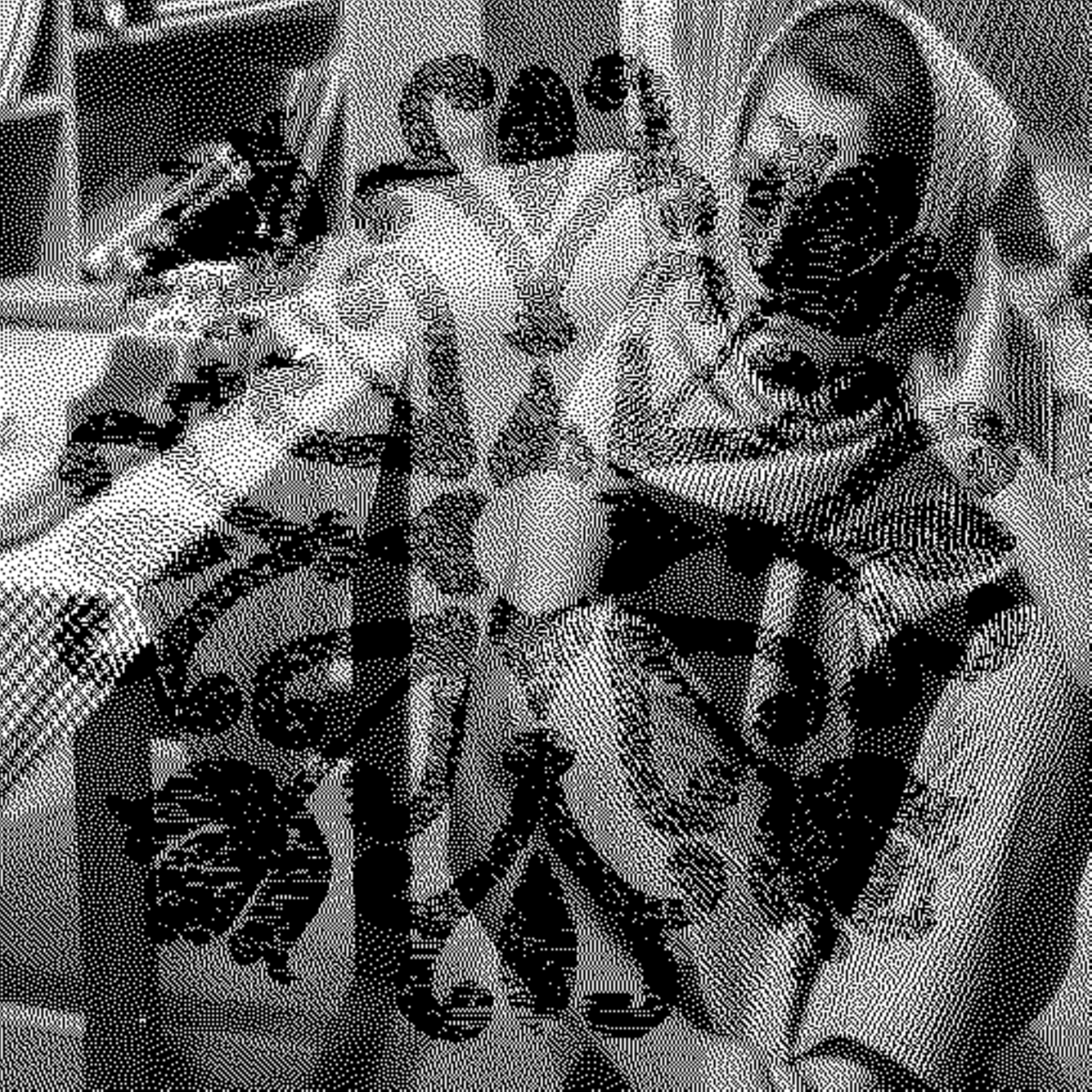}\label{fig:dhcedoverlay}}
  \subfigure[]{
    \includegraphics[width=3.0cm]{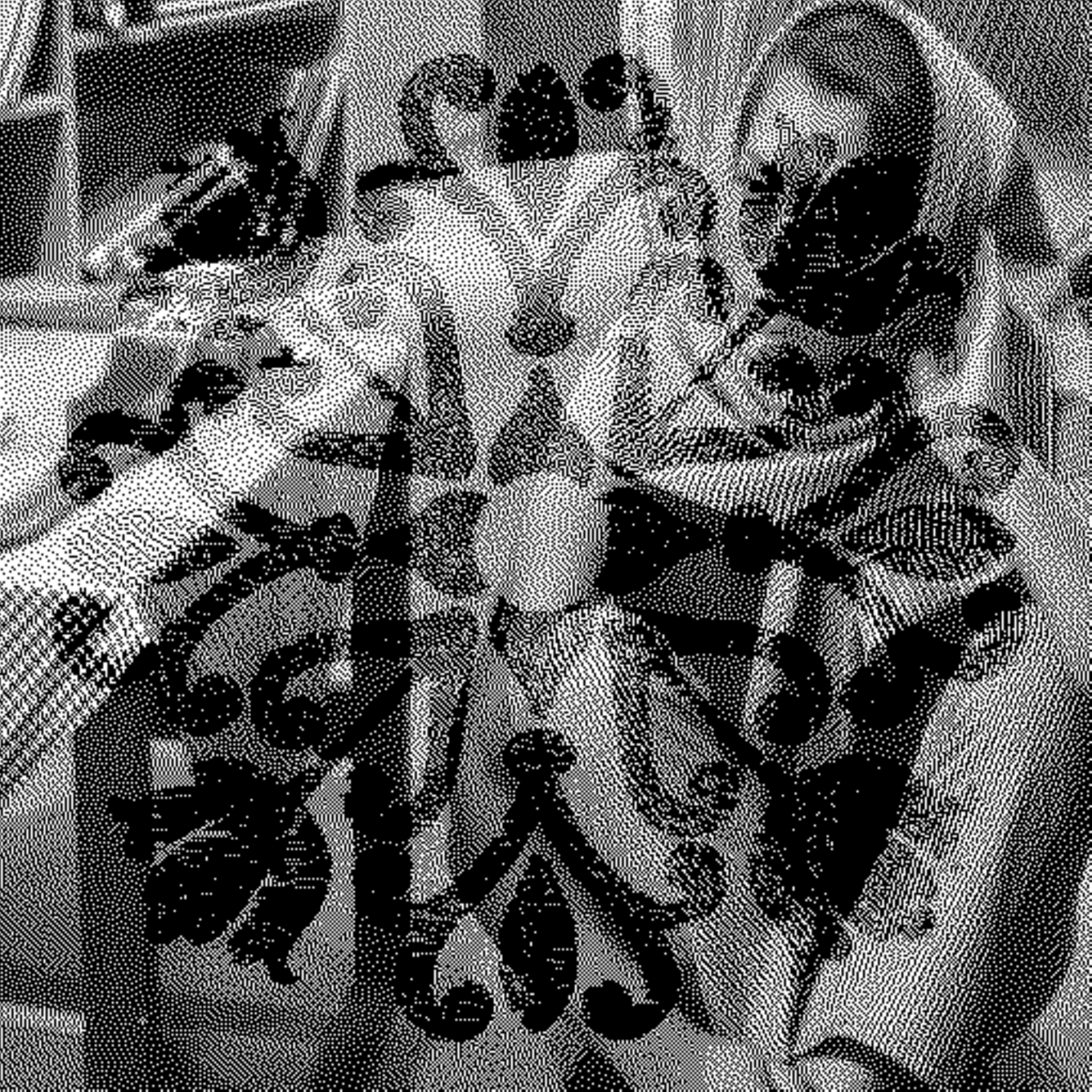}\label{fig:dhdcedoverlay}}
  \subfigure[]{
    \includegraphics[width=3.0cm]{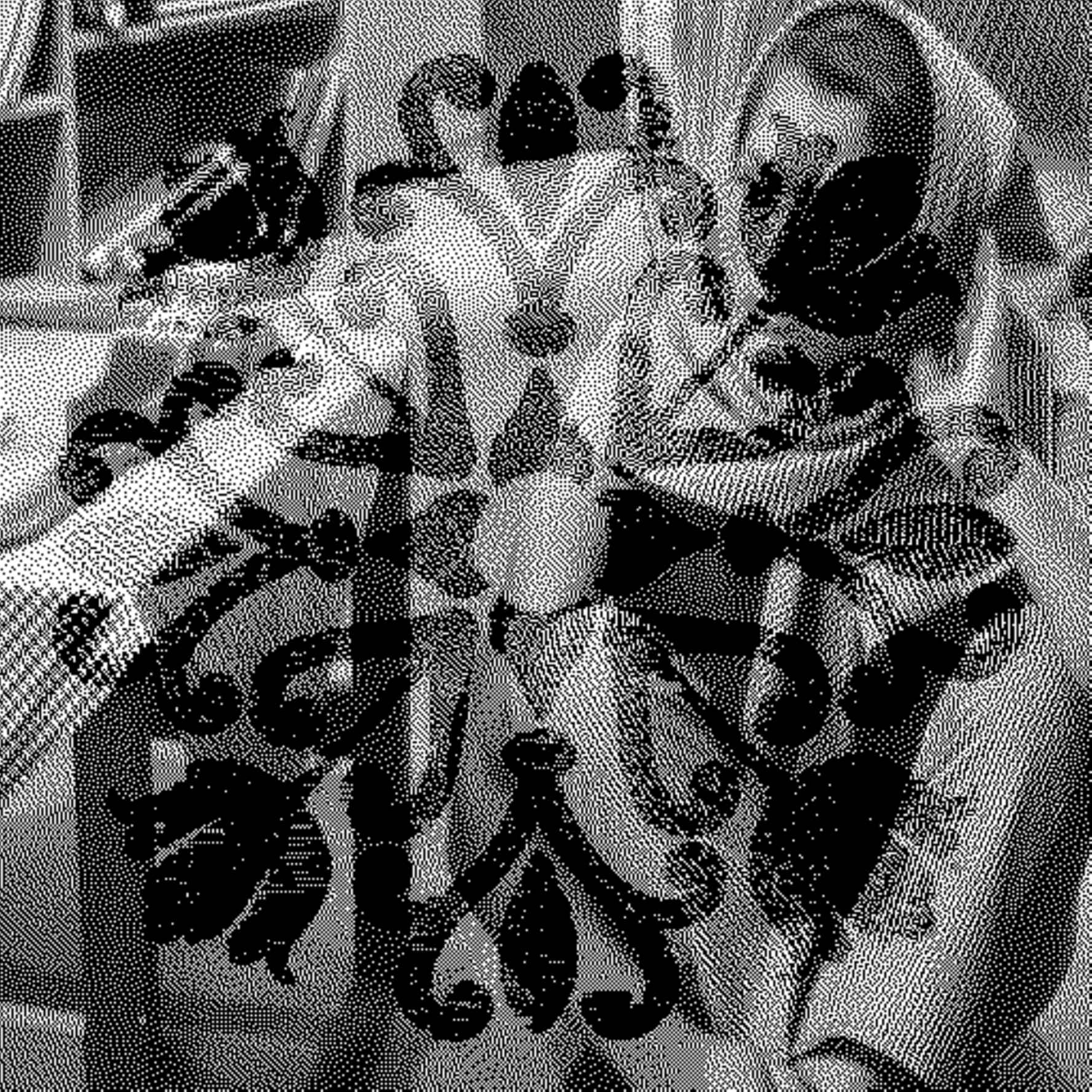}\label{fig:deedl2overlay}}
  \subfigure[]{
    \includegraphics[width=3.0cm]{0_0016barbara_cadeedecoverlays.pdf}\label{fig:cadeedecoverlay}}
  \subfigure[]{
    \includegraphics[width=3.0cm]{0_0008barbara_cadeednioverlays.pdf}\label{fig:cadeednioverlay}}
\vspace{-0.2cm}
\caption{AND operation decoded images, $X_1=X_2=`Barbara'$, Steinberg kernel. (a)DHCED; (b)DHDCED; (c)DEED(L-2); (d)CaDEED-EC; (e)CaDEED-N\&I.
\label{fig:subjectiveanddecoded}}
\vspace{-0.3cm}
\end{figure*}

\begin{figure*}
  \centering
  \subfigure[]{
    \setlength{\fboxrule}{1pt} \setlength{\fboxsep}{0cm}
    \fbox{\includegraphics[width=3.0cm]{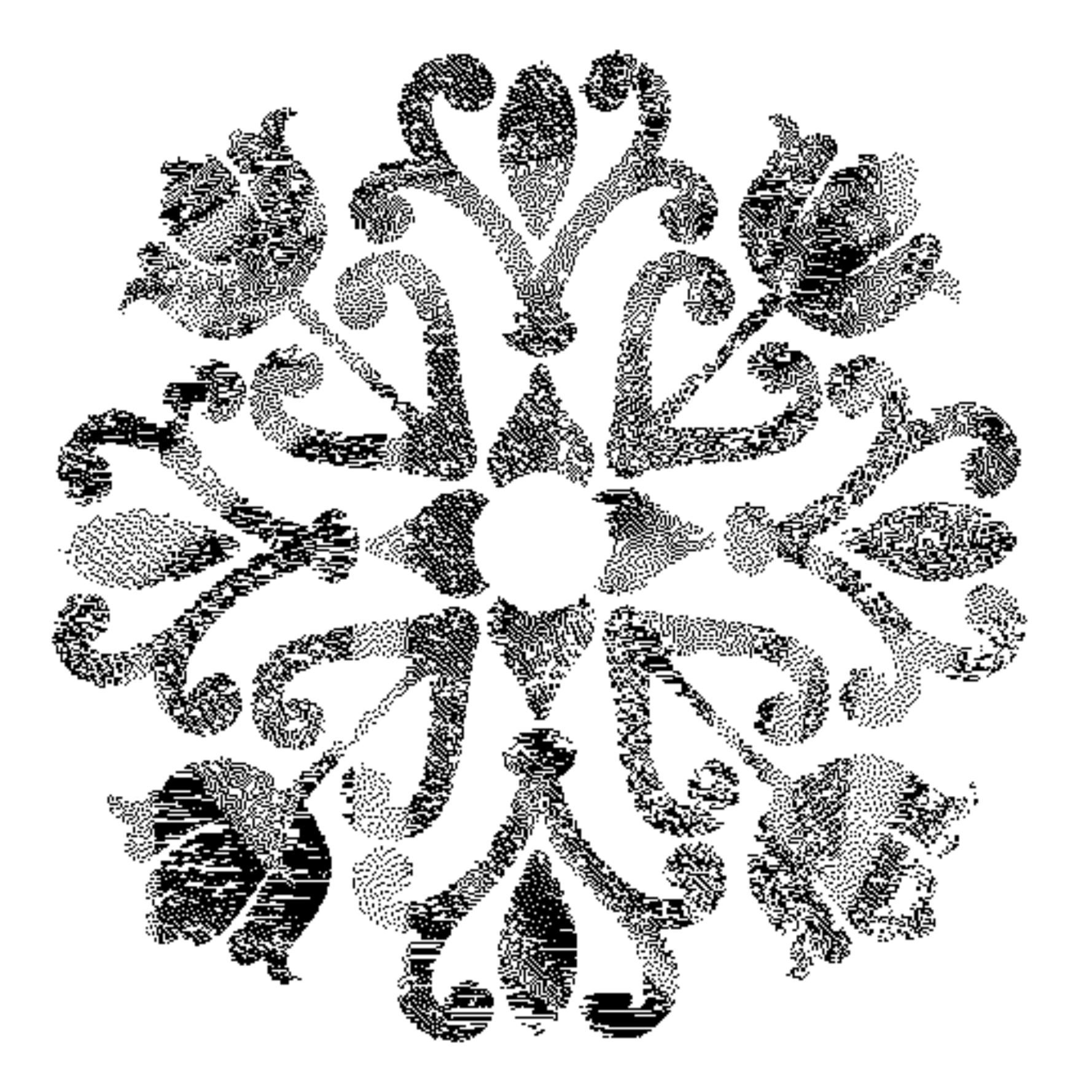}\label{fig:dhcedxnor}}}
  \subfigure[]{
    \setlength{\fboxrule}{1pt} \setlength{\fboxsep}{0cm}
    \fbox{\includegraphics[width=3.0cm]{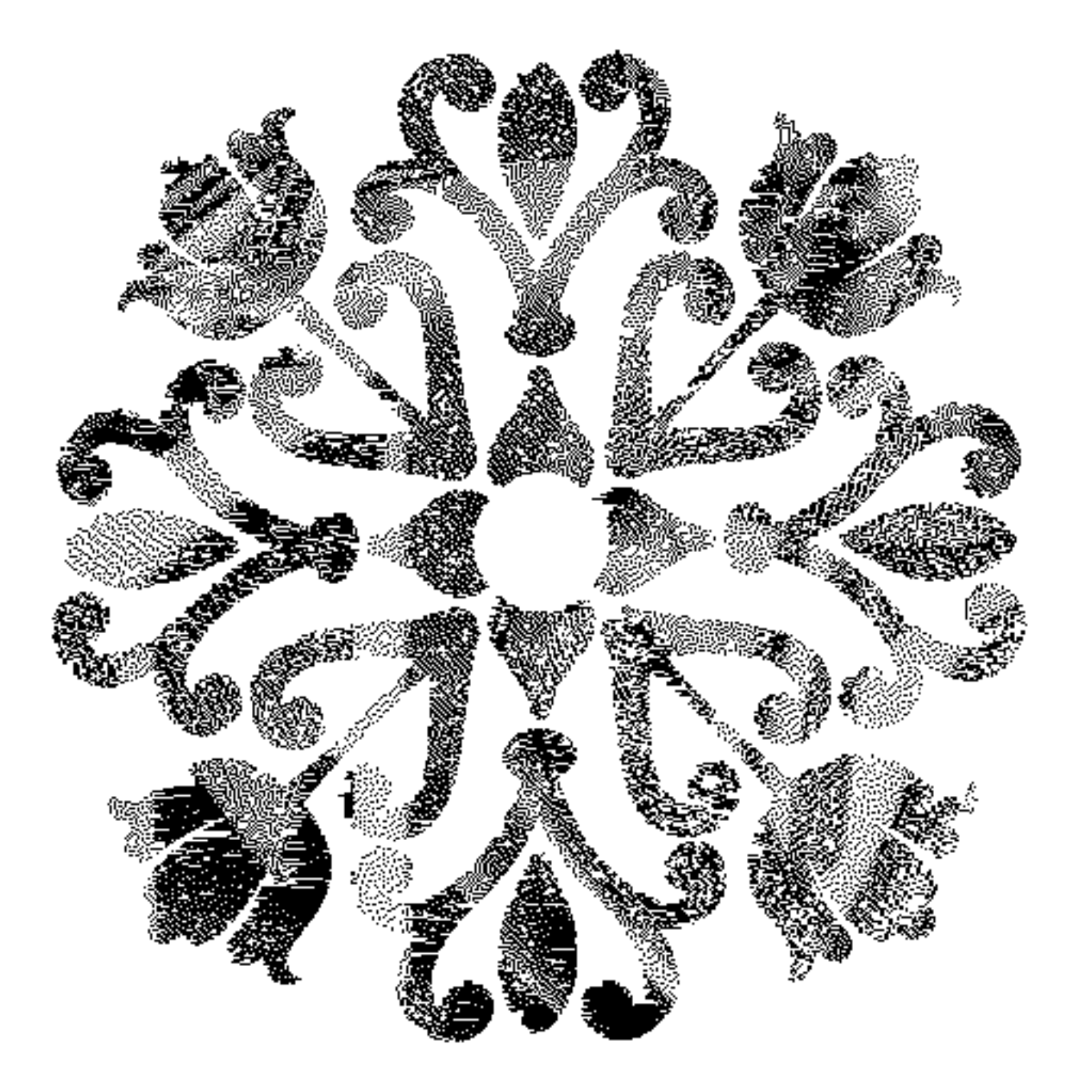}\label{fig:dhdcedxnor}}}
  \subfigure[]{
    \setlength{\fboxrule}{1pt} \setlength{\fboxsep}{0cm}
    \fbox{\includegraphics[width=3.0cm]{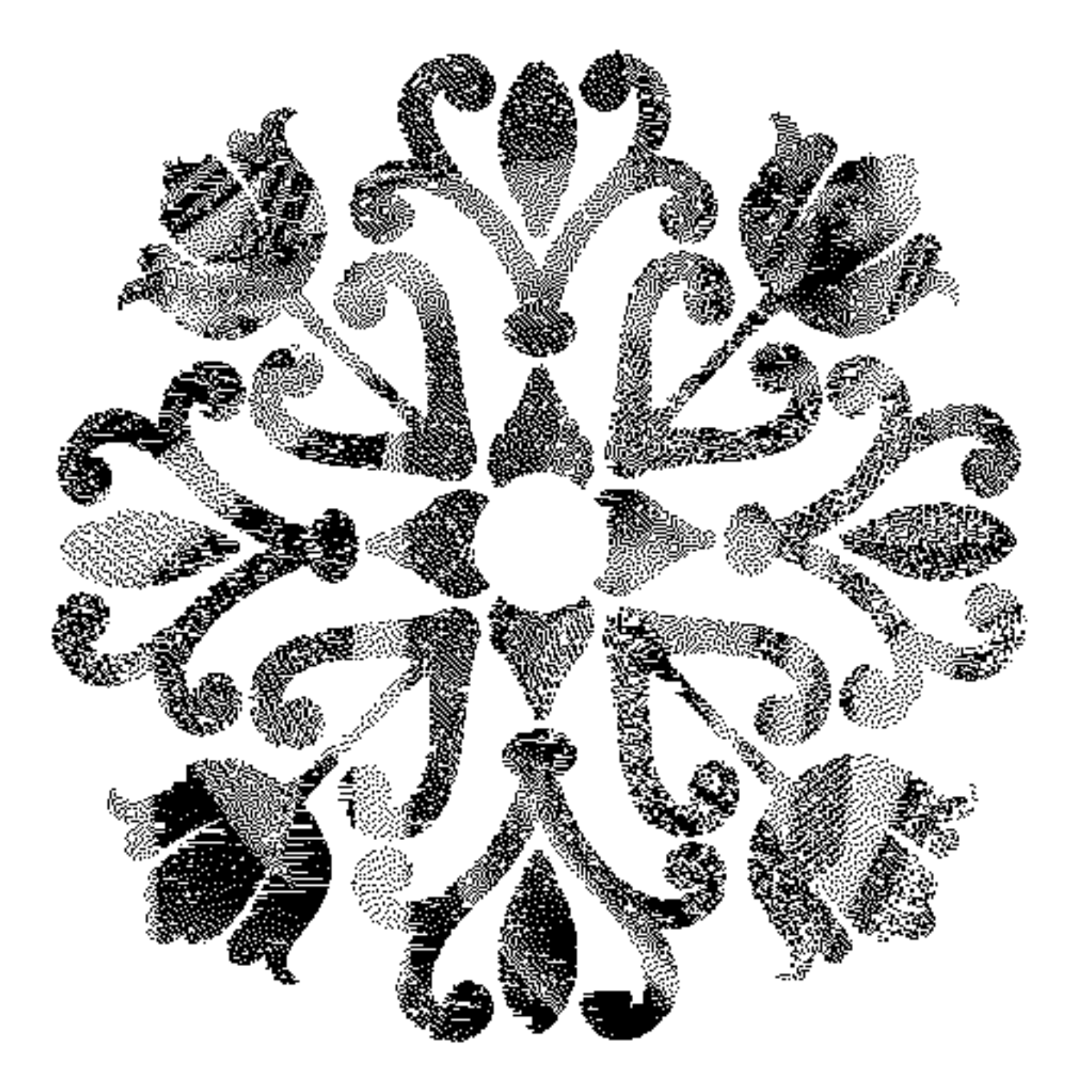}\label{fig:deedl2xnor}}}
  \subfigure[]{
    \setlength{\fboxrule}{1pt} \setlength{\fboxsep}{0cm}
    \fbox{\includegraphics[width=3.0cm]{0_0016barbara_cadeedecxnoroverlays.pdf}\label{fig:cadeedecxnor}}}
  \subfigure[]{
    \setlength{\fboxrule}{1pt} \setlength{\fboxsep}{0cm}
    \fbox{\includegraphics[width=3.0cm]{0_0008barbara_cadeednixnoroverlays.pdf}\label{fig:cadeednixnor}}}
\vspace{-0.2cm}
\caption{XNOR operation decoded image, $X_1=X_2=`Barbara'$, Steinberg kernel. (a)DHCED; (b)DHDCED; (c)DEED(L-2); (d)CaDEED-EC; (e)CaDEED-N\&I.
\label{fig:subjectivexnordecoded}}
\vspace{-0.3cm}
\end{figure*}

After the numerical comparisons, some visual comparisons will be carried out.
Figs. \ref{fig:subjectiveanddecoded}-\ref{fig:subjectivexnordecoded} presents the revealed secret patterns of DHCED, DHDCED, DEED(L-2), CaDEED-EC and CaDEED-N\&I, respectively, with the cover images $X_1=X_2=`Barbara'$. With approximately equivalent NT-PSNRs, Fig. \ref{fig:subjectiveanddecoded} gives the AND operation decoded results, while Fig. \ref{fig:subjectivexnordecoded} reveals the XNOR operation decoded results. As we can observe, CaDEED-EC and CaDEED-N\&I give better contrasts of the revealed secret pattern. Note that CaDEED-N\&I preserves better fine structures of the watermark. For example, in the lower right region of each decoded image where the lady's left leg co-located, CaDEED-N\&I presents the most clear revealed flower pattern among all the results.

\subsection{Discussion of Robustness}

In real distribution and transmission process, halftone images mainly suffer from cropping, human-marking attacks and print-and-scan distortions. For cropping or human-marking attacks, part of the decoded secret pattern which has been cropped or human-marked will be lost and others can still be preserved, as the experiment shows in [\ref{au2003dhced}].

For the print-and-scan distortions, different watermark decoding approach suffers differently. If the users employ the overlaying(AND) operation for decoding, HVW methods will usually be affected by the print distortions only, because no scanning is involved during the decoding process. On the other hand, if the users choose to decode the watermark with the XNOR operation, HVW methods will suffer both the print and scan distortions.

\begin{figure*}
  \centering
  \subfigure[]{
    \includegraphics[width=5.8cm]{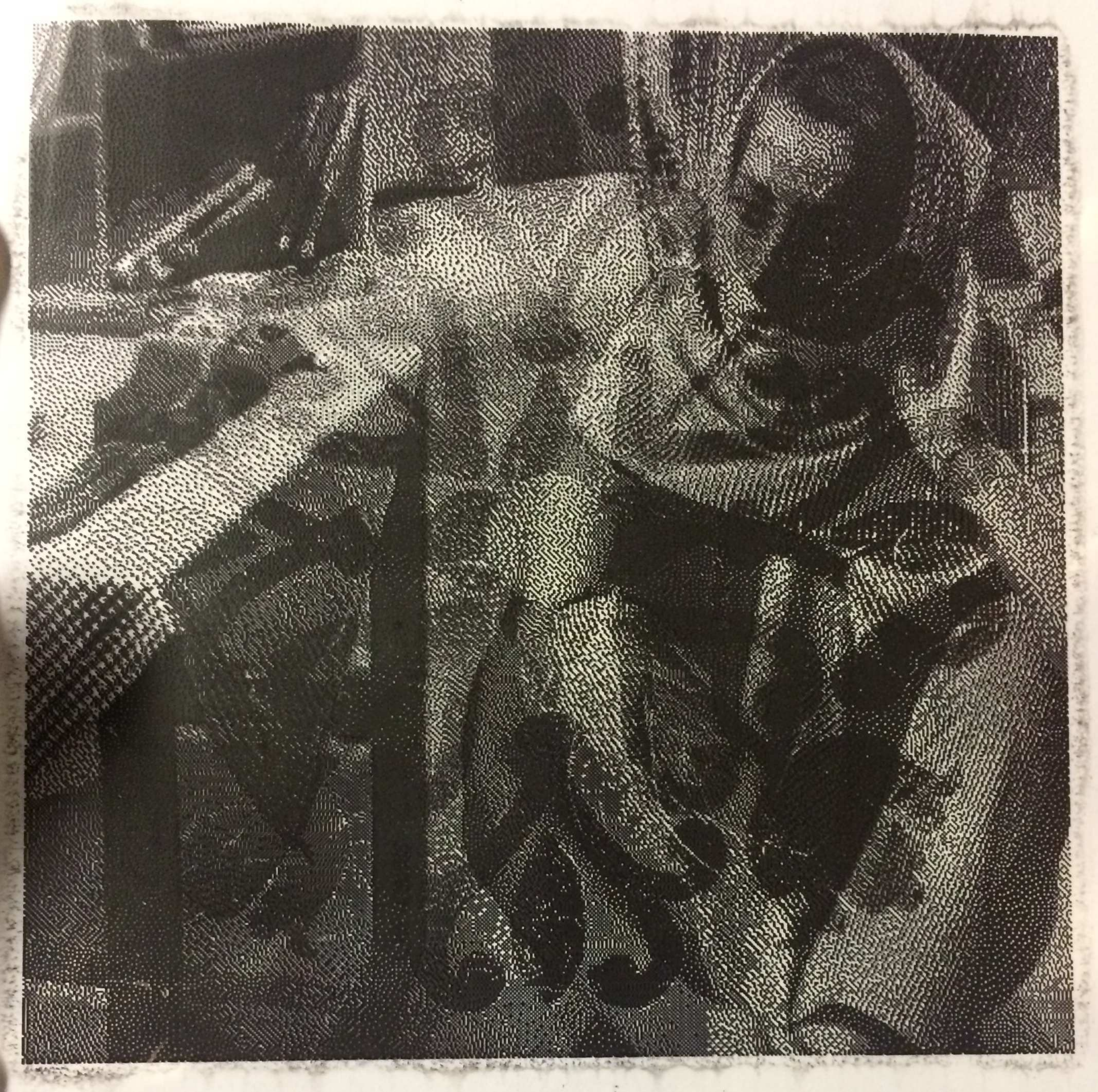}\label{fig:psanddecoded}}
  \subfigure[]{
    \setlength{\fboxrule}{1pt} \setlength{\fboxsep}{0cm}
    \fbox{\includegraphics[width=5.8cm]{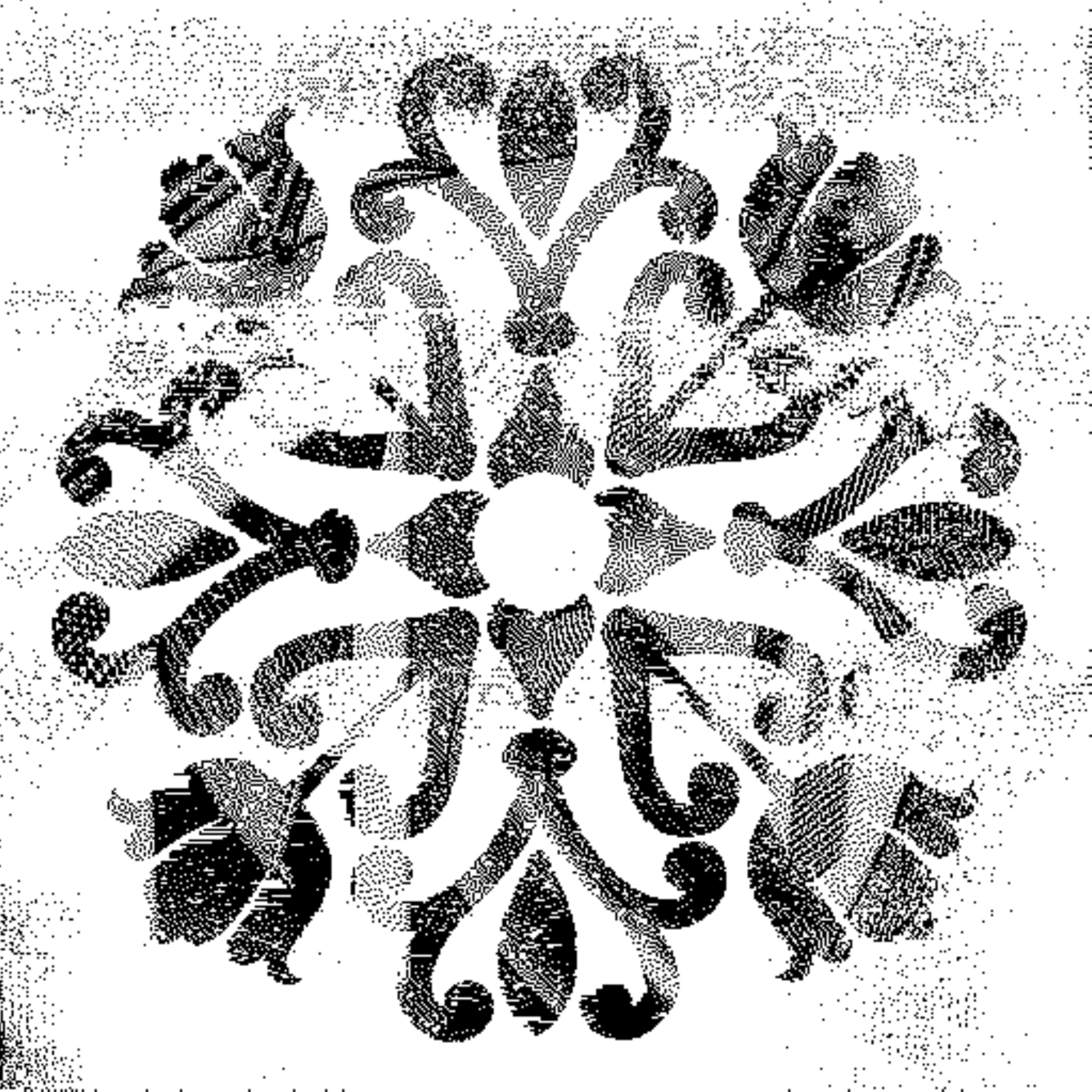}\label{fig:psxnordecoded}}}
\vspace{-0.2cm}
\caption{A real example for CaDEED-N\&I. $Y_1$ and $Y_2$ are from Figs. \ref{fig:cbdcedy12} and \ref{fig:cbdcedy22}, respectively. (a) AND operation decoded image (which only suffered from print distortions).
(b) XNOR operation decoded image (which suffered from print and scan distortions).
\label{fig:psdecoded}}
\vspace{-0.3cm}
\end{figure*}

Figs. \ref{fig:psdecoded} gives a real example, where the stego halftone images $Y_1$ and $Y_2$ are from Figs. \ref{fig:cbdcedy12} and \ref{fig:cbdcedy22}, respectively. For Fig. \ref{fig:psanddecoded}, $Y_1$ and $Y_2$ are printed with a resolution of 96 dot-per-inch(DPI), where the printer model HP LaserJet 4250 is employed, on the transparencies for the convenience of demonstration. Note that we place a white paper below as the background to better show the overlaid result. As we can observe, the overlaid result, i.e. the AND operation decoded result possesses decent decoded watermark result and we can clearly distinguish the majority of the secret pattern. For Fig. \ref{fig:psxnordecoded}, $Y_1$ and $Y_2$ are printed on regular A4 papers with the resolution of 96 DPI and the printer model HP LaserJet 4250. After printing, HP ScanJet G2410 is employed for scanning $Y_1$ and $Y_2$ with a resolution of 1200 pixel-per-inch(PPI). Before the XNOR decoding operation is performed, a series of preprocessing steps [\ref{guo2016hvw}] including rotation, resizing and 1-bit quantization are performed with the help of Adobe Photoshop CS2, to reverse the print-and-scan distortions. As Fig. \ref{fig:psxnordecoded} shows, the decoded result reveals most of the secret pattern, though some noise like regions exist. We also measure Fig. \ref{fig:psxnordecoded} with CDR and CBCDR. It turns out that Fig. \ref{fig:psxnordecoded} achieves excellent decoding performances with $\textrm{CDR}=82.69\%$ and $\textrm{CBCDR}=82.09\%$.

\section{Conclusion}\label{conclusion}

In this paper, we firstly analyzed the expected performances and limitations of the EDHVW methods. Based on the analysis, we proposed a new general EDHVW method, Content aware Double-sided Embedding Error Diffusion, via considering the expected performances which is affected by the content of the cover images and watermark (secret pattern), the different noise tolerance abilities of different cover image content and the different importance levels of different pixels (when being perceived) in the secret pattern. To demonstrate the effectiveness of the proposed CaDEED, CaDEED-EC and CaDEED-N\&I are proposed. CaDEED-EC considered the expected performances only. CaDEED-N\&I exploited more by adopting the noise visibility function [\ref{volo2000nvf}] and proposing the importance factor (IF) for different watermark pixels. In the experiments, the validation tests for CaDEED-EC and CaDEED-N\&I were performed first. Then, after selecting the optimal local region sizes for CaDEED-N\&I, extensive comparison tests were carried out. The performances of the proposed methods and the previous methods were not only measured by the traditional PSNR and CDR measurement, but also measured by our proposed measurement, CB-CDR and NT-PSNR, to further illustrate the significance of the proposed method. Both the numerical and visual comparisons indicated that our proposed work outperforms the classical and latest EDHVW methods.

\end{document}